\documentclass[twoside,openright]{report}

\usepackage[dvips]{graphicx}
\usepackage{amsmath}
\usepackage{amssymb}
\usepackage{eepic}
\usepackage{color}
\begin{document}
\title{{\Huge Near Field Speckles}}
\author {Doriano Brogioli}
\maketitle

%
\chapter{Introduction.}

Elastic light scattering (LS) has been extensively used to study samples
showing a non uniform refraction index on lengthscales from a fraction
of a micrometer to a fraction of a millimeter. Basically, a wide laser
beam is sent through the sample, and the light scattered at any angle
is measured by a detector in the far field. Many phenomena have been 
studied with this technique; among them, thermodynamical fluctuations
of concentration in solutions, of temperature, of pressure, convective 
instabilities and turbulence, colloids and colloidal gels, systems
showing critical opalescence. 

The measure of the intensity of the beams scattered by a target is a
well known way for analysing physical properties of a sample, and
is used in many fields of physics: examples are X ray diffraction 
in crystallographic analysis and particle scattering by nuclei in the
Rutherford experiment. In these examples, as in light scattering,
the intensity of the scattered beams gives informations on
the square modulus of the Fourier transform of a quantity of interest
of the sample, the so called power spectrum, evaluated
in the transferred wavevector. In the case of light
scattering, the quantity of interest
is the refraction index. Since refraction index shows variations due
to concentration, temperature and pressure fluctuations, all these
quantities can be investigated using light scattering.

In principle, measuring the intensity of the light scattered at an arbitrary
small angle allows to obtain informations on arbitrary long wavelength
Fourier components of the sample. In facts, the measurement of the
intensity of the scattered light becomes more and more difficult as
the scattering angle becomes small, mainly due to the stray
light, that is the light scattered by the imperfections of the optical
system, which is mainly scattered at small angles. Thus light scattering
cannot give informations on the features
of a sample, if their associated wavelenght is longer than a given
value. On the other hand, image forming techniques, such as Schlieren,
dark field, phase contrast microscopy, have no limitations on the
size of the features they can observe. The limitation of
many image forming techniques is the difficulty to quantitatively relate the
observed images to the physical properties, mainly when dealing 
with three dimensional samples.

In the present work, we describe three new techniques, which
allow to measure the scattering intensities, overcoming the 
difficulties associated to small angles.
The first of these techniques,
the hOmoyne Near Field Speckles (ONFS), has been
presented very recentely \cite{carpineti2000}; in the present work,
we show new results, obtained with a slightly improved optical setup
\cite{carpineti2001}.
The other techniques,
the hEterodyne Near Field Speckles (ENFS) and the Schlieren-like Near
Field Speckles (SNFS)
are improvements based on that. The first has been recently patented by us
\cite{brevetto,giglio2001}; the second is presented here for the first time.
Moreover, in Chapt. \ref{chap_theory}, we present for the first time a 
mathematical derivation of the working formulas for the three techniques.

Basically, the experimental setup
consists in a wide laser beam passing through the sample; a lens forms
an image of a plane at a given distance from the cell on a CCD sensor.
The image, in the near field, shows speckles, since it
is formed by the stochastical interference of the light coming from a
random sample: the electric field has a gaussian probability distribution.
We will show that, under suitable conditions, the correlation function of
such a field closely mirrors the correlation function of the investigated
sample; moreover, in general, from the correlation function
of the speckle field we can calculate the scattered intensity.

The lens that forms the image on the CCD focuses the transmitted beam
around a given point. In ONFS, a beam stop is placed in that point, in
order to dispose of the transmitted beam; in SNFS a blade stops half
transmitted beam, along with one half of the scattered light, like in
Schlieren technique; in ENFS no opaque element is introduced in the
optical system. In ONFS, the CCD sees the speckles given by the
interference of the scattered beams with themselves: ONFS is an
homodyne technique . In ENFS and SNFS,
the speckle field is heterodyned with the much more intense
transmitted beam, that acts as a reference beam: the measured
intensity is linear in the speckle electric field.  We acquire a set
of images, from one of the three techniques, by using a CCD camera,
connected to a frame grabber; the images are then elaborated by a PC,
to obtain $I\left(\vec{q}\right)$, the scattered intensity as a
function of the transferred wavevector, the same information
obtained by LS. For each technique, we developed algorithms
which allow to evaluate the scattered intensities, including the
corrections for the stray light and for the shot and read noise of the 
CCD camera. The algorithms are described in Chapt. 
\ref{chap_onfs_data_processing} and \ref{chap_enfs_data_processing}.

We used ONFS and ENFS to measure the scattered intensity of some
colloids, and we compared the results with those made by a
state-of-the-art classical Small Angle Light Scattering (SALS)
device. The agreement is very good, notwithstanding a much simpler
and stable layout. In Chapt. \ref{capitolo_confronto_ONFS_ENFS} we compare 
ONFS, ENFS and SALS measurements, and discuss the main sources of errors.
Scattering intensities measured with ENFS show a better quality; we
used this technique to evaluate the diameter distribution of some
known colloids, by using an inversion algorithm based on Mie theory.
The results are
presented in Chapt. \ref{capitolo_particle_sizing_ENFS}; this
shows that ENFS is a simple and  powerful alternative to 
SALS, suited for industrial applications like particle sizing.
Moreover, SNFS has been used to evaluate the power spectrum of non-equilibrium
fluctuations in a free diffusion process, thus showing that such
techniques have interesting applications in fundamental physics; results
are shown in Chapt. \ref{capitolo_dinamico_SNFS}.

%
\chapter{Qualitative description of the technique.}

The intensity of the light scattered from a spatially disordered
sample has a speckled appearance, the speckles being generated by
the random interference of the scattered elementary spherical
waves. While the study of the one point intensity time correlations
has 
proven very useful, and it has generated the technique of Intensity
Fluctuation Spectroscopy (IFS) \cite{benedek}, the measurement of the
two point, equal time, intensity space correlation function, that is
the size and the
shape of the speckles, does not provide any useful information.
Indeed the Van Cittert and Zernike theorem states that the
\emph{far field}  space correlation function depends only on the
intensity distribution of the scattering volume, and in no way
depends on the physical properties of the sample.

In this chapter we will present qualitative elements showing that for
fluctuations the size of the wavelength of light or larger, in the
\emph{near field} we obtain a speckle field, that is, a gaussian field;
moreover its statistics is directly related to the scattered intensity
distribution. We will derive the working formulas for
three tecniques, hOmodyne Near Field Speckles (ONFS), hEterodyne NFS (ENFS)
and Schlieren-like NFS (SNFS);
analogies with the IFS will be pointed out. Advantages
with respect to the more conventional Small Angle 
Light Scattering (SALS) technique
will be discussed.

First of all, we will describe ONFS setup; many considerations hold
also for ENFS and SNFS. The experimental set-up is very unorthodox,
with respect to a conventional SALS device. It
consists of a wide laser beam and of a Charge Coupled Device (CCD)
detector positioned so to be flooded with light coming from any
scattering direction the system can scatter at. 

The Van Cittert and Zernike theorem states that the
field correlation function is \cite{goodman_2}:

\begin{eqnarray}
C_E (\Delta x, \Delta y) = \left< E \left(x,y\right)
E^*(x+\Delta x,y+\Delta y)\right>
= \int{ \int I(\xi ,
\eta)exp \left[ i \frac{2 \pi}{\lambda z} \left( \xi \Delta x +
\eta \Delta y \right) \right] d \xi d \eta}
\label{VCZ}
\end{eqnarray}

where $E\left(x,y\right)$ is
the field in the observation plane $x - y$, $\lambda$ is the
wavelength and $I \left(\xi, \eta \right)$ is the actual intensity
distribution of the source in the plane $\xi - \eta$ at a distance
$z$ from the observation plane. The theorem holds for sources
consisting of point emitters, like atoms. The intensity
correlation function $C_I \left( \Delta x, \Delta y \right)=
\left< I\left(x,y\right) I\left(x+\Delta x,y+\Delta y\right)\right>$ is
then derived by applying the so called Siegert relation
\cite{dainty}:

\begin{eqnarray}
C_I \left( \Delta x, \Delta y \right)= \left< I \right>^2 +
|C_E \left( \Delta x, \Delta y \right)|^2 
\label{SGT}
\end{eqnarray}

Equations (\ref{VCZ}) and (\ref{SGT}) specify that the intensity
correlation function is related to the space Fourier transform of
the source. In practice, this implies that a source of size $D$
will generate speckles of size $\frac{\lambda}{D} z$ on a screen
positioned at a distance $z$ \cite{dainty}.

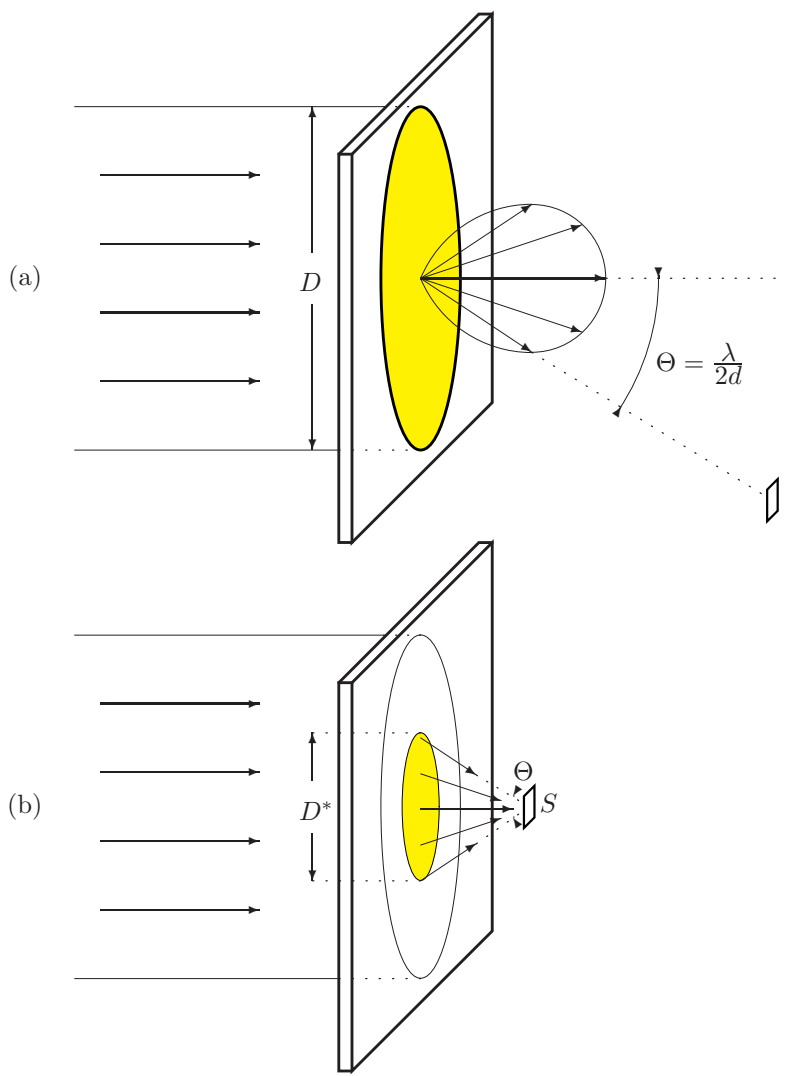
\begin{figure}
\begin{tabular}{cc}
(a)&
\parbox[c]{10cm}{
\begin{picture}(250,200)(-100,0)
\path(-100,165)(17,165)
\path(-100,35)(0,35)
\dottedline{5}(0,35)(31,35)
\dottedline{5}(17,165)(31,165)
\multiput(-90,61)(0,26){4}{\vector(1,0){60}}
\put(-10,90){\vector(0,-1){55}}
\put(-10,110){\vector(0,1){55}}
\put(-15,95){$D$}
\dottedline{5}(101,100)(170,100)
\put(31,100){\arc{180}{0.0}{0.5880026}}
\put(121,101){\vector(0,-1){1}}
\put(105.88453,50.076982){\vector(2,3){1}}
\put(120,65){$\Theta=\frac{\displaystyle \lambda}{\displaystyle 2d}$}
\Thicklines
\put(31,100){\blacken\color{yellow}\ellipse{30}{130}}
\path(5,0)(58,53)(58,200)(5,147)(5,0)
\path(5,0)(0,0)(0,147)(5,147)
\path(0,147)(53,200)(58,200)
\thinlines
\put(31,100){\vector(3,-2){42}}
\put(31,100){\vector(3,2){42}}
\put(31,100){\vector(1,0){70}}
\put(31,100){\vector(3,1){61}}
\put(31,100){\vector(3,-1){61}}
\put(73,100){\arc{56}{-1.5707963}{1.5707963}}
\put(73,82.5){\arc{91}{-2.7468015}{-1.5707963}}
\put(73,117.5){\arc{91}{1.5707963}{2.7468015}}
\dottedline{5}(73,72)(164,16)
\thicklines     
\blacken\color{white}\path(162,8)(162,20)(166,24)(166,12)(162,8)
\end{picture}
}
\\
(b)&
\parbox[c]{10cm}{
\begin{picture}(250,200)(-100,0)
\path(-100,165)(17,165)
\path(-100,35)(0,35)
\dottedline{5}(0,35)(31,35)
\dottedline{5}(17,165)(31,165)
\multiput(-90,61)(0,26){4}{\vector(1,0){60}}
\put(31,100){\ellipse{30}{130}}
\put(31,100){\blacken\color{yellow}\ellipse{14}{56}}
\put(31,72){\vector(3,2){21}}
\put(31,126){\vector(3,-2){21}}
\put(31,99){\vector(1,0){35}}
\put(31,112.5){\vector(3,-1){30.5}}
\put(31,85.5){\vector(3,1){30.5}}
\dottedline{5}(-10,128)(31,128)
\dottedline{5}(-10,72)(31,72)
\put(-10,90){\vector(0,-1){18}}
\put(-10,110){\vector(0,1){18}}
\put(-15,95){$D^*$}
\dottedline{5}(52,112)(72,100)
\dottedline{5}(52,86)(72,100)
\put(66,104){\vector(-2,-3){0.01}}
\put(66,96){\vector(-2,3){0.01}}
\put(66,110){$\Theta$}
\put(76,98){$S$}
\Thicklines
\path(5,0)(58,53)(58,200)(5,147)(5,0)
\path(5,0)(0,0)(0,147)(5,147)
\path(0,147)(53,200)(58,200)
\thicklines
\blacken\color{white}\path(70,92)(70,104)(74,108)(74,96)(70,92)
\end{picture}
}
\end{tabular}
\caption [Formation of near field speckles]{(a) Small angle scattering. 
A beam of diameter $D$
impinges onto a sample composed of particles of diameter $d$. Any
zone within $D$ will scatter light into a lobe of angular width
$\Theta = \lambda/d$ (the length of arrows indicates scattered
intensity). (b) Same sample, as in part (a). A sensor $S$ close
enough to the sample will draw light only from a zone of radius
$D^* < D$. Regions outside, even if illuminated by the main beam,
do not feed light to $S$. Notice that again $\Theta = \lambda/d$.}
\label{euristico_fig_1}
\end{figure}

We will start introducing simple euristic arguments and crude
evaluations for the near field speckles of the scattered light.
Let us consider the case of a large beam diameter $D$,
impinging onto a sample of particles of diameter $d$ larger than
the wavelength of light: see Figure \ref{euristico_fig_1}(a). Most of
the power will be
scattered in a forward lobe of angular width $ \Theta \approx
\frac{\lambda}{d} $.
%
%
Let us consider a small area $S$, for example a
multi-element sensor array, in the immediate vicinity of the
scattering volume: see Figure \ref{euristico_fig_1}(b). Let us assume
that we can ignore
the transmitted beam: we will take care of this problem later on.
Although the sample is illuminated over the entire surface of
diameter $D$, the light falling onto the sensing area will come
only from a smaller area of diameter $D^*$ . In fact the
brightness of the scattering volume will change as a function of
the observation angle in a way that mirrors the scattered
intensity distribution. Consequently, for the sensing area, the
source region from which light is drawn is a circle with a
diameter $D^*=\frac{\lambda}{d}z $, $z$ being the distance of the
sensing area from the scattering surface; source regions outside
do not contribute appreciably. We say that the near field condition
is met if $D^*\ll D$. One can then immediately estimate
the size of the speckles $d_{sp} = \frac{\lambda}{D^* } z \approx
d$, a remarkable result in many respects! The speckles have the
size of the particle diameter, and this value does not depend on
the distance $z$ from the sample, provided the near field condition
$ D^* \ll D$. This
has to be compared with far field speckles, whose diameter scales
linearly with the distance from the source. Also notice that the
actual sample thickness does not matter, provided that the near
field condition is met, and that the speckle size does not depend
on the light wavelength, an unexpected feature for an interference
pattern.

Notice that all the above applies under conditions that are
more stringent than the usual ``near field'' condition \cite{yariv}
for a source of size $D$ , namely $\frac{ \lambda z}{D^2} \ll 1$.
In the present case the condition is $D^*\ll D$ which implies
$\frac{ \lambda z}{D d}\ll 1$.

To put things in a more quantitative way, we will determine the
near field intensity correlation by first re-writing the Van Cittert and
Zernike
theorem in a more appropriate form. We notice that Eq. (\ref{VCZ})
may be rewritten in the following way:
\begin{eqnarray}
C_E ( \vec{r}) = \int{I\left(\vec{q}\right) e^{i\vec{q} \cdot
\vec{r} }\mathrm{d}\vec{q}}, \label{FT}
\end{eqnarray}
where $\vec{r} =\left( \Delta x, \Delta y \right)$, and
$\vec{q}$ is a vector whose components are $ q_x =
\frac{2\pi }{\lambda z} \xi$ and $q_y = \frac{2\pi }{\lambda z}
\eta$, which equals the scattering wave vector for small
scattering angles.

Equation (\ref{FT}) is only a different way of writing Eq. (\ref{VCZ}), and
$I\left(\vec{q}\right)$ is the intensity distribution of the source as seen
from the observation plane as a function of the scaled angles
$\left( 2 \pi /\lambda\right)\left(\xi /z\right)$, and $\left( 2 \pi
/\lambda\right)\left(\eta /z\right)$.
As discussed in the introductory
remarks, in the very near field $I\left(\vec{q}\right)$ equals the
scattered intensity distribution, which is proportional to the Fourier
transform of
the sample density correlation function $g\left(\vec{r}\right)=
\left<\delta l\left(\vec{r}\right)
\delta l\left(0\right) \right>$, where $\delta l$ is the local
fluctuation of the particle number density, integrated over the light
path. Then, from Eq. (\ref{SGT}), it follows that:
\begin{eqnarray}
C_I\left(\vec{r}\right) = \left< I\right>^2\left[1 + 
\left|g\left(\vec{r}\right)\right|^2 \right] .
\label{RI}
\end{eqnarray}
We would like to point out that Eq. (\ref{RI}) closely duplicates
the well known relation that holds for the IFS
$<I(0)I(t)>=<I>^2[1+|g(t)|^2]$ , where $g(t)$ is the
time correlation function (see for example \cite{pecora}). It
should be noted that for some scatterers the Rayleigh Gans
approximation is invalid, for example for larger spheres where the
Mie theory applies, and therefore the pair correlation function
$g(r)$ cannot be extracted from the scattered light. It remains
true however that even in those cases the correlation method
permits the determination of the scattered intensity distribution
$I\left(q\right)$, by using the relation:
\begin{eqnarray}
C_I\left(r\right) = \left< I\right>^2\left[1 + \left|\int{I\left(q\right)
e^{i\vec{q} \cdot \vec{r} }\mathrm{d}\vec{q}} 
\right|^2 \right] .
\label{qualitativa_eq_ONFS_C_I_to_I_di_q}
\end{eqnarray}

To determine the spatial intensity correlation of Eq. (\ref{RI}), one
must first obtain experimentally the instantaneous intensity
distribution  of the near field scattered light.  In order to
evaluate the intensity correlation function with reasonable
statistical accuracy it is also imperative to gather intensity
distributions over a substantial number of points. To this end a
CCD is ideal, the number of pixel being larger than $10^5$.  As we
shall see, it actually turns out that one frame is enough for a
fair acquisition of the correlation function.

\begin{figure}
\includegraphics[scale=0.5]{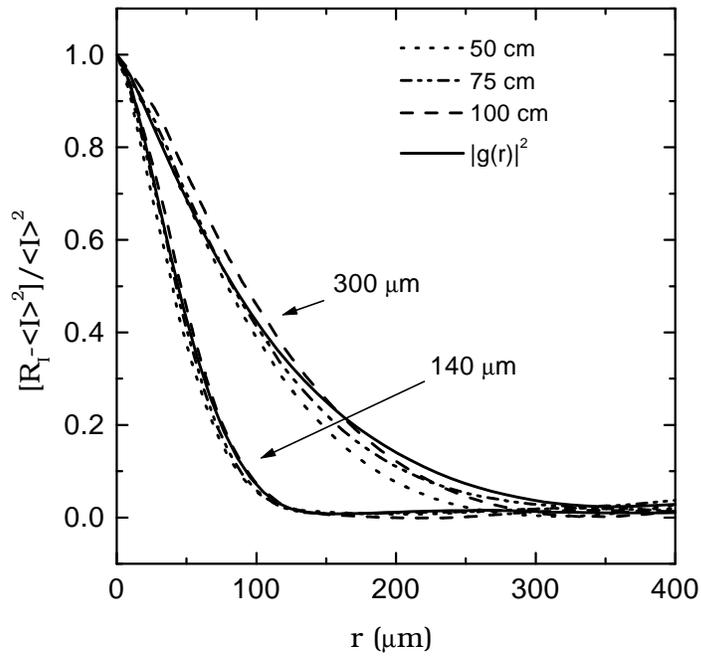}
\caption {Measured intensity auto-correlations as a function of
displacement $r$ for two sets of randomly positioned pinholes, of
$140\mathrm{\mu m}$ and $300\mathrm{\mu m}$ in diameter. For both the samples,
measurements at three distances are reported, together with
$\left|g\left(r\right)\right|^2$, calculated from the digitized images
of the two samples. } 
\label{euristico_fig_2}
\end{figure}

In a previous work \cite{carpineti2000}, 
some measurements have been performed on a  scattering model, an opaque
metallic screens with pinholes of 140 and 300 microns chemically
etched in random positions. The surface fraction occupied by the
pinholes was around 10\% and 20\% respectively. Experimentally this
greatly simplifies the problem, since the scattered field is
stationary and also there is no transmitted beam.
We call this configuration hOmodyne Near Field Speckles, since the signal
is given by the interference of different scattered beams.
Being a two
dimensional sample, the scattered intensity was simply related to
the correlation function of the transparency function  $T\left(x,y\right)$
with $T=1$ inside the pinholes and zero outside \cite{goodman_2}. A
Helium Neon parallel beam with diameter ($\frac{1}{e^2}$ points)
$D= 15\mathrm{mm}$ was sent onto the samples, and the speckle field was
recorded with a CCD at various distances $z=50\mathrm{cm}$, 
$z=75\mathrm{cm}$  and
$z= 100\mathrm{cm}$ 
\footnote{For very large objects the near field is not really near,
since one must let the diffraction figures from various objects interfere. This
leads to the additional condition $D^* = \frac{\lambda }{d} z\gg
\delta$, where $\delta$ is the typical distance between
scatterers.}.
The corresponding values for $D^*$ ranged
from $1\mathrm{mm}$ to $4.3\mathrm{mm}$
so that the very near field condition was
always met. The rather large dimension of the pinholes was chosen
so that the speckles were appreciably larger than the CCD pixel
size (typically $9\mathrm{\mu m}$). For each type of pinholes, the
measurements performed at the three distances showed minute
differences. The results are shown in Fig \ref{euristico_fig_2}, where
the data
are compared with the correlation functions of digitised images of
the set of pinholes on the metallic screen 
\footnote{Experimental correlation functions shown in figure are
calculated according to the following procedure. First,
the Fast Fourier Transform of the intensity distribution is
calculated.  Then the result is squared. The auto-correlation is then
obtained by anti-transforming the power
spectrum. \label{qualitativa_nota_2}}.
Since in this case the sample is two-dimensional,
$g\left(\vec{r}\right)$ is the correlation function of
$T\left(\vec{x}\right)$.
The width
and shape of the main peak are fairly well reproduced, in spite of
the limited number of frames used (four frames on statistically
equivalent samples for each type of pinholes).

While the data obtained with the screens prove that near field
speckles do mirror the properties of the scatterers, we feel that
to assess the desirability of the technique for realistic
applications (for example in colloid physics) measurements had to
be taken with particle solutions down in the micron range. 
In order to do this, three problems had to be solved. The speckles in
the near field close to the cell have dimensions around one micron
and therefore are too small for the available CCD pixel size.
Also, one must dispose of the transmitted beam. Finally, the
speckle intensity distribution must be frozen at a given instant.

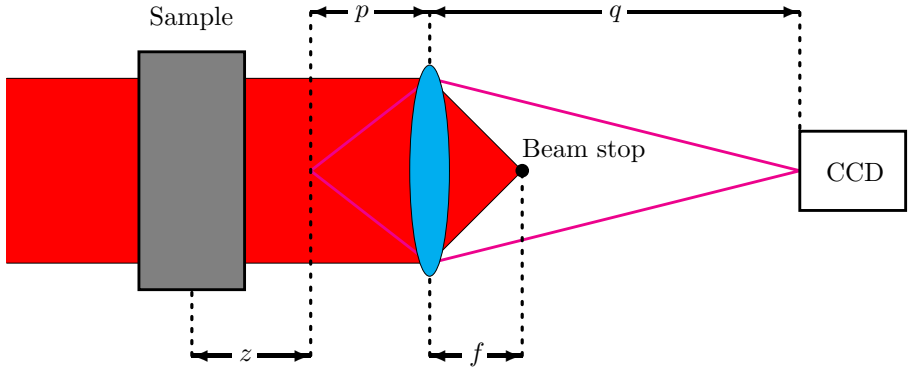
\begin{figure}
\begin{center}
\begin{picture}(350,150)(0,0)
\blacken\color{red}\path(0,40)(160,40)(160,110)(0,110)
\blacken\color{red}\path(160,40)(195,75)(160,110)
\Thicklines
\color{magenta}\path(160,40)(115,75)(160,110)(300,75)(160,40)
\thinlines
\put(160,75){\blacken\color{cyan}\ellipse{15}{80}}
\Thicklines
\shade\path(50,30)(90,30)(90,120)(50,120)(50,30)
\put(54,130){Sample}
\dottedline{5}(70,5)(70,29)
\dottedline{5}(115,5)(115,135)
\dottedline{5}(160,5)(160,34)
\dottedline{5}(195,5)(195,74)
\dottedline{5}(160,116)(160,135)
\dottedline{5}(300,91)(300,135)
\whiten\path(300,60)(340,60)(340,90)(300,90)(300,60)
\put(310,71){CCD}
\put(88,3){$z$}
\put(175,3){$f$}
\put(132,133){$p$}
\put(228,133){$q$}
\put(85,5){\vector(-1,0){15}}
\put(95,5){\vector(1,0){20}}
\put(172,5){\vector(-1,0){12}}
\put(182,5){\vector(1,0){13}}
\put(129,135){\vector(-1,0){14}}
\put(139,135){\vector(1,0){21}}
\put(225,135){\vector(-1,0){65}}
\put(235,135){\vector(1,0){65}}
\put(195,75){\circle*{4}}
\put(195,80){Beam stop}
\end{picture}
\end{center}
\caption {Optical layout for ONFS. The main transmitted beam is
blocked by a stop in the focal plane. Almost all the scattered light 
is sent to the CCD.}
\label{euristico_fig_3}
\end{figure}

The first two problems have been solved with the simple optical
arrangement shown in Fig. \ref{euristico_fig_3}.
A wide parallel beam is sent onto the sample, placed against a large
aperture lens of focal length $f$. A wire is stretched in the focal plane to
intercept the main beam. The CCD is placed a distance $q$ away
from the lens. The system magnifies the speckle size by a factor
$M=(q-f)/f >1$, and the scattering angles are decreased accordingly.
It is illuminating to point out that the technique can be
considered as a scaled down version of the classical Hanbury Brown
and Twiss \cite{hanbury} experiment where the star intensity
distribution is mimicked by the scattered light intensity patch in
the focal plane, and the ground based intensity correlation are
the CCD intensity correlations. The unavoidable presence of the
lens and its finite aperture introduces some complication with
respect to the lensless arrangement used for the pinholes. In a
previous work \cite{carpineti2000}, the poor numerical aperture
of the lens introduced a non-uniform transfer function, which had to
be evaluated by measuring a known sample. In the present work, care
has been taken in order to avoid such problems.

When the scattered speckles
are observed with the CCD in real time, one notices quite vividly
that the speckle size changes as the size of the scatterers is
changed. Also, for a given sample the speckles boil with the
same time constant on the whole screen, the time constant getting
larger for samples with larger diameter particles. With regard to
the third problem mentioned above, these observations also
indicate that even with a conventional CCD and a small power He-Ne
laser there is no problem in getting instantaneous pattern
distributions. Indeed even for the smallest particles that can be
studied with present experimental set-up, with diameters
down to $1\mathrm{\mu m}$, and
assuming diffusive motion, the shortest time constant associated
to the smallest scattering wavevector yields $\tau_{min}= 0.125\mathrm{s}$,
a time long compared with the shortest frame exposure
available with standard frame grabbers, typically $1/16000\mathrm{s}$.

Let us compare the Near Field Speckles technique with the more traditional
Small Angle Light Scattering. The essential feature of a scattering layout
\cite{carpineti1990,ferri1997} is that the light scattered at a given angle
hits the sensors along a circle of given diameter around the
optical axis. We believe that the correlation method of NFS offers some
distinct advantages over the scattering technique. First, there is
no need for accurate positioning  of the CCD, that can be rather
casually placed at a distance $z$ from the focal plane (see Fig.
\ref{euristico_fig_3}). At variance, in SALS one has to know
the precise relation between pixels and scattering angles and this
is troublesome when the distance $z$ is changed to select a new
particle diameter instrumental range. Also, and more important,
SALS is plagued by stray light. To mitigate its
effects, one has to rely on blank measurements to be subtracted
from raw scattering data. The trouble is that stray light is worst
at smaller angles, where the sensing elements are necessarily in
small number and crowded close to the optical axis. With the
present technique, on the contrary, all the pixels are used in
calculating the correlation function for any value of the
displacement $r$ and this allows more accurate stray light
subtraction; the algorithms to subtract the stray light will be described
in Chapter \ref{chap_onfs_data_processing}.

The results of the measurements on some colloid samples are presented
in Chapter \ref{capitolo_confronto_ONFS_ENFS}. The ONFS technique in the
present form has only one tight requirement, 
namely the clean disposal of the transmitted beam that requires
accurate focusing and a proper diffraction limited beam stop. It is
both conceptually and in practice very simple, and it capitalizes on
the high statistical accuracy permitted by the large number of pixels
of a CCD and  by the good handling capabilities of PCs.

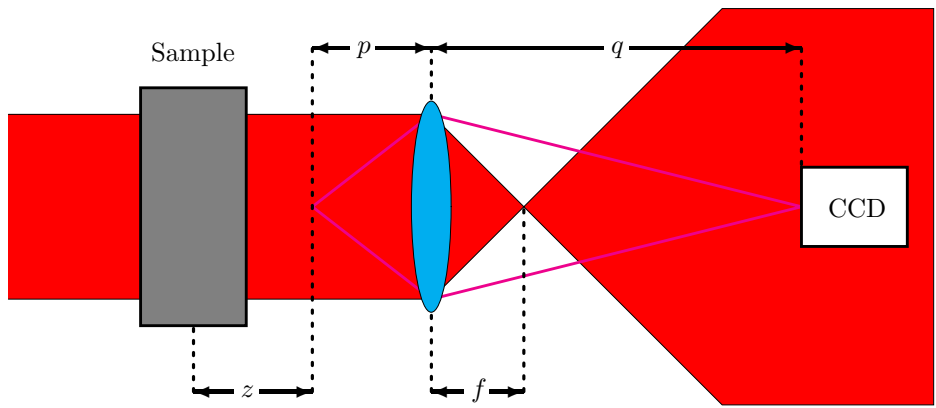
\begin{figure}
\begin{center}
\begin{picture}(350,150)(0,0)
\blacken\color{red}\path(0,40)(160,40)(160,110)(0,110)
\blacken\color{red}\path(160,40)(270,150)(350,150)(350,0)(270,0)(160,110)
\Thicklines
\color{magenta}\path(160,40)(115,75)(160,110)(300,75)(160,40)
\thinlines
\put(160,75){\blacken\color{cyan}\ellipse{15}{80}}
\Thicklines
\shade\path(50,30)(90,30)(90,120)(50,120)(50,30)
\put(54,130){Sample}
\dottedline{5}(70,5)(70,29)
\dottedline{5}(115,5)(115,135)
\dottedline{5}(160,5)(160,34)
\dottedline{5}(195,5)(195,74)
\dottedline{5}(160,116)(160,135)
\dottedline{5}(300,91)(300,135)
\whiten\path(300,60)(340,60)(340,90)(300,90)(300,60)
\put(310,71){CCD}
\put(88,3){$z$}
\put(175,3){$f$}
\put(132,133){$p$}
\put(228,133){$q$}
\put(85,5){\vector(-1,0){15}}
\put(95,5){\vector(1,0){20}}
\put(172,5){\vector(-1,0){12}}
\put(182,5){\vector(1,0){13}}
\put(129,135){\vector(-1,0){14}}
\put(139,135){\vector(1,0){21}}
\put(225,135){\vector(-1,0){65}}
\put(235,135){\vector(1,0){65}}
\end{picture}
\end{center}
\caption {Optical layout for ENFS. The main transmitted beam is
not blocked by the stop in the focal plane. Both the scattered 
and the transmitted beams are sent to the CCD.}
\label{euristico_setup_ENFS}
\end{figure}

It became soon appearent that the main problem with ONFS comes from
the poor statistical quality of the calculated $I\left(q\right)$.
In Chapter \ref{capitolo_confronto_ONFS_ENFS} we
will show that the statistical quality increases only as the fourth
root of the number of processed images. We experimented a different
optical setup (ENFS), drawn in Fig. \ref{euristico_setup_ENFS}.
In ENFS, there's no beam stop: the main beam is let interfere with the
scattered light. This is basically an heterodyne version of NFS, thus
we call it hEterodyne Near Field Speckles (ENFS).

Basically, ONFS data processing consists in evaluating the field
correlation function $C_E \left(\vec{r}\right)$ by using Siegert
relation (\ref{SGT}), then evaluating $I\left(q\right)$ by applying
the inverse Fourier transform to
(\ref{FT}). In ENFS, we measure the interference between the speckle
field of ONFS with the much more intense transmitted beam. We directly
measure a quantity linearly related to the field. The intensity
correlation function of an ENFS image equals $C_E
\left(\vec{r}\right)$, provided that all the conditions needed by ONFS
are met, that is, if the field is circular gaussian. We thus obtain
$C_E\left(\vec{r}\right)$ without the data inversion needed to apply
Siegert relation, and this greatly enhances the statistical accuracy of
the results.

In Chapter \ref{capitolo_confronto_ONFS_ENFS} we show a comparison
between data taken with ONFS and ENFS; data taken with ENFS are
evidently  much less noisy. The quality is comparable with the SALS
one. This good quality allowed to try a Mie-based inversion algorithm,
to obtain an histogram of the distribution of the diameters of some
colloidal samples; the measurements are shown in Chapter 
\ref{capitolo_particle_sizing_ENFS}.

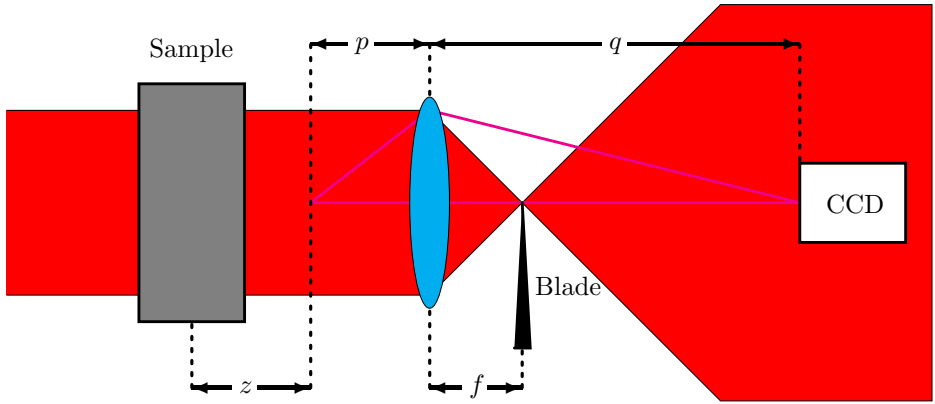
\begin{figure}
\begin{center}
\begin{picture}(350,150)(0,0)
\blacken\color{red}\path(0,40)(160,40)(160,110)(0,110)
\blacken\color{red}\path(160,40)(270,150)(350,150)(350,0)(270,0)(160,110)
\Thicklines
\color{magenta}\path(160,110)(115,75)(300,75)(160,110)
\thinlines
\put(160,75){\blacken\color{cyan}\ellipse{15}{80}}
\Thicklines
\shade\path(50,30)(90,30)(90,120)(50,120)(50,30)
\put(54,130){Sample}
\dottedline{5}(70,5)(70,29)
\dottedline{5}(115,5)(115,135)
\dottedline{5}(160,5)(160,34)
\dottedline{5}(195,5)(195,19)
\dottedline{5}(160,116)(160,135)
\dottedline{5}(300,91)(300,135)
\whiten\path(300,60)(340,60)(340,90)(300,90)(300,60)
\put(310,71){CCD}
\put(88,3){$z$}
\put(175,3){$f$}
\put(132,133){$p$}
\put(228,133){$q$}
\put(85,5){\vector(-1,0){15}}
\put(95,5){\vector(1,0){20}}
\put(172,5){\vector(-1,0){12}}
\put(182,5){\vector(1,0){13}}
\put(129,135){\vector(-1,0){14}}
\put(139,135){\vector(1,0){21}}
\put(225,135){\vector(-1,0){65}}
\put(235,135){\vector(1,0){65}}
\blacken\path(195,75)(198,20)(192,20)
\put(200,40){Blade}
\end{picture}
\end{center}
\caption {Optical layout for SNFS. Part of the transmitted beam is
blocked by a blade in the focal plane, along with half of the
scattered light. Only one half of the scattered light is sent to the CCD,
along with a part of the transmitted beam.}
\label{euristico_setup_SNFS}
\end{figure}

Both ONFS and ENFS are quite sample wasting techniques. They require
a sample much bigger than the statistical quality needs. For
example, consider a non-equilibrium fluctuation measurement in a free
diffusion experiment \cite{vailati1997}. The biggest fluctuations we want to
measure are
about $0.5\mathrm{mm}$. A good statistical sample should be so big to contain
some hundred of the biggest fluctuations: it can be a square with a
$5\mathrm{mm}$ side. This is enough for SALS, but not for ONFS nor ENFS. 
In Chapt. \ref{chap_theory} we will show that, if we
want to cover two decades in wavevectors, we must use a sample with
side $5\mathrm{mm} \cdot 10^2$. To cover two decades, we need a 
half a meter wide
cell, while with SALS we can work with a half a centimeter wide cell!
This is not a difficulty for particle sizing applications, but can
become a serious problem when we want to analyze many
lenghtscales, since NFS is particularly suited for big objects. 
This problem is essentially due to the fact that big objects need long
values of $z$ in order that their scattered field is gaussian; on the
other hand, we need a big sample, so that the sensor collect the light
scattered at high 
angles by small particles. This fact is quite unusual, since in
general big objects are good subjects for classical microscopy
techniques.
The difficulty can be easily circumvented, introducing a new instrumental
setup, called SNFS: see Fig. \ref{euristico_setup_SNFS}.
This setup closely mirrors the Schlieren setup: a blade
stops half focused trasmitted beam, along with one half of the scattered light.
In the case of a two
dimensional sample, we can create an image of it on the CCD
sensor, as in classical Schlieren imaging technique. But in the case of
three dimensional samples, the smaller objects create a speckle field,
while the bigger ones are completely resolved:
this technique passes continuously from the deterministic image
formation of the big objects, like in microscopy techniques, to the
analysis of stochastical interference patterns of NFS, without the
need to know at what lengthscale the passage takes place. We will show
that SNFS 
doesn't require any condition on the position of the focal plane: it
can be made as near to the sample as we need; moreover, the dimensions
of the sample can be quite small, as in SALS.

SNFS requires an additional element with respect to ENFS, the blade, but it
allows easy measurements on many lengthscales, on big objects. We used
such a technique to measure the power spectrum of non-equilibrium
fluctuations in a free diffusion experiment, described in Chapter
\ref{capitolo_dinamico_SNFS}, thus showing that this technique can be
applied to researches in fundamental physics.

%
\chapter{Theory.}
\label{chap_theory}

In this chapter we describe quantitatively the relations between the
scattered intensity and the properties of the field on a given
plane. We will derive the transfer function of various techniques based
on the light scattering. We will show that, under suitable conditions,
the field is gaussian. In this case, we will derive the relation
between the power spectra of the NFS images and the scattered
intensity.

\section{Scattered intensity and field power spectrum.}

From Maxwell equations, we can derive the wave equation for a
transverse component of the electric field in the vacuum \cite{jackson}:
%
%
\begin{equation}
\label{teoria_eq_onde}
\frac{\partial^2}{\partial t^2} E\left(\vec{x},z,t\right) = 
c^2 \left[ 
\frac{\partial^2}{\partial z^2} E\left(\vec{x},z,t\right) +
\nabla_{\vec{x}}^2 E\left(\vec{x},z,t\right)
\right]
\end{equation}
where $\nabla_{\vec{x}}^2$ is the Laplacian oeprator, with respect to
the horizontal coordinate $\vec{x}$. Since we are working with a
laser, and we consider only elastic scattering, the only temporal
frequency involved is $kc$:
%
%
\begin{equation}
E\left(\vec{x},z,t\right) = E\left(\vec{x},z\right) e^
{\displaystyle -i kct}
\end{equation}
where $k$ is the wave vector. Eq. (\ref{teoria_eq_onde}) becomes:
%
%
\begin{equation}
\frac{\partial^2}{\partial z^2} E\left(\vec{x},z\right) +
\nabla_{\vec{x}}^2 E\left(\vec{x},z\right) +
k^2 E\left(\vec{x},z\right) = 0
\end{equation}

We define $E_z\left(\vec{q}\right)$ as the Fourier transform of
$E\left(\vec{x},z\right)$ with respect to $\vec{x}$:
%
%
\begin{equation}
\label{teoria_eq_evoluzione_z_esatta}
\frac{\partial^2}{\partial z^2} E_z\left(\vec{q}\right) =
-\left(k^2-q^2\right) E_z\left(\vec{q}\right)
\end{equation}
The solution is:
%
%
\begin{equation}
\label{teoria_evoluzione_z_campo}
E_z\left(\vec{q}\right) = E_0\left(\vec{q}\right) e^
{\displaystyle i \sqrt{k^2-q^2} z }
\end{equation}
In order that this solution exists, a condition must be fulfilled:
%
%
\begin{equation}
q^2 < k^2,
\end{equation}
This condition is alwais met if we consider only propagating waves.

The quantity $E_z\left(\vec{q}\right)$ is closely related to the
intensity of the light crossing the plane $z=\mathrm{cost}$. Each
two-dimensional Fourier mode of amplitude $E_z\left(\vec{q}\right)$,
on a given $z$, and wavevector $\vec{q}$ is generated by a
three-dimensional plane wave of wavevector $\left[q_x,q_y,k_z\right]$,
where the only value of $k_z$ is obtained by imposing that the
wavevector of any plane wave has length $k$:
\begin{equation}
\label{teoria_definizione_unico_kz}
k_z\left(q\right) = \sqrt{k^2 - q^2}
\end{equation}
Given the values of the two-dimensional Fourier modes
$E_z\left(\vec{q}\right)$ on a given $z=0$, we can evaluate
$E\left(\vec{x},z\right)$ for each $\vec{x}$ and $z$, by using
Eq. (\ref{teoria_evoluzione_z_campo}) and
(\ref{teoria_definizione_unico_kz}). Expressing it by its
three-dimensional Fourier transform:
\begin{equation}
E\left(\vec{q},k_z\right)=2\pi E_0\left(\vec{q}\right)
\delta \left[ k_z - k_z\left(q\right) \right]
\label{teoria_equ_valore_Eqxqyqz_in_Eqxqyz}
\end{equation}
Each three-dimensional component with amplitude
$E\left(\vec{q},k_z\right)$ of the electric field represents a plane
wave travelling in a different direction. 
We define $S_E\left(\vec{q},q_z\right)$, the two-dimensional power
spectrum of $E\left(\vec{x},z\right)$:
\begin{equation}
\left< E_z\left(\vec{q}\right)
E_z^*\left(\vec{q}'\right) \right> = 
\delta \left(\vec{q}-\vec{q}'\right) S_E\left(\vec{q}\right)
\label{teoria_eq_definizione_S_E}
\end{equation}
Light intensity, for each scattering direction, can be defined on the basis
of $E\left(\vec{q},k_z\right)$:
\begin{equation}
\label{teoria_relazione_I_E_di_q3}
\left< E\left(\vec{q},k_z\right)
E^*\left(\vec{q}',k_z'\right) \right> =
4\pi^2
\delta \left(\vec{q}-\vec{q}'\right)
\delta \left(k_z-k_z'\right)
\delta \left[k_z-k_z\left(q\right) \right]
I\left(\vec{q},q_z\right)
\end{equation}
where $I\left(\vec{q},q_z\right)$ has been expressed in terms of the
transferred wavevector $\left[q_x,q_y,q_z\right]=
\left[k_x,k_y,k_z\right] - \left[0,0,k\right]$. From
Eq. (\ref{teoria_definizione_unico_kz}):
\begin{equation}
\label{teoria_definizione_qz}
q_z\left(\vec{q}\right) = \sqrt{k^2 - q^2} - k
\end{equation}

Substituting $E\left(\vec{q},k_z\right)$ of Eq. 
(\ref{teoria_equ_valore_Eqxqyqz_in_Eqxqyz}) in 
Eq. (\ref{teoria_relazione_I_E_di_q3}),
and comparing the result with Eq. (\ref{teoria_eq_definizione_S_E}),
we can relate the scattered intensity $I\left(\vec{q},q_z\right)$ to
the power spectrum of the field $S_E\left(\vec{q}\right)$:
\begin{equation}
\label{teoria_relazione_I_Ez_di_q}
I\left(\vec{q},q_z\right)=S_E\left(\vec{q}\right)
\end{equation}

From Eq. (\ref{teoria_relazione_I_Ez_di_q}) we notice that
$I\left(q\right)$ can be measured by evaluating
$E_z\left(\vec{q}\right)$ on any $z$.

\begin{figure}
\begin{center}
\begin{picture}(200,200)(-100,-100)
\dottedline{4}(-20,60)(80,60)
\dottedline{4}(-100,0)(-20,60)
\thinlines
\put(-100,0){\arc{200}{-1.5707963}{1.5707963}}
\thicklines
\put(0,0){\circle{200}}
\put(-100,0){\vector(1,0){100}}
\put(0,0){\vector(4,3){80}}
\put(0,0){\vector(-1,3){20}}
\put(0,0){\vector(0,1){60}}
\put(-70,-12){$\left[0,0,k\right]$}
\put(40,20){$k'$}
\put(4,33){$q$}
\put(-20,23){$Q$}
\end{picture}
\end{center}
\caption[Relation between the wave vector of the near field and the
transferred wave vector.]{Relation between $q$ and
$Q$. Geometrical interpretation of Eq.
(\ref{teoria_trasferito_vettore_onda})} 
\label{euristico_disegno_q_NFS}
\end{figure}
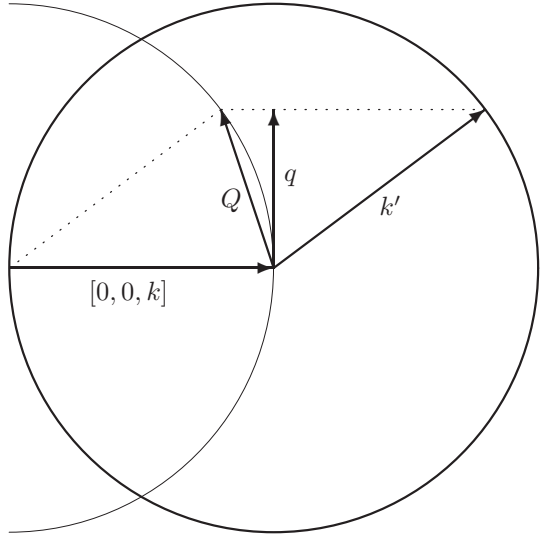

If the sample is isotropic, $I\left(\vec{q},q_z\right)$ depends only on
$Q=\left|q_x,q_y,q_z\right|$:
%
%
\begin{equation}
\label{teoria_trasferito_vettore_onda}
Q\left(q\right) = \sqrt{2}k\sqrt{1-\sqrt{1-\left(\frac{q}{k}\right)^2}}
\end{equation}
In this case, Eq. (\ref{teoria_relazione_I_Ez_di_q}) can be written in
terms on $q$ and $Q$:
\begin{equation}
\label{teoria_relazione_IQ_Ez_di_q}
I\left[Q\left(q\right)\right]=S_E\left(q\right)
\end{equation}
The geometrical meaning of Eq. (\ref{teoria_trasferito_vettore_onda})
is explained in Fig. \ref{euristico_disegno_q_NFS}. For $q\ll k$,
Eq. (\ref{teoria_trasferito_vettore_onda}) can be approximated by
$Q\left(q\right)=q$.
Moreover, if Rayleigh Gans approximation holds, $I\left(q\right)$
represents the power spectrum of the refraction index of the sample.
From these two considerations, we obtain the result that, for scattering
on small angles and under Rayleigh Gans condition, the 
two dimensional correlation function of the electric field is proportional
to the correlation function of the light path through the sample.

%
%
%
%
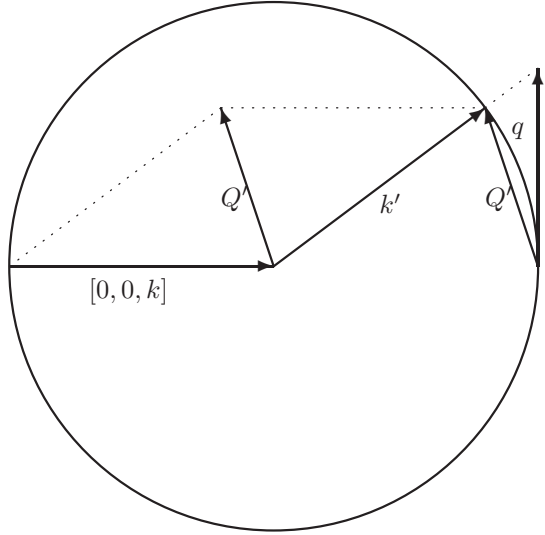
\begin{figure}
\begin{center}
\begin{picture}(205,200)(-100,-100)
\dottedline{4}(-20,60)(80,60)
\dottedline{4}(-100,0)(-20,60)
\dottedline{4}(80,60)(100,75)
\thicklines
\put(0,0){\circle{200}}
\put(-100,0){\vector(1,0){100}}
\put(0,0){\vector(4,3){80}}
\put(0,0){\vector(-1,3){20}}
\put(100,0){\vector(-1,3){20}}
\put(100,0){\vector(0,1){75}}
\put(-70,-12){$\left[0,0,k\right]$}
\put(40,20){$k'$}
\put(90,50){$q$}
\put(-20,23){$Q'$}
\put(80,23){$Q'$}
\end{picture}
\end{center}
\caption[Relation between the coordinate $q$ on a screen, in a far
field experiment, and the tranferred wave vector.]{Relation between
the coordinate $q$ on a screen, in a far field experiment, and the
transferred wave vector $Q'$. Geometrical interpretation of Eq.
(\ref{euristico_eq_q_FFS})}
\label{euristico_disegno_q_scattering}
\end{figure}
%
%
\begin{figure}
\begin{center}
\includegraphics{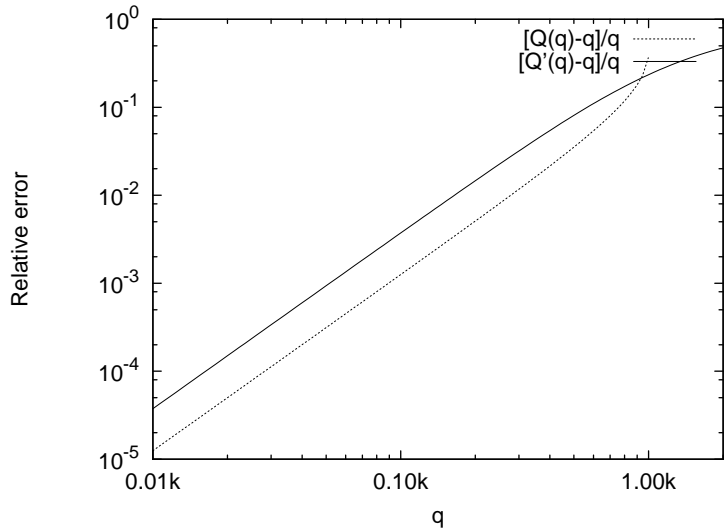}
\end{center}
\caption[Relative error obtained neglecting the non linearity of the
ralation between the sample wave vector and the near field wave
vector.]
{Relative error obtained neglecting the non linearity of the
ralation between the sample wave vector and the near field wave
vector. The graph is obtained from Eqs.
(\ref{teoria_trasferito_vettore_onda}) and (\ref{euristico_eq_q_FFS}). }
\label{euristico_disegno_errore_NFS_scattering}
\end{figure}
When performing a far field, small angle scattering measurement, the
scattered beams are focused on a screen. In suitable units, each point
of the screen has a coordinate $q$. For small values of the wave
vector, $q$ approximates $Q'$, the transferred wavevector.
The exact relation is:
%
%
\begin{equation}
Q'\left(q\right)
=\sqrt{2}k\sqrt{1-\frac{1}{\sqrt{1+\left(\frac{q}{k}\right)^2}}}
\label{euristico_eq_q_FFS}
\end{equation}
Equation (\ref{teoria_trasferito_vettore_onda}) can be used to correct
the results of a Near Field Speckles measurement. Figures 
\ref{euristico_disegno_q_NFS} and \ref{euristico_disegno_q_scattering}
show the geometrical meaning of equations
(\ref{teoria_trasferito_vettore_onda}) and
(\ref{euristico_eq_q_FFS}). For small values of $q$, that is $q/k\ll 1$,
the two 
equations can be approximated with $Q=Q'=q$; the error
due to this approximation is shown in Fig.
\ref{euristico_disegno_errore_NFS_scattering}: it's quite small, and 
it can often  be neglected.

\section {Scattering from a thin sample.}

Let us consider a thin sample, with a non homogeneous refraction
index, and a light plane wave, moving in the direction of the $z$
axis. For $z=0$, at the surface of the sample, the field will be:
%
%
\begin{equation}
\label{teoria_campione_sottile_esatta}
E\left(\vec{x},z=0\right) = E_0 e^{\displaystyle
i \delta l \left(\vec{x}\right) k}
\end{equation}
where $\delta l \left(\vec{x}\right)$ is the difference between the
light path, the integral of the refraction index along $z$, for a
given point $\vec{x}$, and its mean value over the whole sample. If
$\delta l$ is small compared to the light wavelength, we can consider
a first order developement:
%
%
\begin{equation}
\label{teoria_campione_sottile_approssimata}
E\left(\vec{x},z=0\right) = E_0 \left[
1 + i \delta l \left(\vec{x}\right) k\right]
\end{equation}
Neglecting the higher order terms means that we are neglecting higher
order diffracted beams than the first.

Using Eq. (\ref{teoria_evoluzione_z_campo}) we can find the field for
every value of $z$:
%
%
\begin{equation}
\label{teoria_campo_scatterato}
E_z\left(\vec{x}\right) = E_0\left(z\right) + \delta E_z\left(\vec{x}\right)
\end{equation}
where
%
%
\begin{equation}
\delta E_z\left(\vec{q}\right) = 
ikE_0\left(z\right) \delta l\left(\vec{q}\right) e^
{\displaystyle i\left(\sqrt{k^2-q^2}-k\right)z}
\label{teoria_eq_campo_campione_sottile}
\end{equation}
is the scattered field, and 
\begin{equation}
E_0\left(z\right) = 
E_0 e^{\displaystyle ikz}
\end{equation}

\section{Image forming techniques}

Microscopy, in its basic form, consists in forming an image of a plane
on a device which measures light intensity, such as a photographic film
or a CCD sensor. Generally it is used to obtain informations about the
intensity of the transmitted light, for example, in the case of an
organic tissue, treated by some dye.

If a microscope objective forms the image of a plane on the CCD,
the image is given by the interference of the transmitted and
the scattered beams.
For such an image, the signal can be defined as the difference of the
measured intensity $I\left(\vec{x}\right)$ and the transmitted beam 
intensity $I_0$, divided by $I_0$. We will call this signal $i_{shadowgraph}$,
for reasons that will be clear later.
We consider the case in which the scattered beams are much less intense than
the transmitted one. At the first order in $\delta E$, the signal is:
\begin{equation}
i_{shadowgraph}\left(\vec{x}\right)=\frac{I\left(\vec{x}\right)-I_0}{I_0}=
\frac{2}{I_0}\Re\left[E_0 \delta E^*\left(\vec{x}\right)\right]
\label{teoria_eq_tecniche_campo_shadowgr_corr_func}
\end{equation}
and the Fourier transform is:
\begin{equation}
i_{shadowgraph}\left(\vec{q}\right)=
\frac{1}{I_0}\left[E_0 \delta E^*\left(-\vec{q}\right)+
E_0^* \delta E\left(\vec{q}\right)
\right]
\label{teoria_eq_tecniche_campo_shadowgraph}
\end{equation}
If the sample is transparent, and we send a plane wave through it,
Eq. (\ref{teoria_eq_campo_campione_sottile}) tells us the value of the
scattered field:
\begin{equation}
i_{shadowgraph}\left(\vec{q},z\right)=
2k\delta l\left(\vec{q}\right) \sin\left[\left(k-\sqrt{k^2-q^2}\right)z\right]
\label{teoria_eq_tecniche_cammino_shadowgraph}
\end{equation}
In order to obtain the previous result, $\delta l\left(\vec{x}\right)$
has been considered real, so that $\delta l^*\left(-\vec{q}\right) = 
\delta l\left(\vec{q}\right)$.

For $z=0$, that is, if the thin sample is in the focal plane,
$i_{shadowgraph}\left(\vec{q},z\right)=0$.
The intensity is completely uniform, and bears no informations on the sample.

Many techniques has been developed in order to make phase modulations
evident: among them, holography and interferometry. A well known way to make
phase modulations evident is
the phase contrast microscopy. Basically, this technique consists in
changing the phase of the transmitted beam by $\pi/2$.
At the first order in $\delta E$:
\begin{equation}
i_{phase\,contrast}\left(\vec{x}\right)=\frac{I\left(\vec{x}\right)-I_0}{I_0}=
\frac{2}{I_0}\Im\left[E_0^* \delta E\left(\vec{x}\right)\right]
\end{equation}
and the Fourier transform is:
\begin{equation}
i_{phase\,contrast}\left(\vec{q}\right)=
\frac{i}{I_0}\left[-E_0^* \delta E\left(-\vec{q}\right)+
E_0 \delta E^*\left(\vec{q}\right)
\right]
\label{teoria_eq_tecniche_campo_phasecontrast}
\end{equation}
Using Eq. (\ref{teoria_eq_campo_campione_sottile}) to evaluate the
scattered field:
\begin{equation}
i_{phase\,contrast}\left(\vec{q},z\right)=
2k\delta l\left(\vec{q}\right) \cos\left[\left(k-\sqrt{k^2-q^2}\right)z\right]
\label{teoria_eq_tecniche_cammino_phasecontrast}
\end{equation}
For $z=0$, that is, with the sample in the focal plane:
\begin{equation}
i_{phase\,contrast}\left(\vec{q},z=0\right)=
2k\delta l\left(\vec{q}\right)
\end{equation}

Another way to make phase modulations evident is the so called dark
field technique. It consists in stopping the transmitted
beam. This is accomplished by focusing the transmitted and the scattered beams
by a lens, and by removing the transmitted beam by some kind of reflecting or
absorbing object. This is an homodyne technique; the signal must be defined
as the ratio between the measured intensity $I\left(\vec{x}\right)$
and the intensity of the transmited beam $I_0$. Since we have only $\delta E$:
\begin{equation}
i_{dark\,field}\left(\vec{x}\right)=\frac{I\left(\vec{x}\right)}{I_0}=
\frac{\left|\delta E\left(\vec{x}\right)\right|^2}{I_0}
\label{teoria_eq_tecniche_campo_darkfield}
\end{equation}
Equation (\ref{teoria_eq_campo_campione_sottile}), for $z=0$, gives:
\begin{equation}
i_{dark\,field}\left(\vec{x},z=0\right)=k^2
\delta l^2\left(\vec{x}\right)
\label{teoria_eq_tecniche_campo_dark_f_light_path}
\end{equation}
In Fourier space:
%
%
\begin{equation}
i_{dark\,fielf}\left(\vec{q},z=0\right) =
\frac{1}{\left(2\pi\right)^2}
k^2 \int{\delta l\left(\vec{q}'\right) 
\delta l\left(\vec{q}-\vec{q}'\right) \mathrm{d}\vec{q}'}
\label{teoria_eq_tecniche_cammino_darkfield}
\end{equation}

The Schlieren technique consists in focusing the beams from the sample
by a lens; in the focal plane, a blade stops half of the transmitted beam,
along with the beams scattered in one half plane.
At the first order in $\delta E$, the signal is again, like in shadowgraph:
\begin{equation}
i_{Schlieren}\left(\vec{x}\right)=
\frac{I\left(\vec{x}\right)-\tilde{I}_0}{\tilde{I}_0}=
\frac{2}{\tilde{I}_0}\Re\left[\tilde{E}_0
\delta \tilde{E}^*\left(\vec{x}\right)\right]
\end{equation}
where the field $\delta \tilde{E}$ is the scattered field, without 
one half plane in Fourier space:
\begin{equation}
\delta \tilde{E}\left(\vec{q}\right)=
\left\{
\begin{array}{ll}
\delta E\left(\vec{q}\right) & \vec{q}\cdot\vec{n}<0
\\
\delta E\left(\vec{q}\right) & \vec{q}\cdot\vec{n}\ge 0
\end{array}
\right. ,
\label{teoria_eq_campo_scatt_senza_semipiano}
\end{equation}
$\vec{n}$ is the vector ortogonal to the direction of the blade,
$\tilde{I}_0=I_0/2$ and $\tilde{E}_0=E_0/\sqrt{2}$
are the intensity and the field of the trasmitted beam, after the
blade.
The Fourier transform is:
\begin{equation}
i_{Schlieren}\left(\vec{q}\right)=
\frac{\sqrt{2}}{I_0}\left[E_0 \delta
\tilde{E}^*\left(-\vec{q}\right)+
E_0^* \delta \tilde{E}\left(\vec{q}\right)
\right]
\end{equation}
Using Eq. (\ref{teoria_eq_campo_scatt_senza_semipiano}):
\begin{equation}
i_{Schlieren}\left(\vec{q}\right)=
\left\{
\begin{array}{ll}
\frac{\sqrt{2}}{I_0}E_0 \delta E^*\left(-\vec{q}\right)
&\vec{q}\cdot\vec{n} <0
\\
\frac{\sqrt{2}}{I_0}E_0^* \delta E\left(\vec{q}\right)
&\vec{q}\cdot\vec{n}\ge 0
\end{array}
\right.
\label{teoria_eq_tecniche_campo_schlieren}
\end{equation}
Using Eq. (\ref{teoria_eq_campo_campione_sottile}):
\begin{equation}
i_{Schlieren}\left(\vec{q},z\right)=
\left\{
\begin{array}{ll}
\sqrt{2}ik \delta l\left(\vec{q}\right) e^
{\displaystyle -i\left(\sqrt{k^2-q^2}-k\right)z}
&\vec{q}\cdot\vec{n}<0
\\
\sqrt{2}ik \delta l\left(\vec{q}\right)e^
{\displaystyle i\left(\sqrt{k^2-q^2}-k\right)z}
&\vec{q}\cdot\vec{n}\ge 0
\end{array}
\right.
\label{teoria_eq_tecniche_cammino_schlieren}
\end{equation}
In order to obtain the previous result, $\delta l\left(\vec{x}\right)$
has been considered real, so that $\delta l^*\left(-\vec{q}\right) = 
\delta l\left(\vec{q}\right)$.
For $z=0$, that is, if the thin sample is in the focal plane:
\begin{equation}
i_{Schlieren}\left(\vec{q},z=0\right)=
\sqrt{2}ik \delta l\left(\vec{q}\right)
\end{equation}
The factor $i$ means that all the Fourier components of $\delta l
\left(\vec{q}\right)$ undergo a rotation of $\pi/2$: a
sine-like light path modulation gives a cosine-like intensity modulation.

\section{Misfocused microscopy and shadowgraph.}

Equations (\ref{teoria_eq_tecniche_cammino_shadowgraph}),
(\ref{teoria_eq_tecniche_cammino_phasecontrast}) and 
(\ref{teoria_eq_tecniche_cammino_schlieren}) allow to evaluate
the evolution of the signals as $z$, the misfocusing, is increased, for 
a simple microscope objective, for phase contrast and dark field.
It should be noted that we are dealing with images formed by laser light:
as $z$ increases, it is possible to recover the original shape
of the observed objects; this is completely different from a white light
microscope, in which the misfocusing simply smears the images.

Equation (\ref{teoria_eq_tecniche_cammino_darkfield}) has not been extended
for $z\ne 0$; in all the above mentioned techniques, however, the 
variation of $z$ strongly influences the relation between the light path
$\delta l\left(\vec{q}\right)$ and the signal $i\left(\vec{q}\right)$.
This is generally a defect: thick objects, or even thin objects
dispersed in a thick volume, are difficult to be analysed. In general,
all the above mentioned techniques are applied to samples that are thin,
and in the focal plane. No improvement is obtained by misfocusing.

A well known exception is shadograph. Shadowgraph technique consists in 
sending a plane wave onto a sample,
and observing the intensity modulations generated by the sample on a
plane placed at a distance $z$ from the sample. 
Using Eq. (\ref{teoria_eq_tecniche_cammino_shadowgraph}), we can derive
the transfer function $T_{shadowgraph}\left(\vec{q},z\right)$
of the shadowgraph technique \cite{cannell1995,cannell1996,brogioli_mth}:
\begin{equation}
T_{shadowgraph}\left(\vec{q},z\right) = 
2k\sin\left[\left(k-\sqrt{k^2-q^2}\right)z\right] \approx
2k\sin\left(\frac{q^2z}{2k}\right)
\label{teoria_eq_transfer_function_shadowgraph}
\end{equation}
The approximation holds for $q\ll k$.
The transfer function is defined as the ratio between the signal
and the light path modulation amplitude:
\begin{equation}
i\left(\vec{q},z\right)= T\left(\vec{q},z\right) \delta l\left(\vec{q}\right)
\end{equation}

For $z=0$, the transfer function vanishes; misfocusing is needed, and is
a simple way to make phase modulations evident.

Looking at Eq. (\ref{teoria_eq_tecniche_cammino_phasecontrast})
we can notice that phase contrast transfer function, as a function of $z$,
has a cosinusoidal behaviour:
\begin{equation}
T_{phase\,contrast}\left(\vec{q},z\right) = 
2k\cos\left[\left(k-\sqrt{k^2-q^2}\right)z\right] \approx
2k\cos\left(\frac{q^2z}{2k}\right)
\end{equation}
When considering opaque objects, with no
phase modulations, the transfer functions are exchanged: cosinusoidal
for shadowgraph, sinusoidal for phase contrast.

The shadowgraph image is created by the interference between every 
scattered beam
and the transmitted beam; it's always possible, in principle, to find
the value of $\delta l$, in every point, simply by a
deconvolution. Shadowgraph allows the measurement of one component of
the field, which, in turns, is the convolution of the light path with
a particular function. The absolute intensity modulation of the
shadowgraph image is proportional to the mean intensity and to the
light path modulation; the constant of proportionality is the
the transfer function. The transfer function vanishes for
some wave vectors, but has maxima for other ones. At the maxima, the
sensibility equals the sensibility of phase contrast and Schlieren
techniques.

Now we evaluate the effect of misfocusing on a dark field microscope. We
obtain an image of a plane a distance $z$ from the cell. 
Using Eq. (\ref{teoria_eq_campo_campione_sottile}), we can derive
the relation between the light path and the measured intensity, at a given $z$:
\begin{equation}
i_{dark\,field}\left(\vec{q},z\right) =
\frac{1}{\left(2\pi\right)^2} k^2 
\int{
\delta l\left(\vec{q}',z\right)
\delta l\left(\vec{q}-\vec{q}',z\right)e^
{\displaystyle -i\left[\sqrt{k^2-\left(\vec{q}-\vec{q}'\right)^2}-k\right]z}
\mathrm{d}\vec{q}' }
\end{equation}

This expression reduces to Eq. (\ref{teoria_eq_tecniche_cammino_darkfield}) for
$z=0$. At this point, there's no appearent reason to use a misfocused
dark field instead of a focused one.

The knowledge of $i_{dark\,field}\left(\vec{q},z\right)$, for every $z$, can
give some informations about the spreading of the scattered light. For
the scattering of a single particle, one can measure the intensity on
planes with increasing values of $z$, and calculate how fast the light
is diverging. This provides informations both on the position of the
particle and on the scattered intensity. For a sample composed by a
great number of particles, this cannot be done, and a different,
statistical approach must be applied.

For Schlieren technique, from Eq. (\ref{teoria_eq_tecniche_cammino_schlieren}):
\begin{equation}
T_{Schlieren}\left(\vec{q},z\right) = 
\left\{
\begin{array}{ll}
\sqrt{2}ik e^
{\displaystyle -i\left(\sqrt{k^2-q^2}-k\right)z}
&\vec{q}\cdot\vec{n}<0
\\
\sqrt{2}ik e^
{\displaystyle i\left(\sqrt{k^2-q^2}-k\right)z}
&\vec{q}\cdot\vec{n}\ge 0
\end{array}
\right.
\end{equation}
By evaluating the square modulus of the transfer function, we obtain:
\begin{equation}
\left|T_{Schlieren}\left(\vec{q},z\right)\right|^2 = 2k^2
\end{equation}
This means that the power spectrum of the electric field is
proportional to the power spectrum of the light path, without any
dependence on spatial wavelength and misfocusing.

\section{Scattering measurements by microscopy techniques}

By using Eq. (\ref{teoria_relazione_I_Ez_di_q}), we can, in principle,
evaluate the scattered intensities by measuring the field on a given plane.
The above described microscopy techniques allow the determination of some
functions of the electric field. Now, we assume that the scattered
electric field is generated by a thin sample. From 
Eq. (\ref{teoria_eq_campo_campione_sottile}), we derive a property of
the electric field:
\begin{equation}
\delta E\left(-\vec{q}\right)=
-\delta E^*\left(\vec{q}\right)
e^{\displaystyle 2i\left(\sqrt{k^2-q^2}-k\right)z}
\end{equation}

For shadowgraph, we use Eq. (\ref{teoria_eq_tecniche_campo_shadowgraph}),
in order to evaluate the power spectrum $S_i$ of $i\left(\vec{x}\right)$:
\begin{equation}
S_i\left(\vec{q}\right) =\frac{4}{I_0}
\sin^2\left[\left(k-\sqrt{k^2-q^2}\right)z\right]
S_E\left(\vec{q}\right)
\end{equation}
Using Eq. (\ref{teoria_relazione_I_Ez_di_q}), we obtain:
\begin{equation}
S_i\left(q\right) =\frac{4}{I_0}
\sin^2\left[\left(k-\sqrt{k^2-q^2}\right)z\right]
I\left[Q\left(q\right)\right]
\label{shadowgraph_transfer_function_scattering_measurement}
\end{equation}
In order to use shadowgraph to evaluate the scattered intensity, the sample
must be out of the focal plane. This technique has some disadvantages:
some wavevectors cannot be seen, since the transfer function vanishes;
moreover, if the sample is thick we cannot define a $z$: the oscillations
of the transfer function are smeared, but it could be hard to know
quantitatively how much.

For phase contrast, we use Eq. (\ref{teoria_eq_tecniche_campo_phasecontrast}),
in order to evaluate the power spectrum $S_i$ of $i\left(\vec{x}\right)$:
\begin{equation}
S_i\left(\vec{q}\right) =\frac{4}{I_0}
\cos^2\left[\left(k-\sqrt{k^2-q^2}\right)z\right]
S_E\left(\vec{q}\right)
\end{equation}
Using Eq. (\ref{teoria_relazione_I_Ez_di_q}), we obtain:
\begin{equation}
S_i\left(q\right) =\frac{4}{I_0}
\cos^2\left[\left(k-\sqrt{k^2-q^2}\right)z\right]
I\left(Q\left(q\right)\right)
\end{equation}
The disadvantages of shadowgraph, as a scattering measurement technique,
are also found in phase contrast. Phase contrast has a flat transfer
function only if $z=0$: it's an ideal technique for thin samples.

Dark field doesn't allow to recover any information about the phase
of the field. It is not suited to make scattered intensity measurements;
a statistical approach, described in the following sections,
will give interesting results.

We will describe in Section \ref{teoria_sezione_SNFS} the
application of Schlieren technique to the measurement of
the scattered intensity.

\section{Gaussian field generated by the sum of many patterns.}

\begin{figure}[p]
\begin{center}
\includegraphics[scale=0.4]{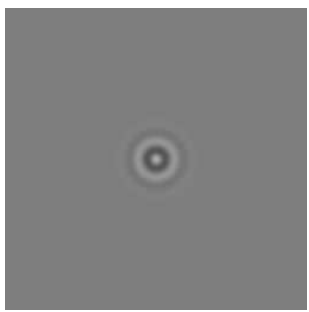}
\end{center}
\caption{Example of the field generated by a particle of the colloid.}
\label{teoria_pattern}
\end{figure}

\begin{figure}[p]
\begin{center}
\includegraphics[scale=0.4]{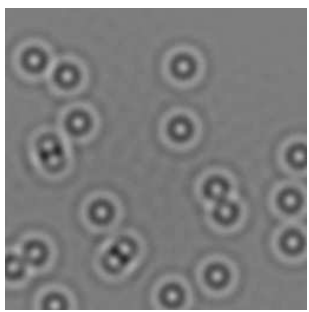}
\end{center}
\caption{Example of the field generated by many particles of the colloid.}
\label{teoria_somma_pattern}
\end{figure}

Consider a monodisperse colloid, and the near field scattered light. The
field can be decomposed into the sum of the waves coming from the different
elements of the colloid. Thus at every distance from the colloid, the field
will be given by the sum of many patterns, each randomly placed. An example 
of this is given in Fig. \ref{teoria_pattern} and 
\ref{teoria_somma_pattern}. Figure \ref{teoria_pattern} shows the pattern:
the intensity is linearily dependent on the field $f\left(\vec{x}\right)$,
which we can consider given by the wave scattered by a particle of the 
colloid. Figure \ref{teoria_somma_pattern} shows the sum of the patterns,
the function  $\rho\left(\vec{x}\right)$:
%
%
\begin{equation}
\rho\left(\vec{x}\right)=\sum_{i=1}^N{f\left(\vec{x}-\vec{x}_i\right)}
\end{equation}
where the $N$ $\vec{x}_i$ are randomly distributed in a surface of measure $S$.

Now we will evaluate the $P$-point correlation functions of the sum of many
patterns, $\rho\left(\vec{x}\right)$. The results will be obteined, first,
for fixed particle density $\mathcal{N}=N/S$, and $N\to \infty$; then
we will show that, in a suitable limit, the $P$-point correlation 
functions, for $P\ge 4$, can be expressed in terms of two-point correlation
function, corresponding to the Wick formula. In other words, every connected
part of the correlation function developement vanishes: the field becomes
gaussian. As a matter of fact, we will prove an extension of the well known
central limit theorem.

In the following,
we will consider only functions with a vanishing average value,
the other cases being easily obtained from this one. This simplifies
the problem, since every odd-$P$-point correlation function will vanish.

The $P$-point correlation function of $\rho\left(\vec{x}\right)$ is:
%
%
\begin{multline}
C\left(\Delta \vec{x}_2,\dots,\Delta \vec{x}_P\right)=
\left<\rho\left(\vec{x}\right) \rho\left(\vec{x}+\Delta\vec{x}_2\right) \dots 
\rho\left(\vec{x}+\Delta\vec{x}_P\right) \right> = 
\frac{1}{S^{N+1}} \\
\sum_{i,j,k,\dots =1}^N \int_S{
f\left(\vec{x}-\vec{x}_i\right)
f\left(\vec{x}-\vec{x}_j+\Delta\vec{x}_2\right)
\dots
f\left(\vec{x}-\vec{x}_k+\Delta\vec{x}_P\right)
\mathrm{d}\vec{x} \mathrm{d}\vec{x}_1 \dots \mathrm{d}\vec{x}_N
}
\end{multline}

The value of the integral does not depend on all the values of the
indices $i,j,k,\dots$, but only on which of them are equal; the
sum involves $NP$ terms, but many of them are equal. For example,
for $P=4$, the term with $i=1, j=2, k=3, l=4$ is equal to
the one with $i=2, j=5, k=7, l=9$, but it is different from
the one with $i=1, j=1, k=3, l=4$. The problem is thus to determine
in how many ways we can obtain a given configuration.

The calculation can be made more easy using graphs. For
evaluating a $P$-point correlation function, we draw
$P$ points on a graph, each one corresponding to one of the
points of the correlation function, 
$0, \Delta \vec{x}_2, \dots, \Delta \vec{x}_P$. Then,
we group the points, so that every set 
contains an even number of points
\footnote{
Sets with odd number of points will give vanishing contributions if 
$f\left(\vec{x}\right)$ has a vanishing mean value.
}.
Each configuration corresponds to
many values of the indices $i,j,k,\dots$; the number of them
is the multiplicity of the graph. Every set corresponds
to an operation of integration on a different $\vec{x}_i$.
We call $G$ the number of sets; we will have only $G$
integration variables, being the integrand independent on the
other $N+1-G$ variables. The integration on these variables
gives a factor $S^{N+1-G}$. Moreover, every integration corresponds
to the evaluation of the correlation function 
$\tilde{C}$ of the single
pattern $f\left(\vec{x}\right)$:
\begin{equation}
\tilde{C}\left(\Delta \vec{x}_2,\dots,\Delta \vec{x}_P\right)=
\frac{1}{S}\int{
f\left(\vec{x}\right)
f\left(\vec{x}+\Delta \vec{x}_2\right)
\dots
f\left(\vec{x}+\Delta \vec{x}_P\right)
\mathrm{d}\vec{x}}
\end{equation}

The multiplicity of the graph depends on $G$; its value is 
$N\left(N-1\right)\left(N-2\right)\dots\left(N-G+1\right)$.
For $N\to \infty$, we can consider only the leading term $N^G$.

We can thus describe the rules for evaluating the correlation
functions, as the sum of all the graphs. The value of every graph
is the product of the factors given by each set. The factor is
the product of $\mathcal{N}$ and the correlation function $\tilde{C}$,
which correlates all the points in the set.

For example, we evaluate the two-point correlation function (Tab. 
\ref{teoria_gaussiano_pattern_2p}) and the four-point correlation function 
(Tab. \ref{teoria_gaussiano_pattern_4p}).
\begin{table}
\begin{tabular}{p{3cm}p{9cm}}
\parbox[c]{2.9cm}{
\begin{picture}(50,10)(0,0)
\put(0,0){\framebox(50,10){}}
\put(2,2){0}
\put(30,2){$\Delta \vec{x}_1$}
\end{picture} }&
$C\left(\Delta \vec{x}_1\right) =
\mathcal{N} \tilde{C}\left(\Delta \vec{x}_1\right)$
\end{tabular}
\caption{Evaluation of the two-point correlation function.}
\label{teoria_gaussiano_pattern_2p}
\end{table}
\begin{table}
\begin{tabular}{p{3cm}p{9cm}}
\multicolumn{2}{c}{
$C\left(\Delta \vec{x}_1,\Delta \vec{x}_2,
\Delta \vec{x}_3\right) =$} \\ 
\parbox[c]{2.9cm}{
\begin{picture}(50,60)(0,-10)
\put(0,0){\framebox(50,40){}}
\put(2,2){0}
\put(30,2){$\Delta \vec{x}_1$}
\put(2,32){$\Delta \vec{x}_2$}
\put(30,32){$\Delta \vec{x}_3$}
\end{picture} }&
$\mathcal{N} \tilde{C}\left(\Delta \vec{x}_1,\Delta \vec{x}_2,
\Delta \vec{x}_3\right)+$ \\
\parbox[c]{2.9cm}{
\begin{picture}(50,60)(0,-10)
\put(0,0){\framebox(50,10){}}
\put(0,30){\framebox(50,10){}}
\put(2,2){0}
\put(30,2){$\Delta \vec{x}_1$}
\put(2,32){$\Delta \vec{x}_2$}
\put(30,32){$\Delta \vec{x}_3$}
\end{picture} }&
$\mathcal{N}^2 \tilde{C}\left(\Delta \vec{x}_1\right)
\tilde{C}\left(\Delta \vec{x}_2-\Delta \vec{x}_3\right)+$ \\
\parbox[c]{2.9cm}{
\begin{picture}(50,60)(0,-10)
\put(0,0){\framebox(22,40){}}
\put(30,0){\framebox(22,40){}}
\put(2,2){0}
\put(30,2){$\Delta \vec{x}_1$}
\put(2,32){$\Delta \vec{x}_2$}
\put(30,32){$\Delta \vec{x}_3$}
\end{picture} }&
$\mathcal{N}^2 \tilde{C}\left(\Delta \vec{x}_2\right)
\tilde{C}\left(\Delta \vec{x}_1-\Delta \vec{x}_3\right)+$ \\
\parbox[c]{2.9cm}{
\begin{picture}(50,60)(0,-10)
\put(-5,5){\line(1,-1){15}}
\put(-5,5){\line(1,1){42}}
\put(10,-10){\line(1,1){42}}
\put(37,47){\line(1,-1){15}}
\put(54,6){\line(-1,-1){15}}
\put(54,6){\line(-1,1){42}}
\put(39,-9){\line(-1,1){42}}
\put(12,48){\line(-1,-1){15}}
\put(2,2){0}
\put(30,2){$\Delta \vec{x}_1$}
\put(2,32){$\Delta \vec{x}_2$}
\put(30,32){$\Delta \vec{x}_3$}
\end{picture} }&
$\mathcal{N}^2 \tilde{C}\left(\Delta \vec{x}_3\right)
\tilde{C}\left(\Delta \vec{x}_1-\Delta \vec{x}_3\right)$ \\
\end{tabular}
\caption{Evaluation of the four-point correlation function.}
\label{teoria_gaussiano_pattern_4p}
\end{table}

For $\mathcal{N}\to \infty$, the leading term in the developement of the
correlation function is the one with the higher power of
$\mathcal N$: it is the one with the higher number of sets.
Since sets with an odd number of elements have a vanishing 
contribution, the greatest number of sets can be obtained
only by making sets of two points. This means that only two point
correlation functions of the single pattern 
$\tilde{C}\left(\Delta \vec{x}\right)$ contribute to any
correlation function $C\left(\Delta \vec{x}_1,\dots \right)$
of the sum.

Since $\mathcal{N}$ is dimensional, it is not possible to state
if it is small or great. This means that we cannot define, in
general, a value of $\mathcal{N}$ so great that the field becomes
gaussian. The following heuristic considerations will show that
the field is gaussian if $\mathcal{N}A\gg 1$, where $A$ is the
area of one pattern, at least if we can define it in some ways.
Consider a pattern $f\left(\vec{x}\right)=\alpha
\chi_A\left(\vec{x}\right)$. The P-point correlation function
of the pattern $f\left(\vec{x}\right)$, evaluated in
$\Delta \vec{x}=0$, has the value 
$\tilde{C}\left(0,\dots\right)\alpha^PA$. Every graph will have
a factor $\alpha^P$ and a factor $A^G$, where $G$ is the number
of sets in the graph. So the factor $\mathcal{N}$ alwais appears
multiplied by $A$. By imposing $\mathcal{N}A\gg 1$, we obtain that
the only contributions to the correlation function of the sum
of patterns comes from the two point correlation function
of the single pattern: all the Wick formulas are valid.
In order that $\mathcal{N}A\gg 1$, the mean number of scattering particles
inside each area $A$ must be large: many pattern must overlap,
in each point.

\section{Siegert relation for the near field speckles.}

Each scatterer of the sample generates a diffraction pattern
which, at least in its far field, becomes larger and larger linearly,
as the distance $z$ from the screen and the sample is made longer.
So, for $z$ longer than a given distance, many diffraction patterns
overlap: the Wick formulas should become valid. Unfortunately, the 
considerations of the previous section cannot be applied directly,
since the area the diffraction pattern has not been already defined
in a quantitative way.

Now we will prove the Wick formula for
the case of Siegert relation, by using the formalism developed
in the previous section.

From the results of the previous section, we can evaluate the
intensity correlation function of the sum of the patterns,
at a given $z$:
\begin{eqnarray}
\left<\left|E_z\left(\vec{x}\right)\right|^2
\left|E_z\left(\vec{x}+\Delta \vec{x}\right)\right|^2\right>
\label{teoria_eq_corr4_somma_pattern}
=\\
\mathcal{N}
\int{
\left|\tilde{E}_z\left(\vec{x}\right)\right|^2
\left|\tilde{E}_z\left(\vec{x}+\Delta \vec{x}\right)\right|^2
d\vec{x}}
+ \nonumber \\
\mathcal{N}^2
\int{
\left|\tilde{E}_z\left(\vec{x}\right)\right|^2
\left|\tilde{E}_z\left(\vec{y}+\Delta \vec{x}\right)\right|^2
d\vec{x} d\vec{y}}
+ \nonumber \\
\mathcal{N}^2
\int{
\tilde{E}_z\left(\vec{x}\right)
\tilde{E}^*_z\left(\vec{x}+\Delta \vec{x}\right)
\tilde{E}^*_z\left(\vec{y}\right)
\tilde{E}_z\left(\vec{y}+\Delta \vec{x}\right)
d\vec{x} d\vec{y}}
+ \nonumber \\
\mathcal{N}^2
\int{
\tilde{E}_z\left(\vec{x}\right)
\tilde{E}_z\left(\vec{x}+\Delta \vec{x}\right)
\tilde{E}^*_z\left(\vec{y}\right)
\tilde{E}^*_z\left(\vec{y}+\Delta \vec{x}\right)
d\vec{x} d\vec{y}} \nonumber
\end{eqnarray}
where $\tilde{E}_z\left(\vec{x}\right)$ is the field from a single
scatterer, at a distance $z$. Assuming that the phases are random,
and considering that the correlation function of any field
does not change with $z$:
\begin{eqnarray}
\left<I_z\left(\vec{x}\right)
I_z\left(\vec{x}+\Delta \vec{x}\right)\right>
=
\mathcal{N}
\int{
\tilde{I}_z\left(\vec{x}\right)
\tilde{I}_z\left(\vec{x}+\Delta \vec{x}\right)
d\vec{x}}
+ \\
\mathcal{N}^2
\left[\int{
\tilde{I}_0\left(\vec{x}\right)
d\vec{x}}\right]^2
+
\mathcal{N}^2
\left|\int{
\tilde{E}_0\left(\vec{x}\right)
\tilde{E}^*_0\left(\vec{x}+\Delta \vec{x}\right)
d\vec{x}}\right|^2 \nonumber
\end{eqnarray}
The first term on the right hand side depends on $z$: it is the
intensity correlation function of the diffraction pattern.
Since the diffraction pattern becomes larger as $z$ increases,
while the total intensity keeps its value, the 
intensity correlation function of the diffraction pattern 
decreases, and vanishes as $z\to \infty$.

In order that Siegert relation holds, for a finite value of $z$,
we must impose that the term with the four point correlation
function is negligible compared to the two point ones:
\begin{equation}
\mathcal{N}
\int{
\tilde{I}_z\left(\vec{x}\right)
\tilde{I}_z\left(\vec{x}+\Delta \vec{x}\right)
d\vec{x}}
\ll
\mathcal{N}^2
\left[\int{
\tilde{I}_0\left(\vec{x}\right)
d\vec{x}}\right]^2
\end{equation}
The first term can be substituted by its higher value, the one
with $\Delta \vec{x}=0$:
\begin{equation}
\mathcal{N}
\frac{ \displaystyle
\left[\int{
\tilde{I}_z\left(\vec{x}\right)
d\vec{x}}\right]^2
}{ \displaystyle
\int{ \tilde{I}^2_z\left(\vec{x}\right)
d\vec{x}}
}
\gg 1
\end{equation}
The fraction represents the area $A$ covered by the diffraction
pattern:
\begin{equation}
\mathcal{N}A \gg 1
\label{teoria_eq_condizione_gaussianita}
\end{equation}
In order that Siegert relation holds, we need that many particles
scatter light inside a single diffraction pattern, that is,
any point of the screen must be hit by light coming from many
particles. This can be obtained without changing $\mathcal{N}$,
but simply increasing $z$, thus increasing $A$.

It should be noted that the validity of Vick formulas for a
given $z$ does not mean that the field is completely gaussian.
For example, we have shown that 
$\tilde{C}^z_I\left(\Delta \vec{x}\right) \to 0$ for 
$z\to \infty$, where 
$\tilde{C}^z_I\left(\Delta \vec{x}\right)$ is the intensity 
correlation function of the diffraction pattern, defined as
$ \int{\tilde{I}_z\left(\vec{x}\right)
\tilde{I}_z\left(\vec{x}+\Delta \vec{x}\right) d\vec{x}}$.
But this does not imply any uniform convergence. Its integral,
$\int {\tilde{C}^z_I\left(\Delta \vec{x}\right) d\Delta \vec{x}}$,
for example, is a constant, and does not vanishes as $z\to \infty$.
This means that we can build suitable linear operators, acting on the field,
yelding
quantities which do not have a gaussian distribution.
A dramatic example can be obtained considering the scattering
from a two dimensional screen, with many holes of a given shape.
As $z\to \infty$, the field meets the Vick formulas ever better.
But it is alwais possible to analyze an area, bigger than the
diffraction pattern of each hole, and to recover the shape
of the holes. This can be done by deconvolving the field by a
suitable function: it's the operation made by a lens, which creates
an image of the holes. The deconvolution gives any information
about the sample, including the fourth order correlations: the
deconvolved field is not gaussian. The gaussianity is only local:
once we defined an area, corresponding to the aperture of a lens,
there's a distance beyond which we are not able to recover the
shape of each hole, and so informations on higher order
correlation functions than second order ones are lost.

We can conclude that Eq. (\ref{teoria_eq_condizione_gaussianita})
implies only a local gaussianity; gaussianity is valid only when
considering points inside an area small compered with the
diffraction pattern of each scatterer. On the other hand, the
knowledge of the field on a whole plane allows to recover
any information on the correlation function of any order.

\section{Vanishing of the $\left<EE\right>$ correlations.}

In Eq. (\ref{teoria_eq_corr4_somma_pattern}) we neglected the 
terms like $\int \tilde{E}\left(\vec{x}\right)
\tilde{E}\left(\vec{x}+\Delta \vec{x}\right) d\vec{x}$. Such terms should give
contributions like $\left<E\left(\vec{x}\right)
E\left(\vec{x}+\Delta \vec{x}\right)\right>$ in the Vick formulas; we
neglected them assuming that the phases are random. In this section we
will analyze the conditions under which this happens.

In general, $\left<EE\right>$ correlations are not negligible. For example,
we can consider an opaque screen, with a transmission coefficient dependent
on the position, with gaussian distribution. We know that Vick formulas
hold. Now, we send a beam through it, and measure the outgoing intensity
correlation function, immediately after the screen. Since all the points
are in phase, we can assume that the field is real. The Vick formula
for the intensity correlation function states that 
$C_I\left(\Delta \vec{x}\right)=\left<I\right>^2+2\left|
C_E\left(\Delta \vec{x}\right)\right|^2$, due to the not
negligible contribution of the term $\left<EE\right>$, which becomes equal
to $\left<EE^*\right>$. Another example can be found in the theory of
shadowgraph: the term $\left<EE\right>$ is responsible for the oscillations
of the transfer function defined in Eq. 
(\ref{teoria_eq_transfer_function_shadowgraph}).

Now we derive the equations giving the evolution of $\left<EE\right>$ as
$z$ increases. We define:
\begin{equation}
F_z\left(\Delta \vec{x}\right)=
\int \tilde{E}_z\left(\vec{x}\right)
\tilde{E}_z\left(\vec{x}+\Delta \vec{x}\right) d\vec{x}
\end{equation}
The Fourier transform of $F\left(\Delta \vec{x}\right)$ is:
\begin{equation}
F_z\left(\vec{q}\right)=
\tilde{E}_z\left(\vec{q}\right)
\tilde{E}_z\left(-\vec{q}\right)
\end{equation}
We can notice that $F\left(\vec{q}\right)$ is the power spectrum if 
$\tilde{E}\left(-\vec{q}\right)$ is the complex conjugate of 
$\tilde{E}\left(\vec{q}\right)$,
that is if $\tilde{E}\left(\vec{x}\right)$ is real.
By using Eq. (\ref{teoria_evoluzione_z_campo}), we obtain the evolution
of $F_z\left(\vec{q}\right)$:
\begin{equation}
F_z\left(\vec{q}\right)=e^{\displaystyle 2i \sqrt{k^2-q^2} z }
\tilde{E}_0\left(\vec{q}\right)\tilde{E}_0\left(-\vec{q}\right)
\end{equation}
This gives the evolution equation of $F_z\left(\vec{q}\right)$:
\begin{equation}
F_z\left(\vec{q}\right)=e^{\displaystyle 2i \sqrt{k^2-q^2} z }
F_0\left(\vec{q}\right)
\label{teoria_eq_evoluzione_F_come_campo}
\end{equation}
Comparing this equation with Eq. (\ref{teoria_evoluzione_z_campo}),
we see that $F_z\left(\vec{q}\right)$ evolves like
the electric field $E_z\left(\vec{q}\right)$, but two times
faster than it, as $z$ increases.

The root mean square amplitude of $F_z\left(\vec{q}\right)$
is a conserved quantity; since $F_z\left(\vec{x}\right)$ gets
larger and larger as $z$ increases, its amplitude must decrease
like $1/z$. We can thus define a condition which is enough to ensure that
the terms $\left<EE\right>$ are negligible: the diffraction pattern 
must be much larger than the correlation lenght. This is implied by
Eq. (\ref{teoria_eq_condizione_gaussianita}).
The gaussianity condition expressed by Eq.
(\ref{teoria_eq_condizione_gaussianita}) is met if many diffraction patterns
overlap in every point. This implies that the diffraction
pattern of each object must be much larger than the object itself, and than
its correlation function, at least if the objects themselves do not
overlap.

Some difficulties arise when we consider the power spectrum, or the
Fourier transform of $\left<EE\right>$ terms. As we already explained,
the root mean square value of $\left<EE\right>$ does not depend on $z$.
A Fourier transform, made over a whole plane at a given $z$,
could be divergent, for some values of $q$, as $z$ increases. 
For example, we consider a Fourier transform made
on a given area $S$, and we evaluate its mode with wavelength 0,
that is, the integral of $F_z\left(\vec{x}\right)$ over $S$.
It is proportional to $S/z$. We can consider a square area $S$, of side
$2\pi/q_1$, where $q_1$ is the wavevector of the longest wavelenght
Fourier mode of the square $S$. So $F_z\left(\vec{q}=0\right) 
\propto 1/q_1^2z$. Once we selected a $q_1$, the lowest wavevector
we will consider, in order that $\left<EE\right>$ is negligible with respect
to a given value, independent on $z$, we must impose a $z \propto 1/q_1^2$.

A more quantitative result can be obtained by considering the 
evaluation of the Fourier transform on $S$ as the evaluation of 
the Fourier transform on the whole plane, followed by the convolution with the 
Fourier transform of $\chi_S\left(\vec{x}\right)$. This is equivalent
to considering the discretization of the allowed wavelengths, due to
a finite area $S$. Near a given value of $q$, the exponential term in Eq. 
(\ref{teoria_eq_evoluzione_F_come_campo}) makes an oscillation
in about $k/qz$. The discretized intervals are spaced by $q_1$:
if $k/qz\ll q_1$ the oscillations are avereged and vanish.
In general, the oscillations will be more visible for small values of $q$.
In order that the oscillations are never visible, $k/q_1z\ll q_1$:
once we have selected $q_1$, that is the side of $S$, we must
provide that:
\begin{equation}
z\gg \frac{k}{q_1^2}.
\label{teoria_eq_condizione_non_overlap}
\end{equation}

In shadowgraph technique, Eq. (\ref{teoria_eq_condizione_non_overlap})
means that the oscillations
of the transfer function are so fast that they cannot be resolved by the
sensor, and are thus averaged.

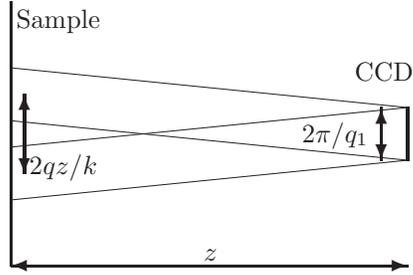
\begin{figure}
\begin{center}
\begin{picture}(150,100)(0,0)
\thicklines
\put(0,0){\line(0,1){100}}
\Thicklines
\put(150,40){\line(0,1){20}}
\thinlines
\path(150,40)(0,25)
\path(150,60)(0,45)
\path(150,40)(0,55)
\path(150,60)(0,75)
\thicklines
\put(5,50){\vector(0,-1){15}}
\put(5,50){\vector(0,1){15}}
\put(7,35){$2qz/k$}
\put(140,50){\vector(0,-1){10}}
\put(140,50){\vector(0,1){10}}
\put(110,47){$2\pi/q_1$}
\put(130,70){CCD}
\put(2,90){Sample}
\put(75,0){\vector(1,0){75}}
\put(75,0){\vector(-1,0){75}}
\put(73,2){$z$}
\end{picture}
\end{center}
\caption{Description of the condition of non overlapping of the
scattering regions.}
\label{teoria_fig_regioni_non_sovrapposte}
\end{figure}

Equation (\ref{teoria_eq_condizione_non_overlap})
has a geometrical interpretation. The vanishing of $\left<EE\right>$
can be expressed in terms of Fourier modes:
\begin{equation}
\left<E\left(\vec{q}\right) E\left(-\vec{q}\right)\right> =0
\label{teoria_eq_corr_angoli_simmetrici}
\end{equation}
The beams,
scattered by a modulation with wavevector $q$, hit
the sensor at an angle $q/k$. Every modulation with  wavevector $q$
scatters at two simmetric angles; the resulting modulation on the sensor
is thus given by light coming from two different regions, of area $S$,
whose distance is about $2qz/k$. If the distance is longer than 
$2\pi/q_1$, the side of $S$, the regions do not overlap: see Fig. 
\ref{teoria_fig_regioni_non_sovrapposte}. Equation 
(\ref{teoria_eq_condizione_non_overlap})
states that we must provide that the regions do not overlap. This
means that light collected at symmetric angles is not correlated, as
required by Eq. (\ref{teoria_eq_corr_angoli_simmetrici}).
This condition ensures that the field is gaussian, only if the density of
scatterers $\mathcal{N}$ is so that $\mathcal{N}S\gg 1$.

For the scattering from a thin sample, the intensities of 
the beams scattered at two symmetric angles are equal, and the phases
are defined. Since the angles are symmetric, the interference of the
scattered beams with the much intense transmitted beam gives two 
interference patterns, sinusoidal modulations, with the same wavevector, and 
a given phase. Changing $z$, the two diffraction patterns change their phase;
at some $z$ they sums, and at other values they cancel out. This is the origin
of the oscillations in the transfer function of shadowgraph technique,
defined in Eq. (\ref{teoria_eq_transfer_function_shadowgraph}). If 
the condition of Eq. (\ref{teoria_eq_condizione_non_overlap}) is met,
the phases of the beams scattered at symmetric directions is random:
on average, the transfer function is constant.

The vanishing of the $\left<EE\right>$ terms can be obtained also
by increasing the thickness of the sample. When we pass from the
two dimensional, Raman Nath scattering to the three dimensional,
Bragg scattering, the correlations between the two beams
scattered at the symmetric angles by a given sinusoidal modulation
are not preserved. In shadowgraph language, the transfer function
oscillations are washed out by superposing many layers, at different
$z$. The thickness of the sample $\delta z$ must meet the condition
$\delta z>k/q_1^2$.

\section{Homodyne near field speckles.}
\label{teoria_sezione_ONFS}

This technique has been presented very recently \cite{carpineti2000,
carpineti2001}.
The device for the measurement of the Near Field Speckles is,
basically, a misfocused dark field microscope. The transmitted
beam is removed, and the image is due only to the light scattered
from the sample.

Some of the parameters of the system must be selected: the
distance $z$ from the sample to the focal plane of the objective,
or from the CCD, if the objective is missing; the diameter $D$
of the sample and of the incident beam;
the superficial particle density of the sample.
The parameters must be selected on the basis of the required
wavevector range $\left[q_{min},q_{max}\right]$. The
ratio $q_{max}/q_{min}$ cannot exceed two decades, due to the
finite size and discretization introduced by the CCD sensor.
The range $\left[q_{min},q_{max}\right]$ is generally selected in
order to cover interesting wavelenghts of the sample: for example,
from one tenth to ten diameters of the particles, in the case of 
the scattering from a monodisperse colloid. Three conditions must be 
fulfilled.

\begin{enumerate}
\item
The misfocusing $z$ must be selected in order to meet the condition
expressed by Eq. (\ref{teoria_eq_condizione_non_overlap}),
with $q_1=q_{min}$,
in order that the correlations $\left<EE\right>$
vanish:
\begin{equation}
z\gg \frac{k}{q_{min}^2}
\label{teoria_eq_condizione_non_overlap_ONFS}
\end{equation}
This condition is stronger than 
$z\gg k/\left(q_{typ}q_{min}\right)$, where
$2\pi/q_{typ}$ is the typical diameter of the particles.
Since $2\pi/q_{min}$ is the side $L$ of the images we take,
Eq. (\ref{teoria_eq_condizione_non_overlap_ONFS})
implies that $q_{typ}z/k\gg L$: the diffraction pattern of
each particle covers a surface much bigger than the observed
one, as required in order that the Fourier transform of the
field can be considered gaussian.
\item
In order that the field is gaussian,
many diffraction patterns must overlap, in each point.
Under the condition of Eq. (\ref{teoria_eq_condizione_non_overlap_ONFS}),
in order to fulfill Eq. (\ref{teoria_eq_condizione_gaussianita}),
we must only provide that there are many particles in the surface
$S$ covered by an image:
\begin{equation}
\mathcal{N}S\gg 1
\label{teoria_eq_condizione_gaussianita_ONFS}
\end{equation}.
For particles suspended in a three dimensional volume, 
we can define the superficial particle density by multipling 
the volumetric particle density by the thickness of the cell.
In order to fulfill condition 
(\ref{teoria_eq_condizione_gaussianita_ONFS}), we can increase
the volumetric particle density, or increase the thickness of the
sample. Care must be teken in order to avoid multiple scattering.
\item
The images we take must
collect light scattered at any angle by the sample. The highest
wavevector we want to measure is $q_{max}$; in order that
the sensor collect light scattered by that wavevector,
coming from any area of the sample, its diameter $D$ must satisfy:
\begin{equation}
D\gg \frac{q_{max}}{k}z
\label{teoria_eq_condizione_diametro}
\end{equation}
This condition ensures that the sensor cannot see the sample 
boundaries: the sample can be considered as infinite. If the diameter
is much less than the one imposed by Eq. 
(\ref{teoria_eq_condizione_diametro}), the speckles 
are governed by the classical, Van Cittert and Zernike
theorem.
\end{enumerate}

Under Eq. (\ref{teoria_eq_condizione_non_overlap_ONFS}),
(\ref{teoria_eq_condizione_gaussianita_ONFS}) and
(\ref{teoria_eq_condizione_diametro}),
Siegert relation holds:
%
%
\begin{equation}
\label{teoria_eq_siegert_2d}
C_I\left(\Delta \vec{x}\right) = \left<
I\left(\vec{x}\right) I\left(\vec{x}+\Delta \vec{x}\right)
\right> = \left< I \right>^2 + \left| 
C_E\left(\Delta \vec{x}\right)
\right|^2
\end{equation}
where the $\left<\cdot\right>$ is the mean over $\vec{x}$ and
$C_E\left(\Delta \vec{x}\right)$ is the field correlation
function.
The intensity we measure in a point is not directly connected with any
physical part of the sample: each speckle is generate by the
superposition of many diffraction patterns.

The measurement of the intensity
allows to recover the modulus of the field correlation function
through Eq. (\ref{teoria_eq_siegert_2d}).
Since $C_E\left(\Delta \vec{x}\right)$ is the Fourier transform of
the power spectrum, which is symmetric and real, 
$C_E\left(\Delta \vec{x}\right)$ is also real.
Moreover, if we think $C_E\left(\Delta \vec{x}\right)$ is alwais
positive, we can calculate it by extracting a square root.
Then,
we Fourier transform $C_E\left(\Delta \vec{x}\right)$,
thus obtaining $S_E\left(\vec{q}\right)$:
\begin{equation}
S_E\left(q\right) =
\mathcal{F}\left[
\sqrt{C_I\left(\vec{x}\right)-\left<I\right>^2}
\right]\left(q\right)
\end{equation}

Using Eq. (\ref{teoria_eq_campo_campione_sottile}),
we obtain, for a thin sample:
\begin{equation}
k^2S_{\delta l}\left(q\right) =
\frac{1}{\left<I\right>}
\mathcal{F}\left[
\sqrt{C_I\left(\vec{x}\right)-\left<I\right>^2}
\right]\left(q\right)
\end{equation}

Using Eq. (\ref{teoria_relazione_IQ_Ez_di_q}), we obtain
the scattered intensity:
\begin{equation}
I\left[Q\left(q\right)\right] =
\mathcal{F}\left[
\sqrt{C_I\left(\vec{x}\right)-\left<I\right>^2}
\right]\left(q\right)
\end{equation}

The results do not depend on $z$. The misfocusing $z$
must be enough, in order that the field is gaussian, but
its value does not affect the results.

The extraction of the square root of the difference between two
experimental data is a dangerous operation, since
the difference could be negative. In general, as any other
inversion of experimental data, it involves an increase and a 
distorsion of noise. Chapter \ref{capitolo_confronto_ONFS_ENFS}
provides a detailed description of this problem.
This kind of problems are avoided by
using ENFS or SNFS, described in Sects.
\ref{teoria_sezione_ENFS} and \ref{teoria_sezione_SNFS}.

We must notice that Eqs. 
(\ref{teoria_eq_condizione_non_overlap_ONFS}) and
(\ref{teoria_eq_condizione_diametro}) give
$D\gg q_{max}/q_{min}^2$. We can think $q_{max}/q_{min}$ as the
wavevector dynamic range we want to measure; hopefully
it can be about one hundred for spatial measurements. On the
other hand, $2\pi/q_{min}$ is of the order of some length
of the particles, for example ten times. For example,
if we consider $10\mathrm{\mu m}$ colloids, each image must cover
about $100\mathrm{\mu m}$. In order to cover two decades in wavelength,
we need a $D$ about one hundred times wider: about $1\mathrm{cm}$.
This is not a huge length; moreover,
from the industrial point of view, there's no problem
in making many acquisitions, with different magnifications and
$z$, for every wavevector range, since each acquisition needs
no accurate positioning.

On the other hand,
in some cases, for scientific purposes, $D$ should be too wide.
This is the case of 
measurements of non-equilibrium fluctuations, described in 
Chapt. \ref{capitolo_dinamico_SNFS}. In this case,
we want to evaluate power spectra on two decades in spatial frequencies.
The dimension of the largest fluctuations is about 
one tenth of millimeter: a good statistical sample is about
a millimeter large, and the whole sample must be two decades
bigger: about $10\mathrm{cm}$.
The building of a Soret cell, or a free diffusion
cell, with such an big diameter can be avoided by using SNFS,
which will be described in Section \ref{teoria_sezione_SNFS}

\section{Heterodyne near field speckles.}
\label{teoria_sezione_ENFS}

We developed this technique very recently; the device has been patented
\cite{brevetto,giglio2001}.
The setup for hEterodyne Near Field Speckles measurement
is identical to ONFS one, but the beam stop is missing.
The transmitted beam is not removed, and the image is due
to the interference of the light scattered from the sample
with the transmitted beam. 

Two parameters of the system must be selected: the
distance $z$ from the sample to the focal plane of the objective,
or from the CCD, if the objective is missing and the diameter $D$
of the sample and of the incident beam. In ENFS,
the superficial particle density of the sample plays no role.
The parameters must be selected on the basis of the required
wavevector range $\left[q_{min},q_{max}\right]$. The
ratio $q_{max}/q_{min}$ cannot exceed two decades, due to the
finite size and discretization introduced by the CCD sensor.
The range $\left[q_{min},q_{max}\right]$ is generally selected in
order to cover interesting wavelenghts of the sample: for example,
from one tenth to ten diameters of the particles, in the case of 
a monodisperse colloid. Two conditions must be fulfilled.

\begin{enumerate}
\item
The misfocusing $z$ must be selected to meet Eq. 
(\ref{teoria_eq_condizione_non_overlap_ONFS}),
in order that the correlations $\left<EE\right>$
vanish. This condition is stronger than 
$z\gg k/\left(q_{typ}q_{min}\right)$, where
$2\pi/q_{typ}$ is the typical diameter of the particles.
Since $2\pi/q_{min}$ is the side $L$ of the images we take,
Eq. (\ref{teoria_eq_condizione_non_overlap_ONFS})
implies that $q_{typ}z/k\gg L$: the diffraction pattern of
each particle covers a surface much bigger than the observed
one, as required in order that the Fourier transform of the
field can be considered gaussian.
\item
The images we take must
collect light scattered at any angle by the sample. The highest
wavevector we want to measure is $q_{max}$; in order that
the sensor collect light scattered by that wavevector,
coming from any area of the sample, its diameter $D$ must satisfy
Eq. (\ref{teoria_eq_condizione_diametro}).
This condition ensures that the sensor cannot see the sample 
boundaries: the cample can be considered as infinite. If the diameter
is much less than the one imposed by Eq. 
(\ref{teoria_eq_condizione_diametro}), the speckles 
are governed by the classical, Van Cittert and Zernike
theorem.
\end{enumerate}

We can notice that the conditions expressed by Eq.
(\ref{teoria_eq_condizione_non_overlap_ONFS}) 
and (\ref{teoria_eq_condizione_diametro}) must hold for both 
ONFS and ENFS. On the contrary, in ENFS
no condition is imposed on the particle density,
in analogy with (\ref{teoria_eq_condizione_gaussianita_ONFS}),
since the field does not need to be gaussian.

In general, Eq. (\ref{teoria_eq_condizione_gaussianita_ONFS})
is fulfilled by the sample; in that case, the field is
gaussian, and the ENFS image represents the interference of
a gaussian field with a plane wave. The particle density
can be so small that the field is not gaussian; this does not
mean that the speckles we see represent real objects in the
sample. Each speckle is due to the interference between 
light scattered by many different particles.

Care must be taken in order to avoid multiple scattering.
Since we want avoid
multiple scattering, the scattered intensity is small compared
to the transmitted beam intensity: the second order effects
in $\delta E/E_0$ can be neglected, and Eq. 
\ref{teoria_eq_tecniche_campo_shadowgr_corr_func} holds.
\begin{equation}
C_i\left(\Delta \vec{x}\right) = 
+\frac{2}{I_0}
\Re C_{\delta E}\left(\Delta \vec{x}\right)
+\frac{2}{I_0^2} \Re \left[
E_0^{*2}\left< \delta E\left(\vec{x}\right)
\delta E\left(\vec{x} + \Delta \vec{x}\right)\right>\right]
\end{equation}
Under Eq. (\ref{teoria_eq_condizione_non_overlap_ONFS}),
the $\left<EE\right>$ correlations vanish:
\begin{equation}
C_i\left(\Delta \vec{x}\right) = 
\frac{2}{I_0}\Re C_{\delta E}\left(\Delta \vec{x}\right)
\end{equation}
This means that $S_i\left(q\right)=\frac{2}{I_0}
S_E\left(q\right)$. The measurement of the intensity
allows to recover the field power spectrum.

Using Eq. (\ref{teoria_eq_campo_campione_sottile}),
we obtain, for a thin sample:
\begin{equation}
2k^2S_{\delta l}\left(q\right) =
S_i\left(q\right)
\end{equation}

Under (\ref{teoria_eq_condizione_diametro}),
Eq. (\ref{teoria_relazione_IQ_Ez_di_q}) holds. We obtain
the scattered intensity:
\begin{equation}
I\left[Q\left(q\right)\right] =
\frac{1}{2}I_0 S_i\left(q\right)
\label{teoria_eq_funzionamento_ENFS}
\end{equation}

The results do not depend on $z$. The misfocusing $z$
must be sufficent, in order that the correlations 
$\left<EE\right>$ vanish, but
its value does not affect the results.

The considerations about the diameter $D$ of the sample
hold also for ENFS: Eq. 
(\ref{teoria_eq_condizione_non_overlap_ONFS}) and
(\ref{teoria_eq_condizione_diametro}) give
$D\gg q_{max}/q_{min}^2$. 
This problem has been discussed in Section
\ref{teoria_sezione_ONFS}. The result is that, in 
some cases, the sample and the laser beam have to be extremely
large. In that cases, SNFS can be used instead of ENFS:
that technique will be described in Section \ref{teoria_sezione_SNFS}.

Equation (\ref{teoria_eq_funzionamento_ENFS}) 
must be compared with Eq. 
(\ref{shadowgraph_transfer_function_scattering_measurement}), that
holds for values of $z$ much less than those imposed by Eq. 
(\ref{teoria_eq_condizione_non_overlap_ONFS}). The oscillations in the
sensibility of shadowgraph technique come from the non vanishing of 
$\left<EE\right>$ correlations, essentially due to the phase relation
of the beams scattered at symmetric angles by a thin sample. For
example, the zeroes of the transfer function are due to the
distructive interference of the symmetrically scattered beams. In
ENFS, the phase relation is destroied, because the light that hit the
sensor at symmetric angles comes from different regions.

\section{Schlieren-like near field speckles.}
\label{teoria_sezione_SNFS}

For Schlieren technique, we can use 
Eq. (\ref{teoria_eq_tecniche_campo_schlieren}),
in order to evaluate the power spectrum $S_i$ of the signal
$i\left(\vec{x}\right)$:
\begin{equation}
S_i\left(\vec{q}\right) =
\left\{
\begin{array}{ll}
\frac{2}{I_0}S_E\left(-\vec{q}\right)
&\vec{q}\cdot\vec{n}<0
\\
\frac{2}{I_0}S_E\left(\vec{q}\right)
&\vec{q}\cdot\vec{n}\ge 0
\end{array}
\right.
\end{equation}
We assume the sample is isotropic, so that $I\left(\vec{q}\right)$
depends only on $\left|\vec{q}\right|$.
Using Eq. (\ref{teoria_relazione_IQ_Ez_di_q}):
\begin{equation}
I\left[Q\left(\vec{q}\right)\right] =
\frac{1}{2} I_0 S_i\left(\vec{q}\right)
\label{teoria_eq_funzionamento_SNFS}
\end{equation}
Schlieren technique can alwais be applied to measure the scatterered
intensity, no matter how long the misfocusing is. The sample can be thick
or thin, in the focal plane or away from it: the result is never affected.

Once the dimension of the image has been selected, in order to
observe an interesting range of wavevectors, the diameter of the
sample must fulfill the condition expressed by Eq. 
(\ref{teoria_eq_condizione_diametro}), as in ONFS and ENFS,
but in this case
there's no limitation on $z$. The diameter will be, in general,
sufficient to give a good statistical sample of the particles
we are measuring; $z$ will be as small as we can.

In general, for a thick sample, some of the objects will be too small,
or too far from the focal plane, to be completely resolved. But their
presence will prodece a speckle field, analogous to that of NFS.
We will call this technique Schlieren-like Near Fiels Speckles, since
it behaves like a true Schlieren technique only for big objects in the
focal plane, while for the other cases it allows to measure
the statistical properties of a speckle field.

Equation (\ref{teoria_eq_funzionamento_SNFS}) 
must be compared with Eq. 
(\ref{shadowgraph_transfer_function_scattering_measurement}), that
holds for values of $z$ much less than those imposed by Eq. 
(\ref{teoria_eq_condizione_non_overlap_ONFS}), and without the blade.
The oscillations in the
sensibility of shadowgraph technique come from the non vanishing of 
$\left<EE\right>$ correlations, essentially due to the phase relation
of the beams scattered at symmetric angles by a thin sample. In
SNFS, the phase relation is destroied, because one of the beams 
scattered at symmetric angles is stopped.

\section{Why using NFS instead of classical scattering
measurement?} 

Classical, high angle light scattering works only for
scattering angles higher than some degrees. Small angle
light scattering (SALS) can measure light scattered by
wavelengths of some microns; its main problem is the stray light.
The new techniques allow an accurate subtraction of the stray
light; in particular, for ENFS and SNFS stray light can be 
evaluated and subtracted point by point.

All SALS instruments must include some device
to stop the transmitted beam, like in ONFS. Moreover, the solid state
sensors must be accurately positioned with reference to the focus of
transmitted beam. In an industrial instrument, the position of both
the beam stop and the sensors must be electronically controlled,
in order to correct the deformations due to mechanical stress and
temperature dilatations. On the contrary,
ENFS has an extremely simple and robust structure, which
does not require any adjustment. This makes ENFS suited for industrial
applications of light scattering measurement, like particle sizing.
In Chapter \ref{capitolo_particle_sizing_ENFS}, we will show some particle
size measurements performed with ENFS.

In Chapter \ref{capitolo_dinamico_SNFS},
we will show that SNFS can be used to make a measurement of the power
spectrum of the non equilibrium fluctuations in a free diffusion process.
Such fluctuations have never been observed by SALS, since they are
extremely weak and involve mainly low wavevectors. This shows that
NFS techniques can be used to obtain measurements not possible with
SALS.

\section{Why using NFS instead of classical microscopy?}

\begin{figure}
\begin{center}
\includegraphics{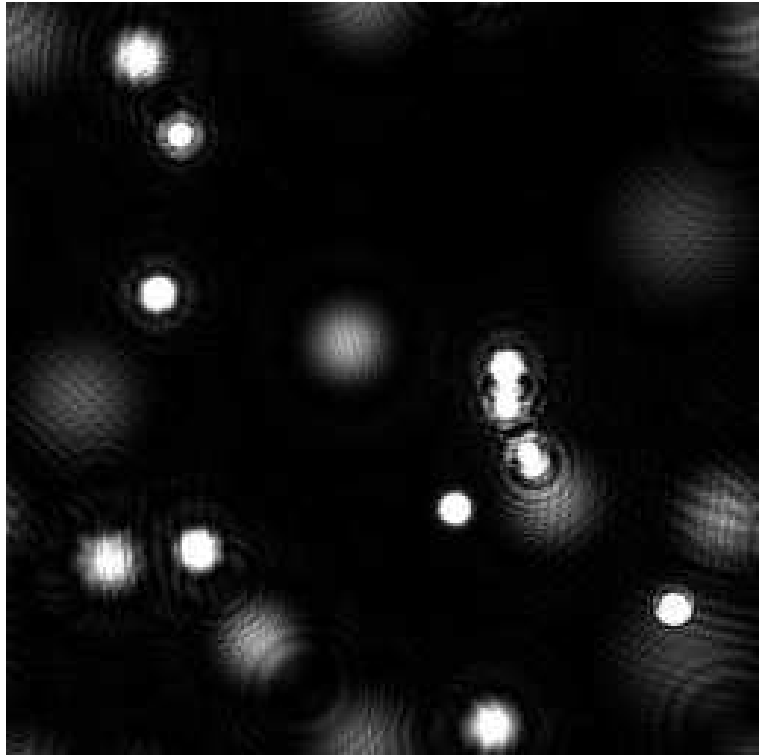}
\end{center}
\caption{Example of dark field image.}
\label{teoria_imm_cerchi_pochi}
\end{figure}
\begin{figure}
\begin{center}
\includegraphics{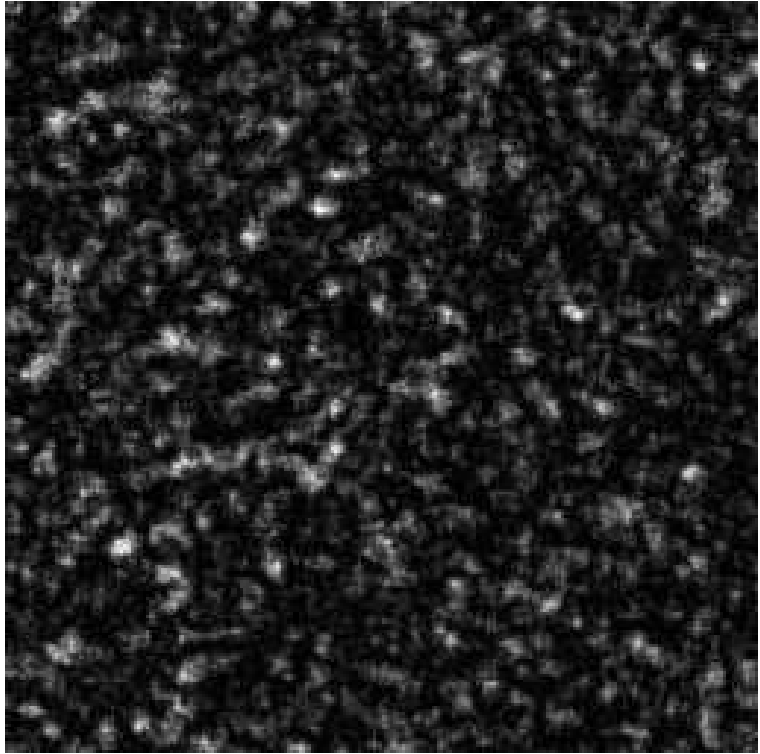}
\end{center}
\caption{Example of Near Field Speckles image.}
\label{teoria_imm_cerchi_scombinata}
\end{figure}
The near scattered field keeps only one feature associated to the
observed sample, the correlation function. For example,
Figure \ref{teoria_imm_cerchi_pochi}
shows a dark field image, 
and \ref{teoria_imm_cerchi_scombinata} shows the corresponding 
NFS image. The correlation function of the two 
fields is the same,
but looking at the second immage we cannot figure that it comes
from a set of discs. With near field scatterig we will never
distinguish an amoeba from a paramecium: it is not a microscopy
technique. Why using NFS instead of dark field?

The first answer comes from the analysis of Figure
\ref{teoria_imm_cerchi_pochi} and
\ref{teoria_imm_cerchi_scombinata}. In Figure
\ref{teoria_imm_cerchi_pochi}, some of the discs are in the focal 
plane, other aren't. If we want to analyze a dark field
image, we must be able to select the particles which are
in the focal plane, and exclude from the analysis all the 
others. On the contrary, NFS gives informations which are
never affected by the misfocusing $z$: it provides 
three dimensional informations, and works well for
thin samples as well as for thick ones.

If we want to analyze, for example, a colloid by a microscopy
technique, we must use a thin sample, in order that the particles can
be focused. If the concentration is low, it could be hard to find even
one particle. Generally, one microscopic image could show only some
particles. On the contrary, NFS can work on thick samples. We can use
a given colloid, with any concentration, and put it in a cell so thick
that it shows a suitable attenuation.

Another reason leads to use ONFS technique
instead of a dark field technique. Dark field image intensity is given by
equation \ref{teoria_eq_tecniche_campo_dark_f_light_path}: every calculation
based on dark field images will concern the square value of the
refraction index fluctuations. In facts, two point correlation
functions of the images will represent four point correlation functions
of the refraction index fluctuations. We consider a fluid,
for which
the refraction index fluctuations has a give distribution.
We are interested in measuring the two point correlation,
but dark field images allow us to work only on four point one. 
Of course, four point correlation function involves the two point one,
but has a non trivial connected contribution. In NFS
images, every connected term in correlation functions vanishes:
we can measure quantities directly connected with two point correlation
functions.

\section{Meaning of the light path correlation function.}

We have shown that NFS allows the measurement
of the light scattered in a quite wide range of angles. If
Reyleight Gans condition is met, the scattered intensity
represents the power spectrum of the sample, evaluated in the
transferred wavevector. For the scattering at small angles, 
the spectrum is evaluated in the direction ortogonal to
the incident beam. A measurement of this componet of the spectrum
leads, through a Fourier transform, to the correlation function
of the light path through the sample:
%
%
\begin{equation}
C_{\delta l}\left(\Delta\vec{x}\right) = \int{
\delta n \left(\vec{x},z\right)
\delta n \left(\vec{x}+\Delta\vec{x},z'\right)
\mathrm{d}z \mathrm{d}z' \mathrm{d}\vec{x}}.
\end{equation}
This quantity is directly accessible from NFS measurements.
Its Fourier transform is the power spectrum for $q_z=0$:
%
%
\begin{multline}
\int{C_{\delta l}\left(\Delta\vec{x}\right)e^{\displaystyle
-i \vec{q}\cdot \Delta\vec{x}}\mathrm{d}\Delta\vec{x}}
=\\
\frac{1}{\left(2\pi\right)^6}
\int \delta n \left(\vec{q}',q_z'\right) \delta n \left(\vec{q}'',q_z''\right)
\\
e^{i\left(\vec{q}'+\vec{q}''\right)\cdot\vec{x}+
i\left(\vec{q}''-\vec{q}\right)\cdot\Delta\vec{x}
+iq_z z+iq_z' z'}
\mathrm{d}\vec{q}' \mathrm{d}\vec{q}''
\mathrm{d}\vec{x} \mathrm{d}\Delta \vec{x}.
\mathrm{d}q_z' \mathrm{d}q_z'' \mathrm{d}z \mathrm{d}z' =\\
\left|\delta n \left(\vec{q},q_z=0\right)\right|^2
\end{multline}

Through a measurement of the scattered light we can know the power spectrum
in the plane perpendicular to the direction of the incident beam. 
This means that the light path correlation function bears less informations
than the refraction index correlation function, but
the light path correlation function is connected to the refraction index
correlation function:
%
%
\begin{multline}
C_{\delta l}\left(\Delta\vec{x}\right) = 
\int{\delta n \left(\vec{x},z\right) 
\delta n \left(\vec{x}+\Delta\vec{x},z+\Delta z\right)
\mathrm{d}\vec{x}\mathrm{d}z\mathrm{d}\Delta z} =\\
\int{C_{\delta n}\left(\Delta\vec{x},\Delta z\right)\mathrm{d}\Delta z}.
\end{multline}

If the sample is isotropic, its power spectrum depends only on the
modulus of the wave vector. If we know the light path correlation
function, we can evaluate its Fourier transform, extend it
to the three dimensions, and then, appling again the Fourier transform,
we obtain the refraction index correlation function. This operation
is generally performed by using the well known Abel transform.

%
\chapter{The experimental system.}
\label{capitolo_sistema_sperimentale}

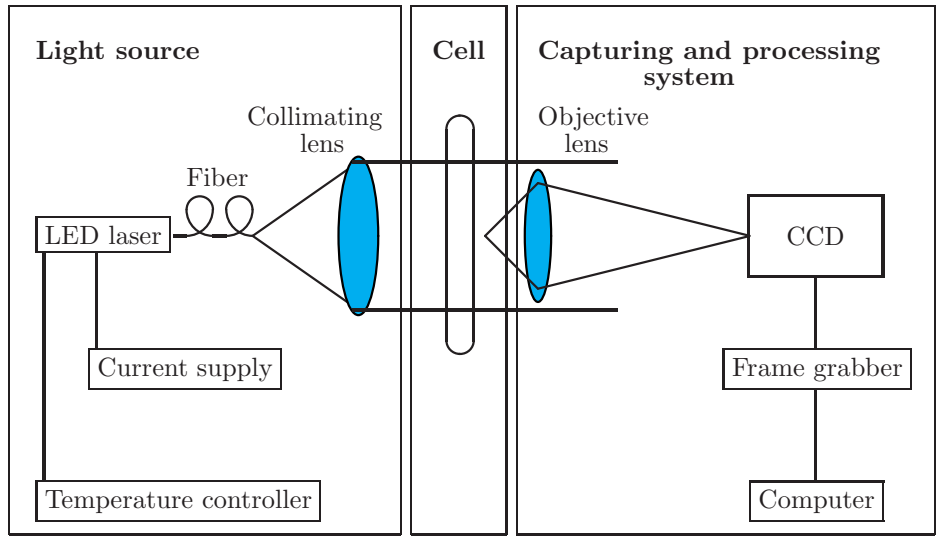
\begin{figure}
\begin{center}
\begin{picture}(350,200)(0,0)
\thinlines
\put(0,0){\line(1,0){148}}
\put(152,0){\line(1,0){36}}
\put(192,0){\line(1,0){158}}
\put(0,200){\line(1,0){148}}
\put(152,200){\line(1,0){36}}
\put(192,200){\line(1,0){158}}
\put(0,0){\line(0,1){200}}
\put(148,200){\line(0,-1){200}}
\put(152,200){\line(0,-1){200}}
\put(188,200){\line(0,-1){200}}
\put(192,200){\line(0,-1){200}}
\put(0,0){\line(0,1){200}}
\put(350,200){\line(0,-1){200}}
\thicklines
\put(10,110){\framebox{LED laser}}
\put(30,60){\framebox{Current supply}}
\put(10,10){\framebox{Temperature controller}}
\put(92,113){\line(10,7){40}}
\put(92,113){\line(10,-7){40}}
\put(62,113){\line(1,0){5}}
\put(67,123){\arc{20}{0.0}{1.5707963}}
\put(72,123){\arc{10}{3.1415927}{6.2831853}}
\put(77,123){\arc{20}{1.5707963}{3.1415927}}
\put(77,113){\line(1,0){5}}
\put(82,123){\arc{20}{0.0}{1.5707963}}
\put(87,123){\arc{10}{3.1415927}{6.2831853}}
\put(92,123){\arc{20}{1.5707963}{3.1415927}}
\put(132,113){\blacken\color{cyan}\ellipse{15}{60}}
\put(33,70){\line(0,1){37}}
\put(13,20){\line(0,1){87}}
\put(67,132){Fiber}
\put(90,155){Collimating}
\put(110,145){lens}
\put(10,180){\bf{Light source}}
\put(130,141){\line(1,0){100}}
\put(130,85){\line(1,0){100}}
\put(170,113){\oval(10,90)}
\put(200,113){\blacken\color{cyan}\ellipse{10}{50}}
\put(160,180){\bf{Cell}}
\put(180,113){\line(1,1){20}}
\put(180,113){\line(1,-1){20}} 
\put(200,133){\line(4,-1){80}}
\put(200,93){\line(4,1){80}}
\put(280,98){\framebox(50,30){CCD}}
\put(200,180){\bf{Capturing and processing}}
\put(240,170){\bf{system}}
\put(200,155){Objective}
\put(210,145){lens}
\put(270,60){\framebox{Frame grabber}}
\put(280,10){\framebox{Computer}}
\put(305,70){\line(0,1){28}}
\put(305,20){\line(0,1){35}}
\end{picture}
\end{center}
\caption{The experimental system.}
\label{exp_sys_fig_schema_generale}
\end{figure}

The whole system is sketched in Fig. \ref{exp_sys_fig_schema_generale}. 
It consists in the
light source, the sample cell and the image forming, capturing and
processing system.
In the first sections we describe all these parts and the criteria we
followed to build our NFS instruments. In Sections
\ref{sect_optical_setup_ENFS_colloidi} and
\ref{misura_snfs_sezione_opt_setup} we describe the three NFS
instruments we built.

\section{The optical system.}

\begin{figure}
\includegraphics[scale=0.82]{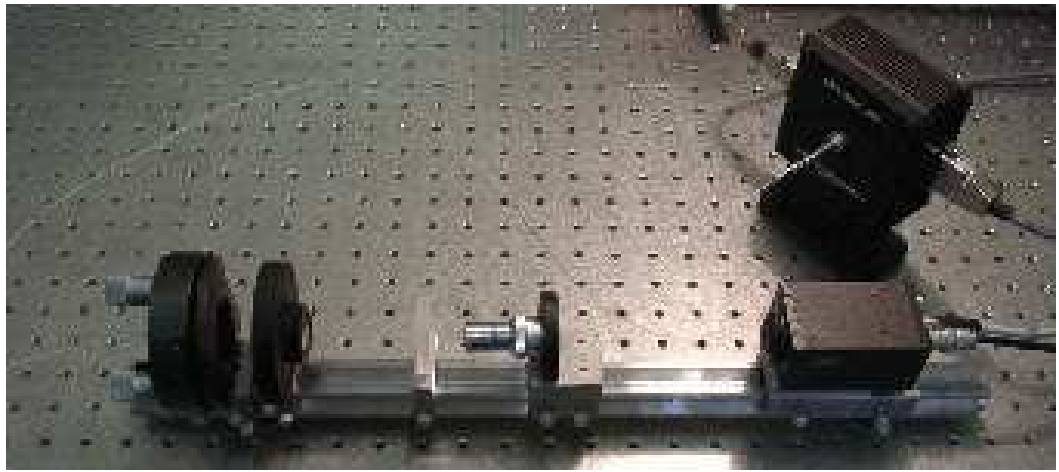}
\caption[Picture of the optical system]
{Picture of the optical system. From left to right, we can see
the adjustable mount that holds the fiber, the collimating lens, the cell,
the microscope objective and the CCD camera.
In the upper right corner there's the laser mount.}
\label{exper_fig_sistema_ottico}
\end{figure}

The optical system is mainly built using elements supplied by Newport.
The optical table is a VH3048 IsoStation. All the elements are mounted on
X26 rails, by using CN26 carriers.
A picture of the system is shown in Fig. \ref{exper_fig_sistema_ottico}.

\begin{figure}
\includegraphics{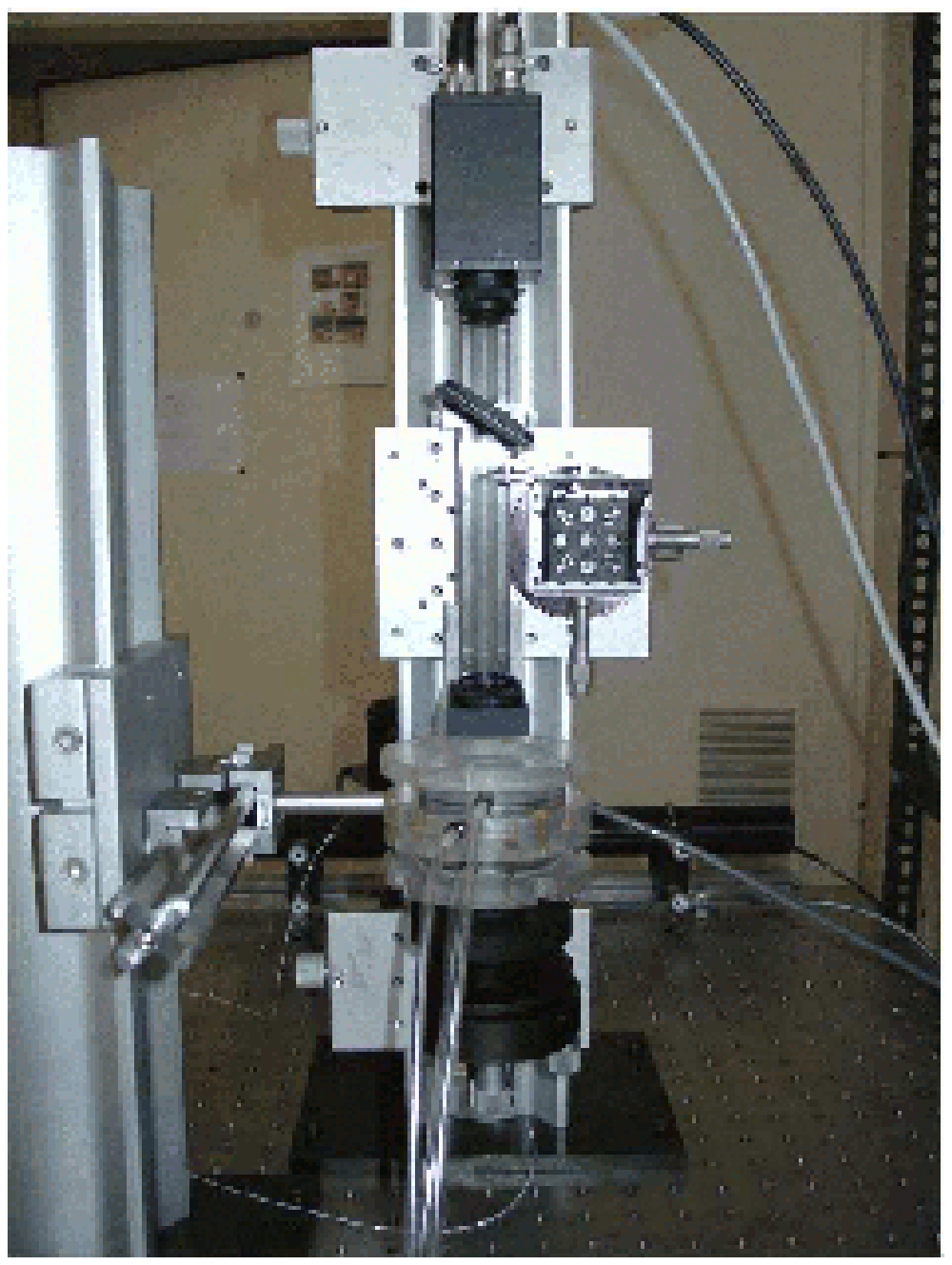}
\caption[The optical system of SNFS]{A view of the optical system for
the experiment described in Chapter \ref{capitolo_dinamico_SNFS}. All
the optical elements are alligned in vertical direction. From
the bottom, we see the optic fiber, held by an adjustable mount, the
collimating lens, the cell, held by the column on the left, the
focusing lens, the blade, the neutral filter, and the CCD camera.}
\label{exp_sys_imm_snfs_setup_tutto}
\end{figure}

For the experiment descibed in Chapter \ref{capitolo_dinamico_SNFS},
the optical axis must be vertical. The X26 rails are held in vertical
position by mounting them on an X95 rail with suitable carriers: see
Fig. \ref{exp_sys_imm_snfs_setup_tutto}.

\section{The light source.}

\begin{figure}
\begin{center}
\includegraphics{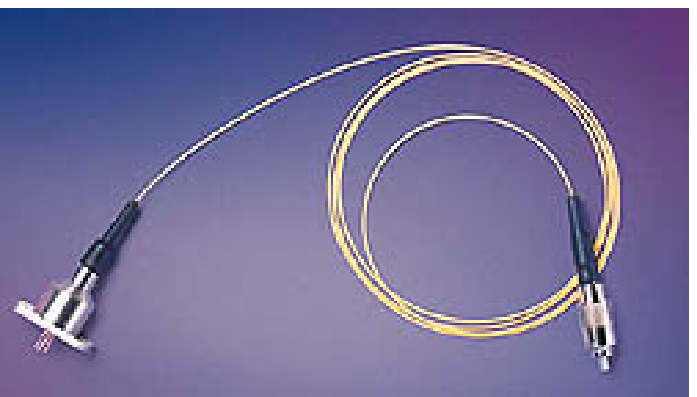}
\end{center}
\caption{The LED laser.}
\label{exp_sys_LED_image1}
\end{figure}

The light source we use is a LED laser, coupled to a single mode
fiber, supplied by Newport. In Tab. \ref{exp_sys_LED_data} we show
the data supplied with the LED. Figure \ref{exp_sys_LED_image1}
shows the LED laser and the fiber.
%
%
\begin{table}
\begin{tabular}{lc}
Item code&LD-635-31A\\
Center wavelength&$635\mathrm{nm}$\\
Wavelength range&$\pm 10\mathrm{nm}$\\
Fiber output power&$1.2\mathrm{mW}$\\
Threshold current&$10\mathrm{mA}$\\
Operating current&$40\mathrm{mA}$\\
Rise/fall time&$1.5\mathrm{ns}$\\
Operating temperature&$-10$ to $40^{\circ}\mathrm{C}$\\
Fiber core/clad diameter&$4/125\mathrm{\mu m}$ - single mode
\end{tabular}
\caption{Data of the LED laser.}
\label{exp_sys_LED_data}
\end{table}
%
%

The LED is housed in a Newport 700P temperature controller mount,
connected to a temperature controller and a current driver.
Since the output wavelenght depends on the temperature,
a temperature controller is needed. The LED laser has a built-in 
monitoring photodiode. The current driver monitors the output power to keep
it constant.

A single mode fiber is directly pigtailed to the LED. The other end of
the fiber is terminated by an FC/PC connector. It is held by a 
Newport SL50BM, a gimbal mount for the adjustment of the azimuth and
elevation, originally built for mounting mirrors or beam splitters. 
The beam output by the fiber is diverging;
we collimate it by sending it through a lens with a focal of 
$5\mathrm{cm}$.

The resulting beam has a diameter of about $2\mathrm{cm}$. This
diameter is enough for our experiments; in general, it must be
selected as a function of the wavevector range
$\left[q_{min},q_{max}\right]$ we want to measure.
For ONFS and ENFS, $D$, the diameter over which the beam intensity is
constant, must be selected in order to fulfill
Eq. (\ref{teoria_eq_condizione_diametro}). The best choice is:
\begin{equation}
D \gtrapprox 50 \frac{q_{max}}{q_{min}^2}
\label{exp_sys_eq_diametro_scelto}
\end{equation}
For SNFS, the beam must have a diameter which covers a good
statistical sample:
\begin{equation}
D \gtrapprox 20/q_{min}.
\label{exp_sys_eq_diametro_scelto_SNFS}
\end{equation}

\begin{figure}
\includegraphics{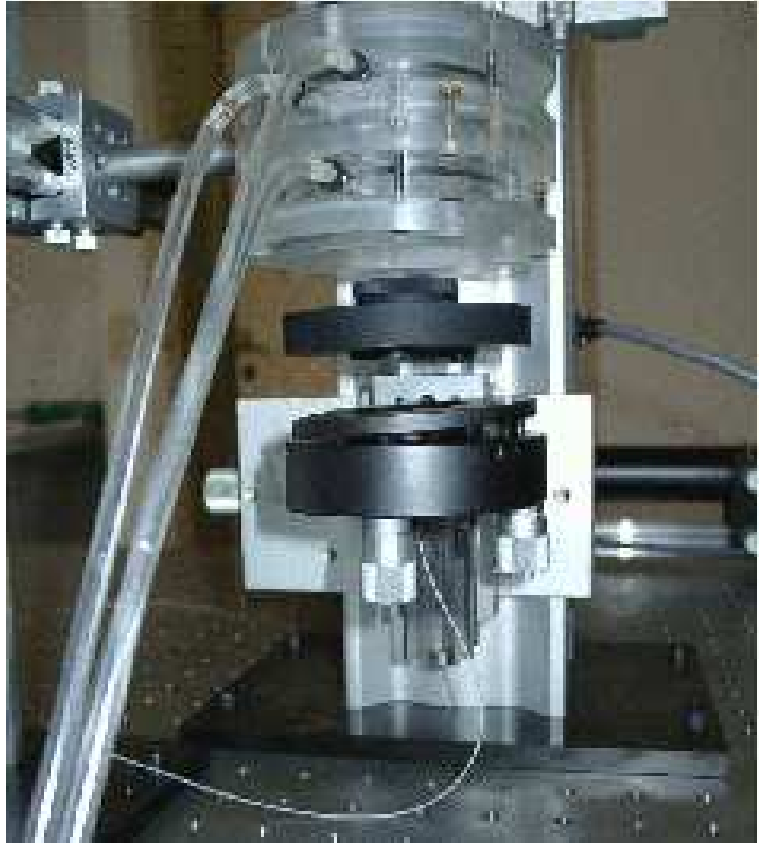}
\caption[The collimator.]{A view of the collimator and the cell.
From the bottom, we see the optic fiber, held by an adjustable
mount,
the collimating lens, and the cell.}
\label{imm_snfs_collimatore}
\end{figure}

Both the adjustment of the direction of the beam and its collimation
are not critical operations. The direction of the beam must be
adjusted to hit a lens, centered at the optical axis, half a meter away
from the fiber end. The 
collimation is checked by measuring the beam diameter on a screen, 
near the lens and one meter away. The collimator is shown in
Fig. \ref{imm_snfs_collimatore}.

The LED laser has been used in order to test its performance 
for industrial applications.
The LED laser is much more compact and robust than a gas laser;
it can operate immediately after it has been powered on
and it does not generate too many heat. For these features
it is ideally suited for industrial applications. Moreover,
the overall cost of a LED laser device included in an industrial
product can be made extremely low, as in the case of CD readers,
though a laboratory LED laser equipement can cost as much as a classical
gas laser.
Moreover, the output of a gas laser must be spatially filtered before
being used. A spatial filter is a critical component in an industrial
equipement, since it must be extremely stable, and must be adjusted by
micrometric actuators controlled by sensors, in order to correct the
deformations due to heating and mechanical stresses. On the contrary,
the single
mode fiber output is more uniform than the output of a spatial filter,
and requires no adjustment.

\section{The cell.}

The liquid samples we measure are held in a cell; the diameter
$D$ must be selected following Eq. (\ref{exp_sys_eq_diametro_scelto})
for ONFS and ENFS, or Eq. (\ref{exp_sys_eq_diametro_scelto_SNFS}) for
SNFS.

For homogeneous samples, like colloids, the thickness must be selected
in order to have a suitable attenuation of the main beam, about 1\%.
For ONFS measurements, the thickness of the cell and the volumetric
particle density must fulfill
Eq. (\ref{teoria_eq_condizione_gaussianita_ONFS}); generally this condition
is spontaneously met.

For ONFS and ENFS measurements, the parallelism
between the windows of the cell is not critical, nor the optical
quality of them. Since the measured scattered light comes from
different regions of the sample, we must provide that it is
homogeneous. This implies that the thickness 
must be uniform, but an optical quality allignment is far beyond
what is needed. On the contrary, SNFS requires optical quality
windows: the well known ``Foucault test'' sees every deformation of
the wavefront, no matter if the associated wavelength is long.

The cells we used are described in detail in Chapters
\ref{capitolo_confronto_ONFS_ENFS} and \ref{capitolo_dinamico_SNFS}.

\section{Objective, beam stop and blade.}

The objective must form an image of a given plane on the CCD
sensor. The magnification $M$ must be selected in order that the
required wavevector range $\left[q_{min},q_{max}\right]$ is inside the
wavevector range the CCD sensor can measure: about
$\left[2\times10^3\mathrm{m}^{-1},2\times10^5\mathrm{m}^{-1}\right]$.
This means that:
\begin{equation}
\left[\frac{q_{min}}{M},\frac{q_{max}}{M}\right]
\subseteq
\left[2\times10^3\mathrm{m}^{-1},2\times10^5\mathrm{m}^{-1}\right]
\label{exp_sys_eq_ingrandimento_scelto}
\end{equation}
Moreover, the numerical aperture of the lens must be enough to resolve
details as small as the smallest wavelength involved,
$2\pi/q_{max}$, or, equivalently, to collect light scattered at an
angle $q_{max}/k$.

In our experiments, we used a 20X microscope objective for high
magnification factors, and an achromatic, $10\mathrm{cm}$ focal length
doublet for magnification factor around $1$.
An achromatic doublet has also been tested for high magnification
factors, since we do not
require the high quality of a microscope objective, nor an extremely wide
numerical aperture.  Experiments proved no different performances of
the doublet compared with the microscope objective, but it was more
difficult to obtain the required magnification.

The objective lens must be placed so that it creates an image of a
given plane on the CCD sensor. For ONFS and ENFS, the plane must be at
a distance $z$ from the sample fulfilling
Eq. (\ref{teoria_eq_condizione_non_overlap}). The best choice is: 
\begin{equation}
z\approx 25 \frac{k}{q_{min}^2}
\label{exp_sys_eq_z_scelto}
\end{equation}
For SNFS:
\begin{equation}
z<\frac{kD}{2q_{max}}
\label{exp_sys_eq_z_scelto_SNFS}
\end{equation}

\begin{figure}
\begin{center}
\includegraphics[scale=0.5,angle=180]{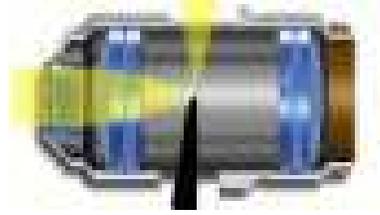}
\end{center}
\caption{Section of the microscope objective and the beam stop.}
\label{exp_sys_objective}
\end{figure}

\begin{figure}
\includegraphics{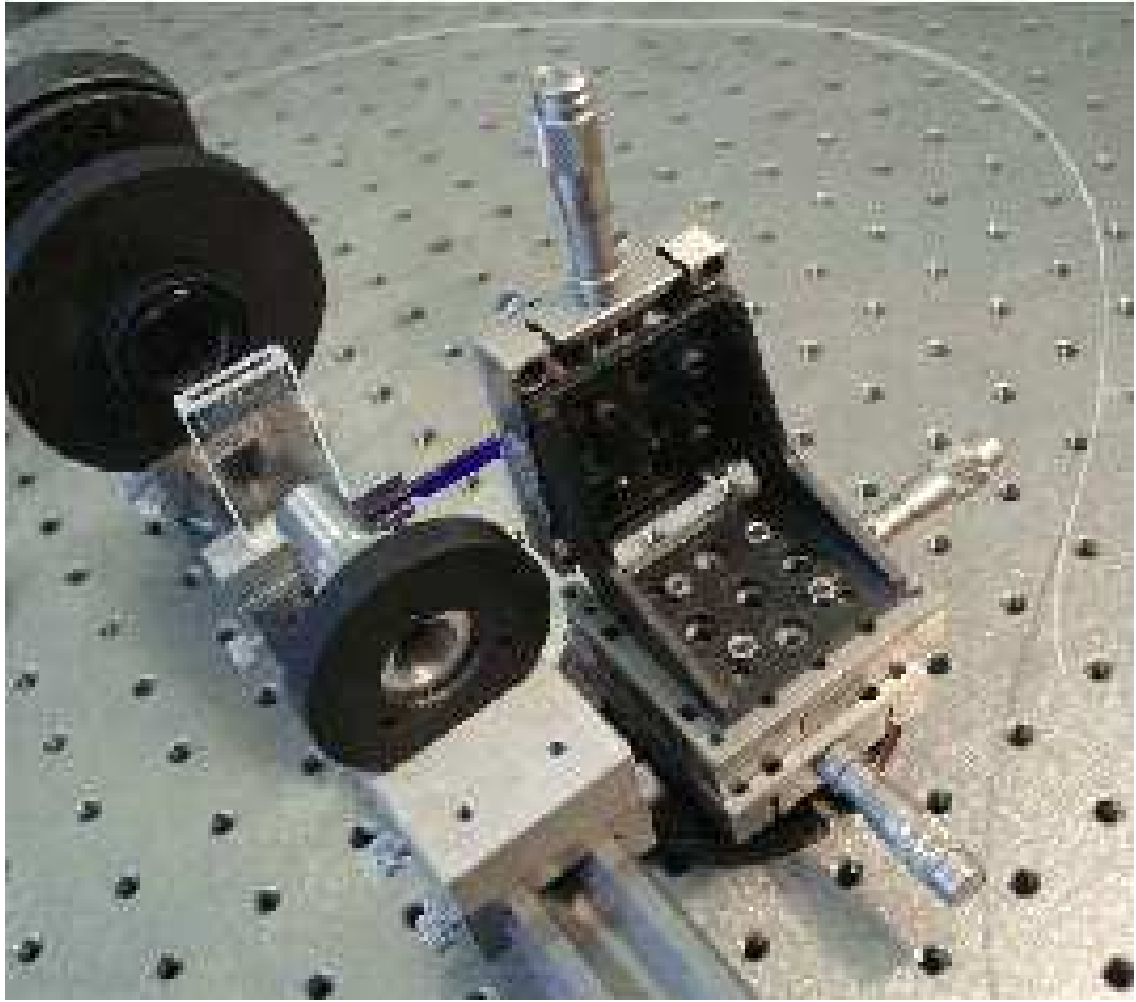}
\caption[Picture of the microscope objective with the beam stop.]
{Picture of the microscope objective with the beam stop. The beam
stop is glued to the blue rod, held by a mount with three micrometric
screws, for the adjustment of the position in every direction.}
\label{exper_fig_beam_stop}
\end{figure}

For ONFS, the transmitted beam, focused by the objective, is stopped
by an opaque or reflective element. In microscope objectives, the
focal plane is inside, between two groups of lenses: we insert the beam
stop through a hole.
We tried three kinds of beam stops: a thin wire, a reflective wedge and
an absorbing disc impressed by on a photographic film.
The wire has a diameter of $70\mathrm{\mu m}$;
 it si stretched in the focal plane
and is positioned by micrometric screws. It reflects the light
inside the objective, and this could, in principle, increase the stray light.
The photographic film we used are high contrast, black and white,
$36\mathrm{mm}$ photographic films. The beam stop is circular, but the 
beam is not completely blocked, thus increasing the stray light.
The wedge was obtained by a steel blade; the edge was kept parallel to the
optic axis. The upper part, in the direction from which the light comes, was
cut at $45^{\circ}$ and polished, in order to obtain a surface that reflects
the main beam outside the 
lens mount, through a second hole. A section of the objective lens is
shown in figure (\ref{exp_sys_objective}).
%
%
This kind of beam stop is not symmetrical with respect to the optical axis.
This could increase the difficulty to process the data.
During the experiments, all the methods showed to be almost equivalent.
Figure \ref{exper_fig_beam_stop} shows the mount that holds the beam stop.

\begin{figure}
\includegraphics{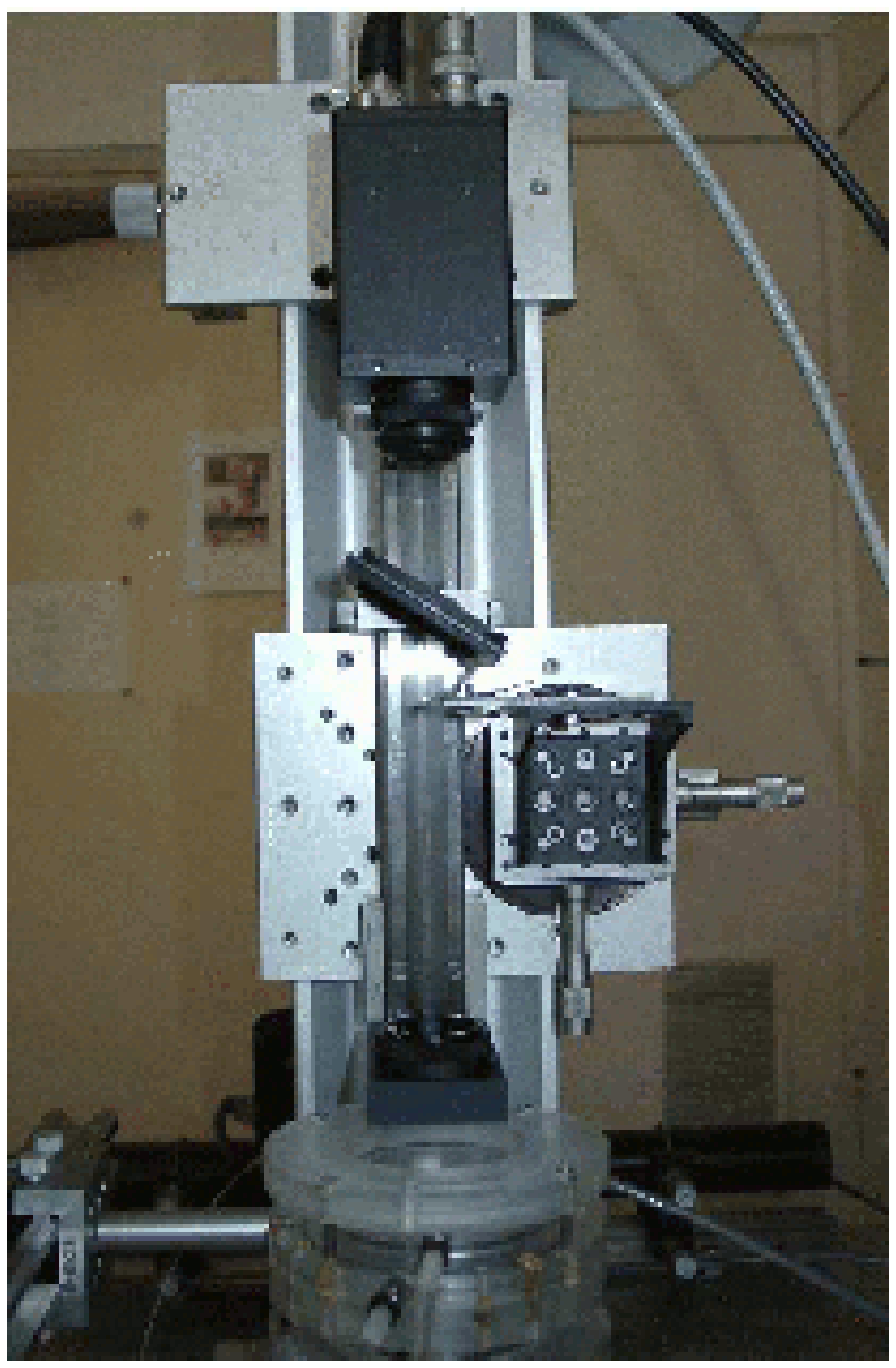}
\caption[The Schlieren system.]{A view of the Schlieren system.
From the bottom, we see the cell, the focusing lens, 
the blade, held by a micrometric mount, the neutral filter
and the CCD camera.}
\label{imm_snfs_obiettivo}
\end{figure}

For SNFS, a blade must be placed in the plane where the transmitted
beam is focused. The blade must be extremely sharp: a razor blade is
required. We mount it on a system with three micrometric screws, in
order to accurately position it in the space. A picture of the
Schlieren system is shown in Fig. \ref{imm_snfs_obiettivo}.

\section{CCD sensor.}

\begin{figure}
\begin{center}
\includegraphics[scale=0.4]{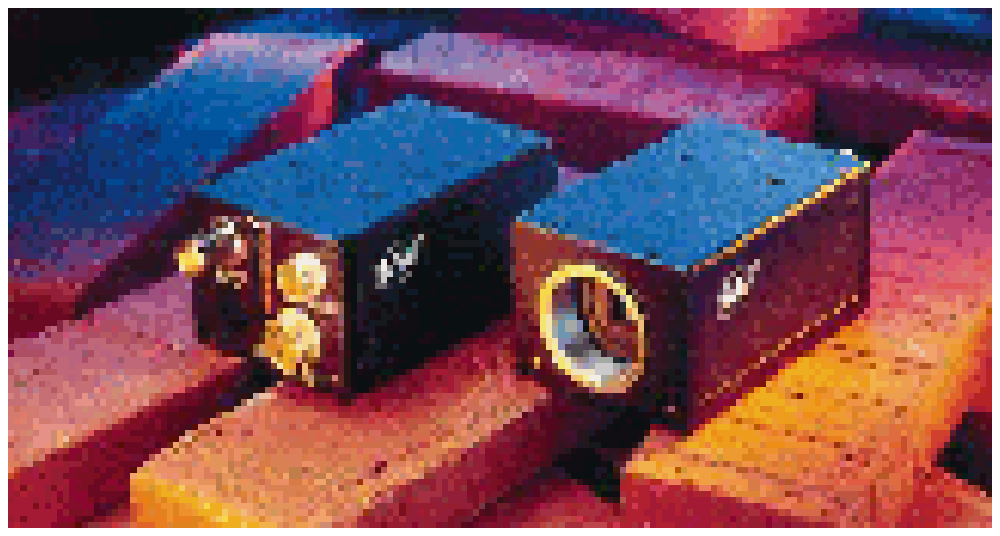}
\end{center}
\caption{Jai CV M50 camera.}
\label{ottica_manuale_jai1}
\end{figure}

We used an industrial CCD camera: JAI CV M50. 
An image is shown in Fig. \ref{ottica_manuale_jai1};
the data are provided in Tab. \ref{ottica_manuale_jai2}.
\begin{table}
\begin{tabular}{l|p{6cm}}
Specifications&CV-M50C\\ \hline
Scanning system&625 lines 25 frames/s\\ \hline
CCD Sensor&Monochrome 1/2" Hyper HAD IT CCD\\ \hline
Sensing area&6.6mm$\times$4.8mm \\ \hline
Picture elements effective&752$\times$582 \\ \hline
Elements in video out&737$\times$575 \\ \hline
Cell size&8.6$\times$8.3 \\ \hline
Resolution (horizontal)&560 TV lines \\ \hline
Resolution (vertical)&575 TV lines \\ \hline
Sensitivity&0.5 lux, f1.4\\ \hline
Sensitivity peak wave length&500nm\\ \hline
Wave length range&400nm - 675nm (sensitivity $>$ 50\%) \\ \hline
S/N ratio&$>$56dB (AGC off, Gamma 1)\\ \hline
Video output& Composite VBS Signal 1.0 $V_{pp}$, 75 $\Omega$\\ \hline
Gamma&0.45 - 1.0\\ \hline
Gain&Manual - Automatic\\ \hline
Scanning&2:1 interlace\\ \hline
Accumulation&Field - Frame\\ \hline
Syncronization&Internal, Xtal-generated - External HD/VD - random trigger\\ \hline
HD Sync input - output&4V, 75 $\Omega$\\ \hline
VD Sync input - output&4V, 75 $\Omega$\\ \hline
Trigger input&4V, 75 $\Omega$\\ \hline
Trigger input duration&$>$ HD interval\\ \hline
WEN output (write enable)&4V, 75 $\Omega$\\ \hline
EEN output (exposure enable)&4V, 75 $\Omega$\\ \hline
Pixel clock out (optional)&4V, 75 $\Omega$\\ \hline
Internal shutter&Off,1/100s, 1/250s, 1/500s, 1/1000s, 1/2000s, 
1/4500s, 1.10000s\\ \hline
Trigger shutter&1/60,1/100s, 1/250s, 1/500s, 1/1000s, 1/2000s, 
1/4500s, 1.10000s\\ \hline
Long time exposures&one field to $+\infty$. Duration between external VD pulses.\\ \hline
Operating temperature&-5$^\circ$C to +45$^\circ$C\\ \hline
Humidity&20\%-80\% non-condensing\\ \hline
Storage temperature&-25$^\circ$C to +60$^\circ$C\\ \hline
Storage humidity&20\%-90\%\\ \hline
Power&12V DC $\pm$10\% 2.5W\\ \hline
Lens mount&C-mount\\ \hline
Dimensions&40mm$\times$50mm$\times$80mm\\ \hline
Weight&245g\\ \hline
\end{tabular}
\caption{Specification of Jai CV M50C camera.}
\label{ottica_manuale_jai2}
\end{table}

The output is a standard CCIR; since it is interlaced, 
an image is alwais formed by two fields acquired with a time delay of
$0.04\mathrm{s}$, although the internal
shutter allows to acquire a single frame in $100\mathrm{ns}$.

The number of pixels and their dimension determine the wavevector
range the CCD can directly measure:
$\left[2\times10^3\mathrm{m}^{-1},2\times10^5\mathrm{m}^{-1}\right]$.
Other wavevector ranges can be covered, by creating on image with a
suitable magnification factor, but we cannot cover more than two
decades.

\section{The acquisition and elaboration system.}

The frame grabber we used is an IC-RGB, from Imaging Technology. It
performs an 8 bit digitalization of three standard composite video
signals from the CCD cameras. It can be used to acquire simultaneously
from two syncronized CCD cameras, for evaluating the
intensity correlation function on two different planes,
or to acquire from a single
camera. 

The syncronyzed acquisition from two cameras can be used to evaluate the
three dimensional correlation function; the meaning of the three
dimensional correlation function is explained in Appendix 
\ref{capitolo_correlazione_tridimensionale}.

The software for image elaboration was developed under Linux,
written in C language. The drivers and the libraries are the
``IC-PCI'' provided by ``GOM Optical Measurements Techniques''.
The algorithms used to process the images are described in Chapt.
\ref{chap_onfs_data_processing} and \ref{chap_enfs_data_processing}.

\section{ONFS and ENFS setup for colloid measurements.}
\label{sect_optical_setup_ENFS_colloidi}

The systems are sketched in Figs. \ref{imm_schema_generale_ENFS} and
\ref{imm_schema_generale_ONFS}.
\begin{figure}
\begin{center}
\begin{picture}(250,150)(0,0)
\blacken\color{red}\path(40,75)(70,95)(160,95)(250,47)(250,103)(160,55)(70,55)(40,75)
\color{black}
\shade\path(82,50)(82,100)(102,100)(102,50)(82,50)
\put(70,75){\blacken\color{cyan}\ellipse{15}{50}}
\Thicklines\color{magenta}\path(235,75)(160,60)(110,75)(160,90)(235,75)
\put(160,75){\blacken\color{cyan}\ellipse{10}{50}}
\blacken\color{white}\path(235,60)(235,90)(250,90)(250,60)(235,60)
\put(239,81)C
\put(239,71)C
\put(239,62)D
\dottedline{5}(110,20)(110,130)
\dottedline{5}(160,20)(160,48)
\dottedline{5}(235,20)(235,48)
\put(127,20){\vector(-1,0){17}}
\put(143,20){\vector(1,0){17}}
\put(190,20){\vector(-1,0){30}}
\put(205,20){\vector(1,0){30}}
\put(132,18){$p$}
\put(196,18){$q$}
\dottedline{5}(92,130)(92,102)
\dottedline{5}(160,130)(160,102)
\dottedline{5}(197,130)(197,80)
\put(97,130){\vector(-1,0){5}}
\put(105,130){\vector(1,0){5}}
\put(171,130){\vector(-1,0){11}}
\put(186,130){\vector(1,0){11}}
\put(99,128){$z$}
\put(175,128){$f$}
\path(36,75)(40,75)
\put(36,47){\arc{56}{3.1415927}{4.712389}}
\put(22,47){\arc{28}{1.5707963}{3.1415927}}
\put(22,40){\arc{14}{-1.5707963}{1.5707963}}
\put(22,33){\arc{28}{3.1415927}{4.712389}}
\path(8,33)(8,25)
\thicklines
\path(0,10)(27,10)(27,25)(0,25)(0,10)
\put(1,15){Laser}
\put(84,40){Cell}
\end{picture}
\end{center}
\caption{The optical setupt for the measurement of scattering from
colloids some microns large with ENFS.}
\label{imm_schema_generale_ENFS}
\end{figure}
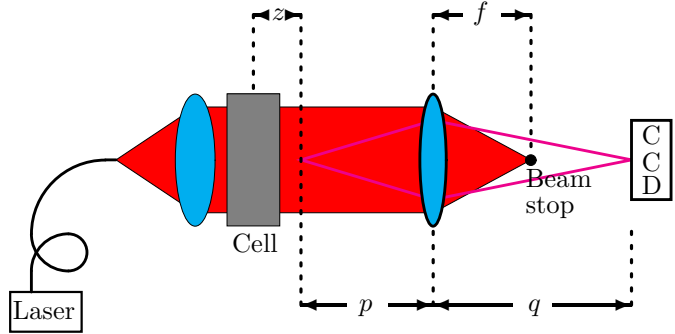
\begin{figure}
\begin{center}
\begin{picture}(250,150)(0,0)
\blacken\color{red}\path(40,75)(70,95)(160,95)(197,75)(160,55)(70,55)(40,75)
\put(197,75){\blacken\circle{4}}
\color{black}
\shade\path(82,50)(82,100)(102,100)(102,50)(82,50)
\put(70,75){\blacken\color{cyan}\ellipse{15}{50}}
\Thicklines\color{magenta}\path(235,75)(160,60)(110,75)(160,90)(235,75)
\put(160,75){\blacken\color{cyan}\ellipse{10}{50}}
\blacken\color{white}\path(235,60)(235,90)(250,90)(250,60)(235,60)
\put(239,81)C
\put(239,71)C
\put(239,62)D
\dottedline{5}(110,20)(110,130)
\dottedline{5}(160,20)(160,48)
\dottedline{5}(235,20)(235,48)
\put(127,20){\vector(-1,0){17}}
\put(143,20){\vector(1,0){17}}
\put(190,20){\vector(-1,0){30}}
\put(205,20){\vector(1,0){30}}
\put(132,18){$p$}
\put(196,18){$q$}
\dottedline{5}(92,130)(92,102)
\dottedline{5}(160,130)(160,102)
\dottedline{5}(197,130)(197,80)
\put(97,130){\vector(-1,0){5}}
\put(105,130){\vector(1,0){5}}
\put(171,130){\vector(-1,0){11}}
\put(186,130){\vector(1,0){11}}
\put(99,128){$z$}
\put(175,128){$f$}
\path(36,75)(40,75)
\put(36,47){\arc{56}{3.1415927}{4.712389}}
\put(22,47){\arc{28}{1.5707963}{3.1415927}}
\put(22,40){\arc{14}{-1.5707963}{1.5707963}}
\put(22,33){\arc{28}{3.1415927}{4.712389}}
\path(8,33)(8,25)
\thicklines
\path(0,10)(27,10)(27,25)(0,25)(0,10)
\put(1,15){Laser}
\put(84,40){Cell}
\put(195,65){Beam}
\put(195,55){stop}
\end{picture}
\end{center}
\caption{The optical setupt for the measurement of scattering from
colloids some microns large with ONFS.}
\label{imm_schema_generale_ONFS}
\end{figure}

The measurements described in Chapters
\ref{capitolo_confronto_ONFS_ENFS} and
\ref{capitolo_particle_sizing_ENFS} cover the wavevector range
$\left[q_{min},q_{max}\right]$ about $\left[2\times
10^5\mathrm{m}^{-1},4\times 10^6\mathrm{m}^{-1} \right]$.
By using Eqs. (\ref{exp_sys_eq_diametro_scelto}) and
(\ref{exp_sys_eq_z_scelto}), we obtain $D\gtrapprox5\mathrm{mm}$
and $z\approx 6\mathrm{mm}$.
A $2\mathrm{cm}$ beam diameter, obtained with
lenses with $25\mathrm{mm}$ diameter, as shown in
Fig. \ref{exper_fig_sistema_ottico}, is enough to ensure 
that the intensity is constant over the length $D$. For the
measurements described in Chapters \ref{capitolo_confronto_ONFS_ENFS} and
\ref{capitolo_particle_sizing_ENFS}, we used larger lenses with
$50\mathrm{mm}$ diameter, with a larger beam diameter, in order to
ensure a better uniformity, and $z$ was increased acordingly.

Following Eq. (\ref{exp_sys_eq_ingrandimento_scelto}), we obtain the
magnification: $M=20$. We used a 20X microscope objective and
numerical aperture of 0.45. 

The whole optical system for ENFS is shown in
Fig. \ref{exper_fig_sistema_ottico}.

For ONFS, we insert a beam stop through a hole inside the lens mount.
The beam stop and its adjustable mount is shown in
Fig. \ref{exper_fig_beam_stop}.

\section{SNFS setup for non equilibrium fluctuation measurements.}
\label{misura_snfs_sezione_opt_setup}

\begin{figure}
\begin{center}
\begin{picture}(150,250)(0,0)
\blacken\color{red}\path(75,40)(95,90)(95,160)(47,250)(103,250)(55,160)(55,90)(75,40)
\color{black}
\shade\path(50,113)(100,113)(100,137)(50,137)(50,113)
\put(75,90){\blacken\color{cyan}\ellipse{50}{15}}
\Thicklines\color{magenta}\path(75,235)(60,160)(75,85)(75,235)
\put(75,160){\blacken\color{cyan}\ellipse{50}{10}}
\blacken\color{white}\path(60,235)(90,235)(90,250)(60,250)(60,235)
\put(65,240){CCD}
\dottedline{5}(20,85)(48,85)
\dottedline{5}(20,160)(48,160)
\dottedline{5}(20,235)(48,235)
\put(20,115){\vector(0,-1){30}}
\put(20,130){\vector(0,1){30}}
\put(20,190){\vector(0,-1){30}}
\put(20,205){\vector(0,1){30}}
\put(18,121){$p$}
\put(18,196){$q$}
\dottedline{5}(130,125)(102,125)
\dottedline{5}(130,85)(102,85)
\dottedline{5}(130,160)(102,160)
\dottedline{5}(130,197)(80,197)
\put(130,98){\vector(0,-1){13}}
\put(130,112){\vector(0,1){13}}
\put(130,171){\vector(0,-1){11}}
\put(130,186){\vector(0,1){11}}
\put(128,103){$z$}
\put(128,175){$f$}
\path(75,36)(75,40)
\put(47,36){\arc{56}{0.0}{1.5707963}}
\put(47,22){\arc{28}{1.5707963}{3.1415927}}
\put(40,22){\arc{14}{3.1415927}{6.2831853}}
\put(33,22){\arc{28}{0}{1.5707963}}
\path(33,8)(25,8)
\thicklines
\path(25,0)(25,16)(0,16)(0,0)(25,0)
\put(0,5){Laser}
\blacken\path(75,197)(110,194)(110,200)(75,197)
\put(65,120){Cell}
\put(83,202){Blade}
\end{picture}
\end{center}
\caption{The optical setupt for the measurement of non equilibrium
fluctuations with SNFS.}
\label{snfs_imm_schema_generale}
\end{figure}
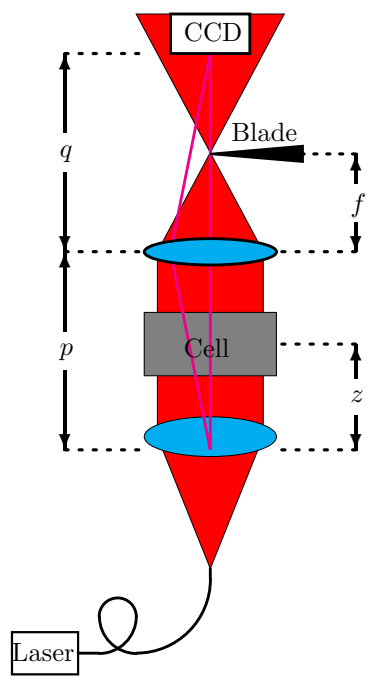

The overall system is sketched in Fig. \ref{snfs_imm_schema_generale}. 

The range $\left[q_{min},q_{max}\right]$ of the fluctuations we
measure is about $\left[2\times 10^3\mathrm{m}^{-1},2\times
10^5\mathrm{m}^{-1} \right]$, that is, the fluctuations range from ten
microns to some millimeters. 
By using Eqs. (\ref{exp_sys_eq_diametro_scelto_SNFS}) and
(\ref{exp_sys_eq_z_scelto_SNFS}), we obtain $D\gtrapprox 10\mathrm{mm}$
and $z<125\mathrm{mm}$. The cell we used, described in Chapter
\ref{capitolo_dinamico_SNFS}, has an internal diameter of about
$25\mathrm{mm}$. 

Following Eq. (\ref{exp_sys_eq_ingrandimento_scelto}), we obtain the
magnification: $M=1$. We used an achromatic doublet with a
$25\mathrm{mm}$ diameter and focal length $f=100\mathrm{mm}$. To
obtain the required magnification, $p=q=200\mathrm{mm}$.
Since SNFS is affected by small inhomogeneous
fluctuations of air temperature, we choose to put the collimating lens
and the objective lens as close as possible to the cell, in order to
prevent air movements. This resulted in a negative $z$.

The whole optical system is shown in
Fig. \ref{exp_sys_imm_snfs_setup_tutto}.

%
\chapter{ONFS data processing.}
\label{chap_onfs_data_processing}

In the chapter \ref{chap_theory}, we showed that, under some given
conditions, the correlation function of ONFS images can be used to
derive the intensity of the light scattered by a sample. The
calcultions were performed in the ideal case, in which the scattered
light comes only from the sample. The presence of non ideal lenses and
optical elements introduces an amount of undesired scattered
light. This problem is common to every kind of scattering
measurement; the undesired light, often referred to as stray light,
is generally scattered at small angles.

In standard scattering measurements, the effect of the undesired light
is additive. It can be subtracted, since the stray light can be
measured by a blank measurement.

Dynamic scattering gives a way to distinguish the effect of the light
scattered from elements that evolve with time from stationary ones. If
the stray light comes from stationary elements, such as imperfections
of the optical elements, its effect is to increase the correlation
function, with no dependence on the delay. Thus the time dependent
informations on the sample will be given by the bell shaped part of
the correlation function, while the pedestal will contain informations
on both the statically and dynamically scattered light.

If a blank measurement is possible, a more refined subtraction of the
stray light becomes possible \cite{cippelletti1997}, provided that the
stray light constitutes a speckle field, that is, the field is
gaussian. Such a data processing can be extended to ONFS too. In the
following sections, we will find a way to subtract the effect of the
stray light, first considering an unlimited number of images, taken at
different times, and then a finite set of images. Then, we will
describe the whole data processing algorithm.

\section{Effect of the stray light.}
\label{sect_stray_light_senza_bessel}

From a set of ONFS images $I\left(\vec{x}\right)$,
we can measure the intensity correlation function:
%
%
\begin{equation}
C_I\left(\Delta \vec{x}\right) = \left\{ \left<
I\left(\vec{x}\right) I\left(\vec{x} + \Delta \vec{x}\right)
\right> \right\},
\end{equation}
where $\left<\cdot\right>$ is the mean over $\vec{x}$, $\{\cdot\}$ is
the mean over different images, and the intensity of the images 
$I\left(\vec{x}\right)$ 
is the intensity of the sum of $\delta E\left(\vec{x}\right)$,
the field scattered by the
sample, and $E_{SL}\left(\vec{x}\right)$, the field of the stray light:
%
%
\begin{equation}
I\left(\vec{x}\right) = \left| 
\delta E\left(\vec{x}\right) + E_{SL}\left(\vec{x}\right)
\right|^2.
\end{equation}
So we obtain:
%
%
\begin{equation}
C_I\left(\Delta \vec{x}\right) = \left\{ \left<
\left[
\left|\delta E\left(\vec{x}\right) \right|^2+
\left|E_{SL}\left(\vec{x}\right) \right|^2+
\delta E\left(\vec{x}\right)E_{SL}^*\left(\vec{x}\right)+
\delta E^*\left(\vec{x}\right)E_{SL}\left(\vec{x}\right)
\right]
\left[
\left|\delta E\left(\vec{x}+\Delta\vec{x}\right) \right|^2+
\left|E_{SL}\left(\vec{x+\Delta\vec{x}}\right) \right|^2+
\delta E\left(\vec{x}+\Delta\vec{x}\right)
E_{SL}^*\left(\vec{x}+\Delta\vec{x}\right)+
\delta E^*\left(\vec{x}+\Delta\vec{x}\right)
E_{SL}\left(\vec{x}+\Delta\vec{x}\right)
\right]
\right> \right\}.
\end{equation}

Since $\delta E\left(\vec{x}\right)$ is a random, circular gaussian field,
the mean over different images of its odd powers
vanishes; since $E_{SL}\left(\vec{x}\right)$ is static, it can be
considered as a costant,
with respect to $\{\cdot\}$, the average on the images:
%
%
\begin{multline}
C_I\left(\Delta \vec{x}\right) = 
\left\{\left<\left|\delta E\left(\vec{x}\right) \right|^2
\left|\delta E\left(\vec{x}+\Delta\vec{x}\right) \right|^2\right>\right\} +
\left<\left|E_{SL}\left(\vec{x}\right) \right|^2
\left|E_{SL}\left(\vec{x}+\Delta\vec{x}\right) \right|^2\right> + \\
\left<\left\{\left|\delta E\left(\vec{x}\right) \right|^2\right\}
\left|E_{SL}\left(\vec{x}+\Delta\vec{x}\right) \right|^2 \right> +
\left< \left|E_{SL}\left(\vec{x}\right) \right|^2
\left\{\left|\delta E\left(\vec{x}+\Delta\vec{x}\right) \right|^2\right\}
\right> + \\
\left< 
\left\{\delta E\left(\vec{x}\right)\delta E^*\left(\vec{x}+\Delta\vec{x}\right)\right\}
E_{SL}^*\left(\vec{x}\right)E_{SL}\left(\vec{x}+\Delta\vec{x}\right) \right>
+ \\
\left< 
\left\{\delta E^*\left(\vec{x}\right)\delta E\left(\vec{x}+\Delta\vec{x}\right)\right\}
E_{SL}\left(\vec{x}\right)E_{SL}^*\left(\vec{x}+\Delta\vec{x}\right) \right>.
\end{multline}
The mean over the images $\{\cdot\}$ equals the mean over $\vec{x}$, 
$\left<\cdot\right>$, for
the field $\delta E\left(\vec{x}\right)$:
%
%
\begin{multline}
C_I\left(\Delta \vec{x}\right) = 
\left<\left|\delta E\left(\vec{x}\right) \right|^2
\left|\delta E\left(\vec{x}+\Delta\vec{x}\right) \right|^2\right> +
\left<\left|E_{SL}\left(\vec{x}\right) \right|^2
\left|E_{SL}\left(\vec{x}+\Delta\vec{x}\right) \right|^2\right> + \\
\left<\left|\delta E\left(\vec{x}\right) \right|^2\right> \left<
\left|E_{SL}\left(\vec{x}+\Delta\vec{x}\right) \right|^2 \right> +
\left< \left|E_{SL}\left(\vec{x}\right) \right|^2 \right>
\left<\left|\delta E\left(\vec{x}+\Delta\vec{x}\right) \right|^2\right> + \\
\left<\delta E\left(\vec{x}\right)\delta E^*\left(\vec{x}+\Delta\vec{x}\right)\right>
\left<
E_{SL}^*\left(\vec{x}\right)E_{SL}\left(\vec{x}+\Delta\vec{x}\right) \right>
+ \\
\left< 
\delta E^*\left(\vec{x}\right)\delta E\left(\vec{x}+\Delta\vec{x}\right)\right>
\left<E_{SL}\left(\vec{x}\right)E_{SL}^*\left(\vec{x}+\Delta\vec{x}\right) \right>.
\end{multline}

Since both $\delta E\left(\vec{x}\right)$ and $E_{SL}\left(\vec{x}\right)$
are gaussian fields, we can use Siegert relation
Eq. (\ref{teoria_eq_siegert_2d}) to express four-point correlation
functions in terms of two-point ones.
%
%
\begin{multline}
C_I\left(\Delta \vec{x}\right) = 
\left<\left|\delta E\left(\vec{x}\right) \right|^2\right>
\left<\left|\delta E\left(\vec{x}+\Delta\vec{x}\right) \right|^2\right> +
\left|\left<\delta E\left(\vec{x}\right)
\delta E^*\left(\vec{x}+\Delta\vec{x}\right) \right>\right|^2 + \\
\left<\left|E_{SL}\left(\vec{x}\right) \right|^2\right>
\left<\left|E_{SL}\left(\vec{x}+\Delta\vec{x}\right) \right|^2\right> +
\left|\left<E_{SL}\left(\vec{x}\right)
E_{SL}^*\left(\vec{x}+\Delta\vec{x}\right) \right>\right|^2 + \\
\left<\left|\delta E\left(\vec{x}\right) \right|^2\right> \left<
\left|E_{SL}\left(\vec{x}+\Delta\vec{x}\right) \right|^2 \right> +
\left< \left|E_{SL}\left(\vec{x}\right) \right|^2 \right>
\left<\left|\delta E\left(\vec{x}+\Delta\vec{x}\right) \right|^2\right> + \\
\left<\delta E\left(\vec{x}\right)\delta E^*\left(\vec{x}+\Delta\vec{x}\right)\right>
\left<
E_{SL}^*\left(\vec{x}\right)E_{SL}\left(\vec{x}+\Delta\vec{x}\right) \right>
+ \\
\left< 
\delta E^*\left(\vec{x}\right)\delta E\left(\vec{x}+\Delta\vec{x}\right)\right>
\left< E_{SL}\left(\vec{x}\right)E_{SL}^*\left(\vec{x}+\Delta\vec{x}\right) \right>.
\end{multline}

We define $\left<\delta I\right> = \left<\left|\delta E\left(\vec{x}\right)
\right|^2\right>$,
$\left<I_{SL}\right> =
\left<\left|E_{SL}\left(\vec{x}\right)\right|^2\right>$, 
$C_{\delta E}\left(\Delta \vec{x}\right) = \left< \delta E\left(\vec{x}\right)
\delta E^*\left(\vec{x}+\Delta\vec{x}\right) \right>$, 
$C_{SL}\left(\Delta \vec{x}\right) = \left< E_{SL}\left(\vec{x}\right) 
E_{SL}^*\left(\vec{x}+\Delta\vec{x}\right) \right>$:
%
%
\begin{multline}
C_I\left(\Delta \vec{x}\right) = 
\left<\delta I\right>^2 +
\left|C_{\delta E}\left(\Delta \vec{x}\right)\right|^2 +
\left<I_{SL}\right>^2 +
\left|C_{SL}\left(\Delta \vec{x}\right)\right|^2 + \\
2\left<\delta I\right>\left<I_{SL}\right> +
C_{\delta E}\left(\Delta \vec{x}\right) C_{SL}^*\left(\Delta \vec{x}\right) +
C_{\delta E}^*\left(\Delta \vec{x}\right) C_{SL}\left(\Delta \vec{x}\right) .
\end{multline}

The result is that 
the stray light field correlation sums to the scattered field correlation:
%
%
\begin{equation}
\label{stray_light_correlazione_intensita}
C_I\left(\Delta \vec{x}\right) = 
\left( \left<\delta I\right> + \left<I_{SL}\right> \right)^2 +
\left|C_{\delta E}\left(\Delta \vec{x}\right) +
C_{SL}\left(\Delta \vec{x}\right)\right|^2.
\end{equation}

In order to obtain informations about the correlation of the stray light field,
we acquire a great number of images, with different scattered field, and
we average them, thus obtaining the correlation function of the mean intensity
$\left\{I\left(\vec{x}\right)\right\}$. Then, we measure the correlation
function of the mean intensity:
%
%
\begin{equation}
C_{\left\{I\right\}}\left(\Delta \vec{x}\right) = \left< \left\{ 
I\left(\vec{x}\right)\right\} \left\{I\left(\vec{x} + \Delta
\vec{x}\right)\right\}\right>.
\end{equation}

We evaluate the mean intensity $\left\{I\left(\vec{x}\right)\right\}$ :
%
%
\begin{multline}
\left\{I\left(\vec{x}\right)\right\} = \left\{\left| 
\delta E\left(\vec{x}\right) + E_{SL}\left(\vec{x}\right)
\right|^2\right\} =\\
\left\{\left| \delta E\left(\vec{x}\right)\right|^2\right\}+
\left\{\left| E_{SL}\left(\vec{x}\right)\right|^2\right\}+
\left\{\delta E\left(\vec{x}\right) E_{SL}^*\left(\vec{x}\right) \right\}+
\left\{\delta E^*\left(\vec{x}\right) E_{SL}\left(\vec{x}\right) \right\}.
\end{multline}

Since $E_{SL}$ does not depend on the image:
%
%
\begin{equation}
\left\{I\left(\vec{x}\right)\right\} =
\left\{\left|\delta E\left(\vec{x}\right)\right|^2\right\}+
\left| E_{SL}\left(\vec{x}\right)\right|^2+
\left\{\delta E\left(\vec{x}\right)\right\} E_{SL}^*\left(\vec{x}\right) +
\left\{\delta E^*\left(\vec{x}\right)\right\} E_{SL}\left(\vec{x}\right) .
\end{equation}

Using the gaussian properties of the scattered light:
%
%
\begin{equation}
\label{stray_light_intensita_media_campioni}
\left\{I\left(\vec{x}\right)\right\} =
\left<\delta I\right> + \left| E_{SL}\left(\vec{x}\right)\right|^2 .
\end{equation}

Now we can evluate the correlation function of the mean intensity:
%
%
\begin{multline}
C_{\left\{I\right\}}\left(\Delta \vec{x}\right) = \left<
\left[\left<\delta I\right> + \left| E_{SL}\left(\vec{x}\right)\right|^2\right]
\left[\left<\delta I\right> + \left| E_{SL}\left(\vec{x} + \Delta \vec{x}\right)
\right|^2\right] \right> = \\
\left<\delta I\right>^2 + 
\left<\delta I\right> \left<\left| E_{SL}\left(\vec{x}\right)\right|^2\right> +
\left<\delta I\right> \left<\left| E_{SL}\left(\vec{x} + \Delta \vec{x}\right)
\right|^2\right> +
\left< \left| E_{SL}\left(\vec{x}\right)\right|^2 \left| E_{SL}\left(\vec{x} +
\Delta \vec{x}\right) \right|^2 \right> .
\end{multline}
Using the gaussian properties of the field $E_{SL}$:
%
%
\begin{equation}
\label{stray_light_correlazione_intensita_media}
C_{\left\{I\right\}}\left(\Delta \vec{x}\right) = 
\left(\left<\delta I\right> + \left<I_{SL}\right> \right)^2 +
\left|C_{SL}\left(\Delta \vec{x}\right) \right|^2
\end{equation}

From eq. (\ref{stray_light_intensita_media_campioni}), we can evaluate
the mean value of the intensity of the images:
%
%
\begin{equation}
\label{stray_light_intensita_media}
\left\{ \left< I\right>\right\} = 
\left<\delta I\right> + \left<I_{SL}\right>
\end{equation}

Eq. (\ref{stray_light_correlazione_intensita}), 
(\ref{stray_light_correlazione_intensita_media}),
(\ref{stray_light_intensita_media}) give some informations about the
field correlation of the scattered and stray light. If both the correlation
functions are real and positive, the best evaluation of the field correlation
function of the scattered field is:
%
%
\begin{equation}
C_E\left(\Delta \vec{x}\right) = 
\sqrt{C_I\left(\Delta \vec{x}\right) - \left\{ \left< I\right>\right\}^2 } 
-\sqrt{C_{\left\{I\right\}}\left(\Delta \vec{x}\right) - 
\left\{ \left< I\right>\right\}^2 } 
\end{equation}

\section{Correction for finite samples.}

In order to evaluate the correlation function of the mean intensity, we
average a given amount of images, then we evaluate the correlation function
of the obtained mean value. Since the number of images we average is finite,
the correlation function will not correspond to that of eq. 
(\ref{stray_light_correlazione_intensita_media}). For example, if the stray
light vanishes, the mean intensity will still present fluctuations, due
to the scattered light. These fluctuations vanish as the square root of
the number of the averaged images, and consequently the correlation function
becomes flat only for infinite samples.

A similar problem arises when working with a stochastic, gaussian variable.
If we have $N$ values of the stocastic variable $x$,
distributed with probability
$P\left(x\right)\propto \exp\left[-\left(x-x_0\right)/
\left(2\sigma^2\right)\right]$, we find that the best value for $x_0$
is the mean of the values $x$, and the best value for $\sigma$ is the 
root mean square displacement of the values $x$ from $x_0$. On the other
hand, the average on a finite number of elements will be displaced from
$x_0$ of an amount, vanishing as the square root of the number of the 
samples $N$, but so that the root mean square displacement of the data from 
the mean is alwais smaller than $\sigma$.  It is thus necessary to use the
Bessel correction, dependent on the number of the samples $N$.

Generally the Bessel correction is obtained in consequence of the 
``maximum likelihood'' condition. This means that, given a set of values of a
stochastic variable, and given a family of probability distributions,
the parameters of the family must be selected in order to maximize
the probability of finding the given data. Another approach is to find
a suitable algorithm which gives the values of the parameters, from 
a set of data. The algorithm will be selected in order that the output
values will be distributed around the true ones, with minimum square
displacement. For a gaussian distribution, the two approaches give
the same result. It is easy to show that, for example for a Heaviside
distribution, the maximum likelihood condition fails to obtain the
best results.

In our case, the distribution function of the intensity is not gaussian.
We will use weak condition, that is, we will look for an algorithm
giving values which average to the true ones. In other words: we will try
to avoid sistematic erroneous evaluations of the correlation function.

We define $\bar{\cdot}$ as the mean over $N$ samples. In particular
$C_{\bar{i}}$ is the correlation function of the averaged $N$ images.
To avoid sistematic errors, we must first evaluate
$\left\{C_{\bar{I}}\right\}$:
%
%
\begin{equation}
\left\{C_{\bar{I}}\left(\Delta \vec{x}\right)\right\} = \left\{ \left<
\bar{I}\left(\vec{x}\right) \bar{I}\left(\vec{x} + \Delta \vec{x}\right)
\right> \right\},
\end{equation}
where the average intensity of $N$ images $\bar{I}$ is given
by the sum of the field scattered by the sample $\delta E\left(\vec{x}\right)$ 
and the field of the stray light $E_{SL}\left(\vec{x}\right)$:
%
%
\begin{equation}
\bar{I}\left(\vec{x}\right) = \frac{1}{N} \sum_n {\left| 
\delta E_n\left(\vec{x}\right) + E_{SL}\left(\vec{x}\right)
\right|^2}.
\end{equation}
So we obtain:
%
%
\begin{equation}
\left\{C_{\bar{I}} \left(\Delta \vec{x}\right)\right\} = 
\frac{1}{N^2} \sum_{n,m} { \left\{ \left<
\left[
\left|\delta E_n\left(\vec{x}\right) \right|^2+
\left|E_{SL}\left(\vec{x}\right) \right|^2+
\delta E_n\left(\vec{x}\right)E_{SL}^*\left(\vec{x}\right)+
\delta E_n^*\left(\vec{x}\right)E_{SL}\left(\vec{x}\right)
\right]
\left[
\left|\delta E_m\left(\vec{x}+\Delta\vec{x}\right) \right|^2+
\left|E_{SL}\left(\vec{x+\Delta\vec{x}}\right) \right|^2+
\delta E_m\left(\vec{x}+\Delta\vec{x}\right)E_{SL}^*\left(\vec{x}+\Delta\vec{x}\right)+
\delta E_m^*\left(\vec{x}+\Delta\vec{x}\right)E_{SL}\left(\vec{x}+\Delta\vec{x}\right)
\right]
\right> \right\} }.
\end{equation}

We can follow the calculations performed in Section 
\ref{sect_stray_light_senza_bessel} to obtain Eq. (
\ref{stray_light_correlazione_intensita}). In this case we obtain:
%
%
\begin{multline}
\left\{C_{\bar{I}}\left(\Delta \vec{x}\right)\right\} =
\frac{1}{N^2}\sum_{n,m}{ 
\left<\left|\delta E_n\left(\vec{x}\right) \right|^2\right>
\left<\left|\delta E_m\left(\vec{x}+\Delta\vec{x}\right) \right|^2\right>} + \\
\frac{1}{N^2}\sum_{n,m}{ 
\left|\left<\delta E_n\left(\vec{x}\right)
\delta E_m^*\left(\vec{x}+\Delta\vec{x}\right) \right>\right|^2} + \\
\left<\left|E_{SL}\left(\vec{x}\right) \right|^2\right>
\left<\left|E_{SL}\left(\vec{x}+\Delta\vec{x}\right) \right|^2\right> +
\left|\left<E_{SL}\left(\vec{x}\right)
E_{SL}^*\left(\vec{x}+\Delta\vec{x}\right) \right>\right|^2 + \\
\frac{1}{N^2}\sum_{n,m}{
\left<\left|\delta E_n\left(\vec{x}\right)\right|^2 \right>} \left<
\left|E_{SL}\left(\vec{x}+\Delta\vec{x}\right) \right|^2 \right> + \\
\left< \left|E_{SL}\left(\vec{x}\right) \right|^2 \right>
\frac{1}{N^2}\sum_{n,m}{
\left<\left|\delta E_m\left(\vec{x}+\Delta\vec{x}\right) \right|^2\right>} + \\
\frac{1}{N^2}\sum_{n,m}{
\left< \delta E_n\left(\vec{x}\right)\delta E_m^*\left(\vec{x}+\Delta\vec{x}\right)\right>}
\left<
E_{SL}^*\left(\vec{x}\right)E_{SL}\left(\vec{x}+\Delta\vec{x}\right) \right>
+ \\
\frac{1}{N^2}\sum_{n,m}{
\left< 
\delta E_n^*\left(\vec{x}\right)\delta E_m\left(\vec{x}+\Delta\vec{x}\right)\right>}
\left< E_{SL}\left(\vec{x}\right)E_{SL}^*\left(\vec{x}+\Delta\vec{x}\right) \right>.
\end{multline}
The mean values can be calculated, provided that the two cases, $n=m$ and
$n\ne m$ are taken into account:
%
%
\begin{multline}
\left\{C_{\bar{I}}\left(\Delta \vec{x}\right)\right\} =
\left<\delta I\right>^2 + 
\frac{1}{N}\left|C_{\delta E}\left(\Delta\vec{x}\right)\right|^2 + 
\left<I_{SL}\right>^2 +
\left|C_{SL}\left(\Delta\vec{x}\right)\right|^2 + 
2\left<\delta I\right> \left<I_{SL}\right> + \\
\frac{1}{N}C_{\delta E}\left(\Delta\vec{x}\right)
C_{SL}^*\left(\Delta\vec{x}\right) +
\frac{1}{N}C_{\delta E}^*\left(\Delta\vec{x}\right)
C_{SL}\left(\Delta\vec{x}\right).
\end{multline}
The result reduces to eq. (\ref{stray_light_correlazione_intensita_media})
in the limit $N\to \infty$:
%
%
\begin{equation}
\label{stray_light_correlazione_intensita_media_bessel}
\left\{C_{\bar{I}}\left(\Delta \vec{x}\right)\right\} = 
\left( \left<\delta I\right> + \left<I_{SL}\right> \right)^2 +
\frac{N-1}{N}\left|C_{SL}\left(\Delta \vec{x}\right)\right|^2 +
\frac{1}{N}\left|C_{\delta E}\left(\Delta \vec{x}\right) +
C_{SL}\left(\Delta \vec{x}\right)\right|^2 .
\end{equation}

From eq. (\ref{stray_light_correlazione_intensita}), 
(\ref{stray_light_correlazione_intensita_media_bessel}) and
(\ref{stray_light_intensita_media}) we can evaluate the field correlation
function:
%
%
\begin{equation}
\label{stray_light_sottrazione_sl_bessel}
C_E\left(\Delta \vec{x}\right) = 
\sqrt{C_I\left(\Delta \vec{x}\right) - \left\{ \left< I\right>\right\}^2 } 
-\sqrt{
\frac{N}{N-1}C_{\bar{I}}\left(\Delta \vec{x}\right) - 
\frac{1}{N-1}C_I\left(\Delta \vec{x}\right) -
\left\{ \left< I\right>\right\}^2 
} 
\end{equation}

\section{Data processing algorithm.}

Once the experimental apparatus has been built, as described in
Chapter \ref{capitolo_sistema_sperimentale}, and the sample is placed
in it, we acquire one hundred images. The electronic shutter of
the CCD and its interlacement time must be so short that no evident
evolution of the system happens during the exposure: for the samples
we measured, that is colloids some microns large, with brownian
movements, and non equilibrium fluctuations in the free diffusion of
simple liquids, an interlacement delay of
$1/25\mathrm{s}$ is sufficient. Moreover, different images must be completely
uncorrelated. For a $10.0\mathrm{\mu m}$ colloid, images must be grabbed at
intervals longer than one minute, if only brownian movements are the
source of decorrelation. For the non equilibrium fluctuations we
studied, the interval was about one second.

\begin{figure}
\begin{center}
\includegraphics[scale=0.4]{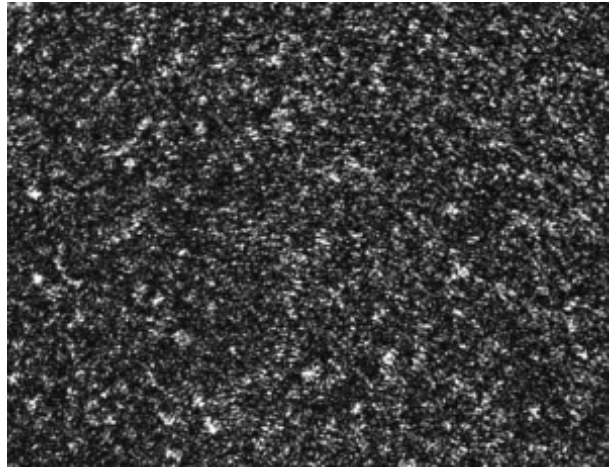}
\end{center}
\caption{Near field intensity of the light scattered by a colloid
of $5.2\mathrm{\mu m}$.}
\label{results_colloid_2d_imm_5um}
\end{figure}

\begin{figure}
\begin{center}
\includegraphics[scale=0.4]{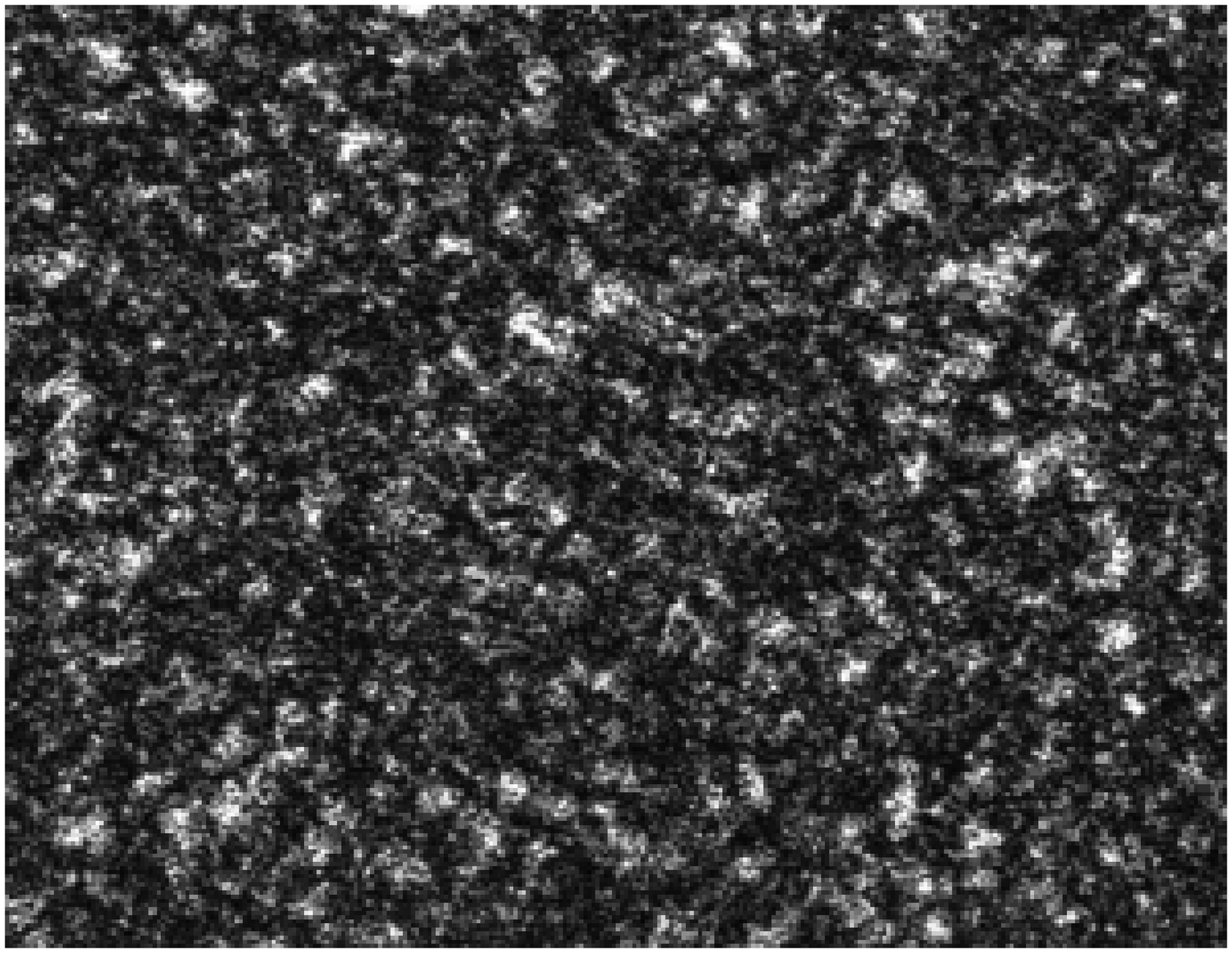}
\end{center}
\caption{Near field intensity of the light scattered by a colloid
of $10.0\mathrm{\mu m}$.}
\label{results_colloid_2d_imm_10um}
\end{figure}

In figure \ref{results_colloid_2d_imm_5um} and 
\ref{results_colloid_2d_imm_10um} we show two typical images of the
near field intensity of the light scattered by colloids of $5.2\mathrm{\mu m}$
and $10.0\mathrm{\mu m}$.
We can notice the different typical size of the speckles.

For each image, we evaluate the correlation function. This operation
is quite fast, since we can use a Fast Fourier Transform (FFT)
algorithm. An FFT algorithm allows to evaluate the Fourier tranform of
an $M\times N$ matrix, with a number of arithmetic operations
proportional to $MN\log\left(MN\right)$. By using Perceval relation,
we can obtain the correlation function by making an FFT, evaluating the
square modulus, and making an Inverse FFT (IFFT). This only requires a
number of operation of the order of $MN\log\left(MN\right)$. By
scanning every value of $\Delta x$, and averaging over every $N\times
M$ pixels, the number of operations would be of the order of
$\left(MN\right)^2$. Using FFT, care must be taken in order to
correcly evaluate the correlations near the boundarys: FFT assumes
periodic boundarys, so the image must be embedded in a bigger
matrix, filled with zeroes. Since the FFT is faster if $N$ and
$M$ are powers of 2 \cite{num_recipes}, we used a matrix of $1024\times 1024$
points. After the correlation function has been evaluated, we normalize
it, by dividing by the number of independent couples used to evaluate
the correlation function.

\begin{figure}
\begin{center}
\includegraphics{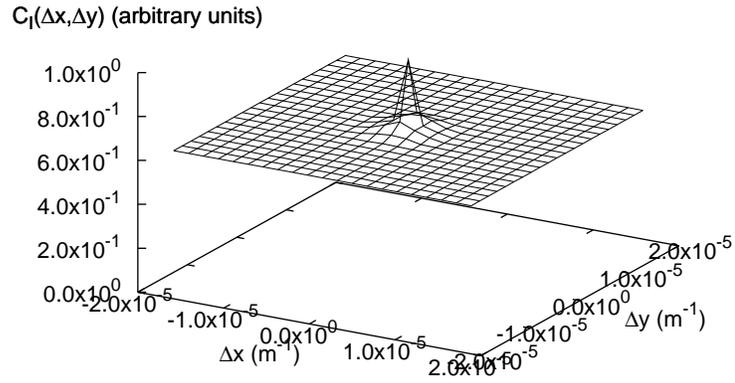}
\end{center}
\caption
[2-d $C_I\left(\Delta \vec{x}\right)$ for $5.2\mathrm{\mu m}$ colloid.]
{Intensity correlation function 
$C_I\left(\Delta \vec{x}\right)$, for a colloid made of polystyrene spheres
with diameter of $\mathrm{5.2\mu m}$}
\label{results_colloid_2d_ci_5um}
\end{figure}

\begin{figure}
\begin{center}
\includegraphics{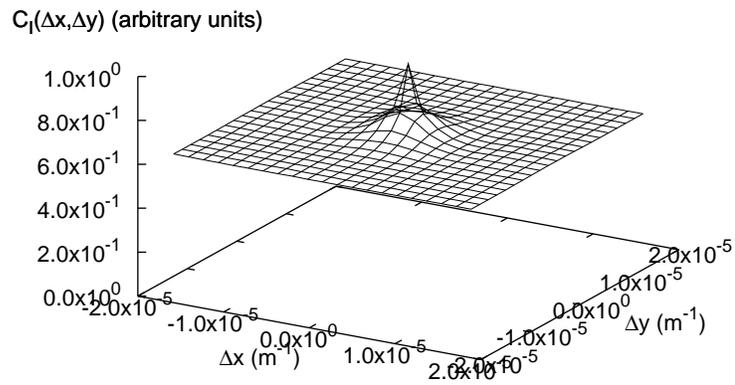}
\end{center}
\caption
[2-d $C_I\left(\Delta \vec{x}\right)$ for $10.0\mathrm{\mu m}$ colloid.]
{Intensity correlation function 
$C_I\left(\Delta \vec{x}\right)$, for a colloid made of polystyrene spheres
with diameter of $10.0\mathrm{\mu m}$}
\label{results_colloid_2d_ci_10um}
\end{figure}

The correlation functions of every image are averaged, thus obtaining
$C_I\left(\Delta \vec{x}\right)$. Fig. \ref{results_colloid_2d_ci_5um} and
\ref{results_colloid_2d_ci_10um} show typical graphs of the intensity
correlation function $C_I\left(\Delta \vec{x}\right)$, for a colloid
made of polystyrene spheres with diameters of $5.2\mathrm{\mu m}$ and
$10.0\mathrm{\mu m}$. We can notice that the correlation function has a maximum
at $\Delta \vec{x}=0$, then decreases, until it reaches the plateau
value, about one half the peak value. This behaviour is typical of
every speckle field.

\begin{figure}
\begin{center}
\includegraphics{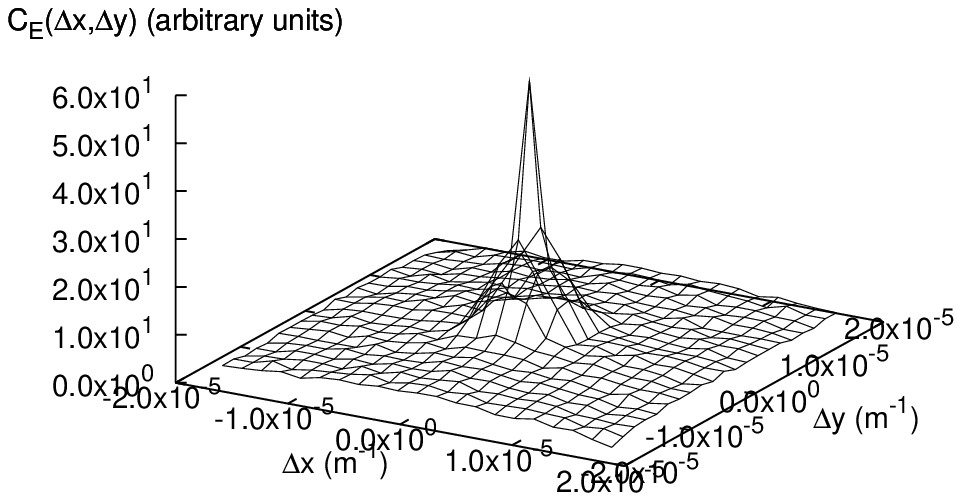}
\end{center}
\caption
[2-d $C_E\left(\Delta \vec{x}\right)$ for $5.2\mathrm{\mu m}$ colloid,
 not corrected.]
{Field correlation function $C_E\left(\Delta \vec{x}\right)$,
not corrected for the stray light,
for the colloid made of polystyrene spheres with diameter of
$5.2\mathrm{\mu m}$}
\label{results_colloid_2d_c_non_corretta_5um}
\end{figure}

\begin{figure}
\begin{center}
\includegraphics{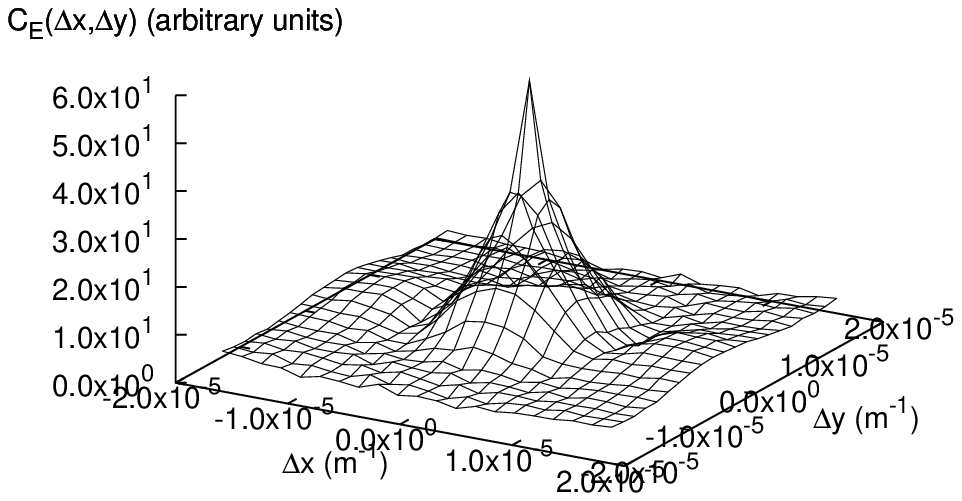}
\end{center}
\caption
[2-d $C_E\left(\Delta \vec{x}\right)$ for $10.0\mathrm{\mu m}$ colloid, not corrected.]
{Field correlation function $C_E\left(\Delta \vec{x}\right)$,
not corrected for the stray light,
for the colloid made of polystyrene spheres 
with diameter of $10.0\mathrm{\mu m}$}
\label{results_colloid_2d_c_non_corretta_10um}
\end{figure}

Neglecting the stray ligth, we could evaluate the field correlation
function by using the Siegert relation, Eq. (\ref{teoria_eq_siegert_2d}):
\begin{equation}
C_E\left(\Delta \vec{x}\right) =
\sqrt{C_I\left(\Delta \vec{x}\right) - \left<I\right>^2}
\end{equation}
where the mean intensity $\left\{\left<I\right>\right\}$, is obtained by
averaging the measured intensity over every pixel of the image and
over every image. In Fig. \ref{results_colloid_2d_c_non_corretta_5um} and 
\ref{results_colloid_2d_c_non_corretta_10um} are shown typical graphs
of the field correlation function, calculated from the intensity correlation
function, without any correction for the stray light. The correlation
should vanish for $\Delta x \to \infty$, in absence of stray light.

\begin{figure}
\begin{center}
\includegraphics{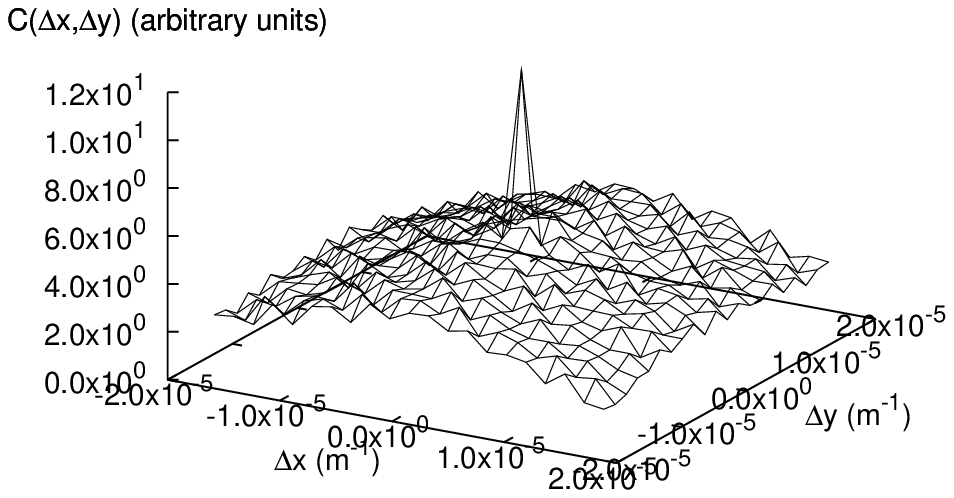}
\end{center}
\caption
[2-d $C_{\bar{I}}\left(\Delta \vec{x}\right)$ for $5.2\mathrm{\mu m}$ colloid.]
{Correlation function of the mean intensity
$C_{\bar{I}}\left(\Delta \vec{x}\right)$,
for the colloid made of polystyrene spheres with diameter of
$5.2\mathrm{\mu m}$}
\label{results_colloid_2d_c_sl_5um}
\end{figure}

\begin{figure}
\begin{center}
\includegraphics{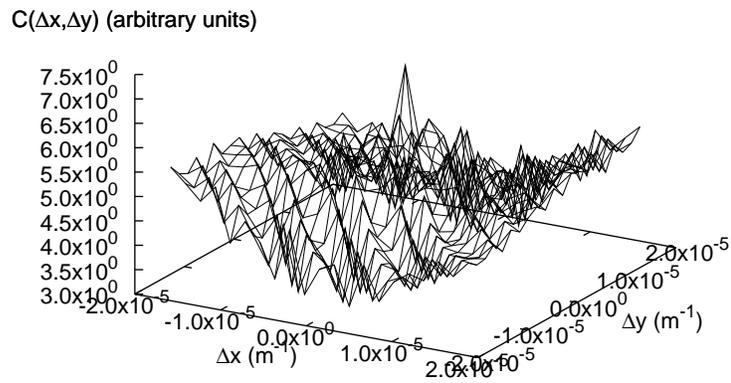}
\end{center}
\caption
[2-d $C_{\bar{I}}\left(\Delta \vec{x}\right)$ for $10.0\mathrm{\mu m}$ colloid.]
{Correlation function of the mean intensity
$C_{\bar{I}}\left(\Delta \vec{x}\right)$,
for the colloid made of polystyrene spheres with diameter of
$10.0\mathrm{\mu m}$}
\label{results_colloid_2d_c_sl_10um}
\end{figure}

In order to subtract the contribution of the stray light, we evaluate
the correlation function of the average of all the images, thus
obtaining $C_{\bar{I}}\left(\Delta \vec{x}\right)$. The evaluation of
the correlation function is obtained with the above described
algorithm. In Fig. \ref{results_colloid_2d_c_sl_5um} and
\ref{results_colloid_2d_c_sl_10um} are shown typical graphs of the
correlation function of the mean intensity, for the two colloids. The
graphs are not flat, due to the stray light. 

\begin{figure}
\begin{center}
\includegraphics{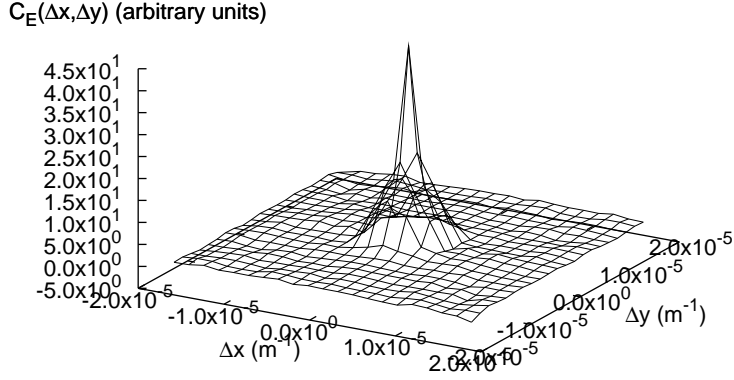}
\end{center}
\caption
[2-d $C_E\left(\Delta \vec{x}\right)$ for $5.2\mathrm{\mu m}$ colloid.]
{Field correlation function $C_E\left(\Delta \vec{x}\right)$
for the colloid made of polystyrene spheres with diameter of
$5.2\mathrm{\mu m}$}
\label{results_colloid_2d_c_5um}
\end{figure}

\begin{figure}
\begin{center}
\includegraphics{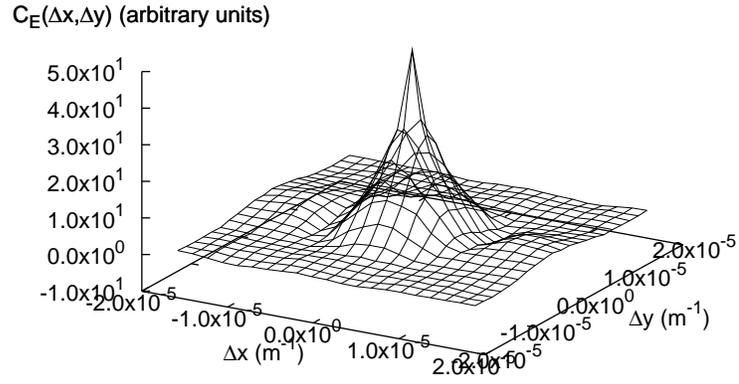}
\end{center}
\caption
[2-d $C_E\left(\Delta \vec{x}\right)$ for $10.0\mathrm{\mu m}$ colloid.]
{Field correlation function $C_E\left(\Delta \vec{x}\right)$
for the colloid made of polystyrene spheres with diameter of
$10.0\mathrm{\mu m}$}
\label{results_colloid_2d_c_10um}
\end{figure}

Through Eq. (\ref{stray_light_sottrazione_sl_bessel}) we evaluate
$C_E\left(\Delta \vec{x}\right)$, under the hypothesis that both the
stray light field and the scattered light field have a real and
positive correlation function. Typical field correlation function,
corrected for the stray light using
Eq. (\ref{stray_light_sottrazione_sl_bessel}), are shown in figure
\ref{results_colloid_2d_c_5um} and \ref{results_colloid_2d_c_10um}: we
can notice a signitificative increase in the smoothness of the graphs,
with respect to Fig. \ref{results_colloid_2d_c_non_corretta_5um} and 
\ref{results_colloid_2d_c_non_corretta_10um}.

\begin{figure}
\begin{center}
\includegraphics{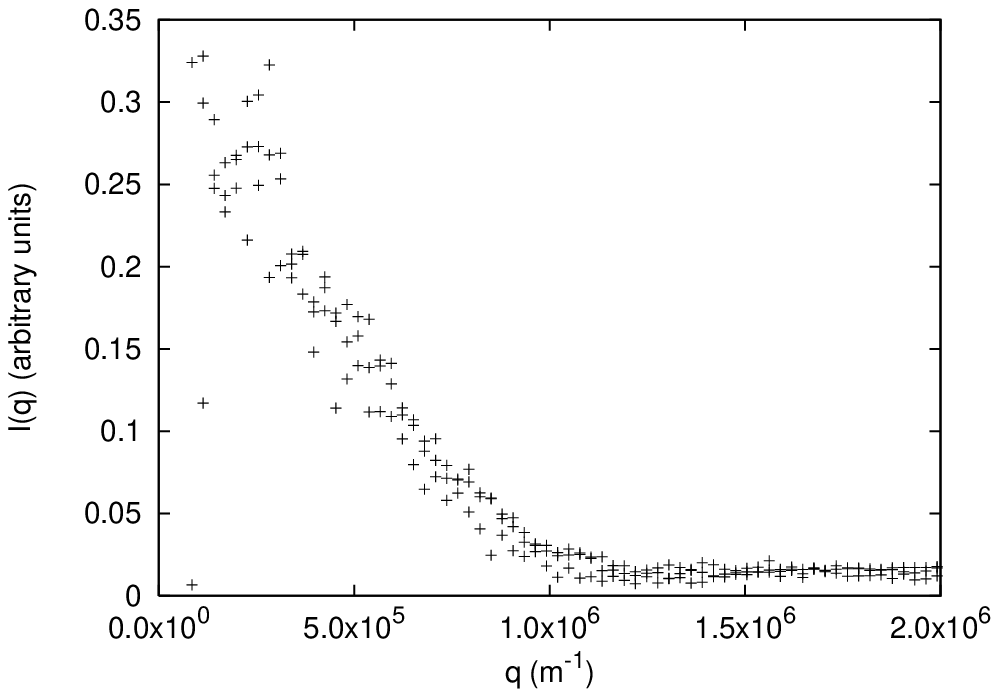}
\end{center}
\caption{ONFS measurement of the scattered intensity of a $5.2\mathrm{\mu m}$ colloid.}
\label{results_colloid_spettro_5um}
\end{figure}

\begin{figure}
\begin{center}
\includegraphics{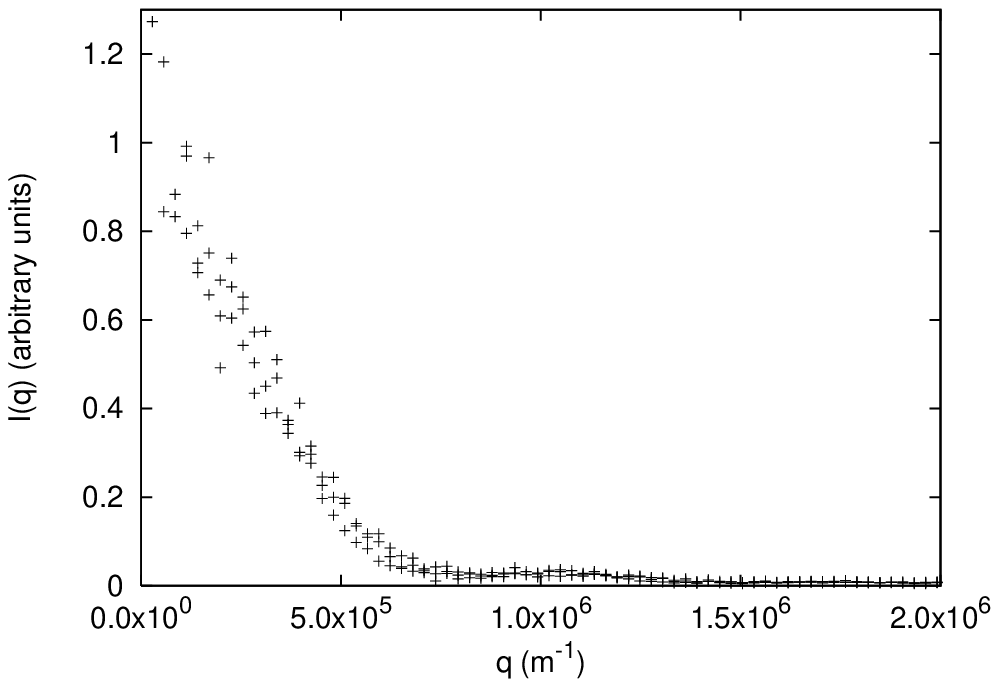}
\end{center}
\caption{ONFS measurement of the scattered intensity of a  $10.0\mathrm{\mu m}$ colloid.}
\label{results_colloid_spettro_10um}
\end{figure}

We apply a Fourier tranform to the two dimensional correlation
function $C_E\left(\Delta \vec{x}\right)$, thus obtaining the field
power spectrum $S_E\left(q\right)$. Since our samples are isotropic,
we make an angular average of the power spectra, and represent our
data as a
function of the modulus $q$ of $\vec{q}$. The scattered intensity
$I\left(q\right)$ is then obtained by using
Eq. (\ref{teoria_relazione_IQ_Ez_di_q}), that is, simply relating each
value of the power spectra, with wavelength $q$ to a value of
$I\left(Q\right)$, where the relation $Q\left(q\right)$ is given by
Eq. (\ref{teoria_trasferito_vettore_onda}). In
Fig. \ref{results_colloid_spettro_5um} and
\ref{results_colloid_spettro_10um} we show the measured $I\left(q\right)$. 

%
\chapter {ENFS and SNFS data processing.}
\label{chap_enfs_data_processing}

In this chapter we describe the algorithm used to process ENFS and SNFS
data, acquired by a CCD, in order to evaluate of the
scattered intensity. From a set of images, taken at different times,
we are able to subtract the stray light, point by point: this is a
noteworthy feature of the heterodyne techniques. The
algorithm we describe is similar to the one which has been applied
to shadowgraph, and should work for every heterodyne technique.

The algorithm has been applied to SNFS, and the results are shown in
Chapter \ref{capitolo_dinamico_SNFS}; in this chapter, all examples
refers to colloid measurements made with ENFS.

\section{Subtraction of the stray light}

As in every heterodyne technique, we measure the overall intensity
$I\left(\vec{x}\right)$, generated by the interference of 
an object field $\delta E\left(\vec{x}\right)$ with
a more intense reference beam with 
amplitude $E_0$. In our 
case, the object field is generated by the scattered beams, and
the reference beam is the transmitted one. We measure the 
heterodyne signal $i\left(\vec{x}\right)$:
\begin{equation}
i\left(\vec{x}\right) = \frac{I\left(\vec{x}\right) - I_0}{I_0},
\end{equation}
where $I_0$ is the mean intensity. At first order in 
$\delta E/E_0$, Eq. (\ref{teoria_eq_tecniche_campo_shadowgraph}) holds:
\begin{equation}
i\left(\vec{x}\right) = 2 
\frac{\Re\left[\delta E\left(\vec{x}\right)\right]}
{E_0},
\end{equation}
where we have assumed that $E_0$ is real. Equation 
(\ref{teoria_eq_funzionamento_ENFS})
states that, under the conditions in which NFS works, the power
spectrum of $i\left(\vec{x}\right)$ is the power spectrum of the 
electric field, the quantity we must measure in order to
evaluate the scattered intensities.

We developed an algorithm to subtract the contribution
of the stray light, directly on each image, point by point. This is a
noteworthy feature of the heterodyne techniques, since in dynamic light
scattering and in ONFS the subtraction is possible only on the
scattered intensity or on the correlation function, averages of
square values. The scattered field can be decomposed into
$E_{SL}\left(\vec{x}\right)$, the stray light field, and
$\delta E\left(\vec{x}\right)$, the field of the light scattered
by the sample. Both $E_{SL}\left(\vec{x}\right)$ and 
$\delta E\left(\vec{x}\right)$ are much less intense than $E_0$,
the reference field. At the first order, the resulting intensity
is:
\begin{equation}
I\left(\vec{x}\right) = E_0^2 +
2 E_0 \Re \left[E_{SL}\left(\vec{x}\right)\right] +
2 E_0 \Re \left[\delta E\left(\vec{x}\right)\right]
\end{equation}
In many cases, $\delta E\left(\vec{x}\right)$ fluctuates in time
and is correlated only on finite delays.
On the contrary, stray light comes mainly from hard surfaces,
and does not change as times go on. This is the case of the samples
we studied. The spatial average of a scattered field is 
alwais wanishing; this property, along with the absence of 
correlation on different images, says that the average over many
images of $\delta E\left(\vec{x}\right)$ vanishes.

In order to separate the contribution of the stray light,
we average $I\left(\vec{x}\right)$ over many different images.
Since the phase of $\delta E\left(\vec{x}\right)$ is random,
its mean vanishes:
\begin{equation}
\left\{I\left(\vec{x}\right)\right\} = E_0^2 +
2 E_0 \Re \left[E_{SL}\left(\vec{x}\right)\right].
\end{equation}
We use the symbol $\left\{\cdot\right\}$ for the mean over many samples,
and the symbol $\left<\cdot\right>$ for the mean over $\vec{x}$.
The fluctuation $I\left(\vec{x}\right) - \left\{I\left(\vec{x}\right)
\right\}$ does not depend on $E_{SL}\left(\vec{x}\right)$:
\begin{equation}
I\left(\vec{x}\right) - \left\{I\left(\vec{x}\right)\right\} = 
2 E_0 \Re \left[\delta E\left(\vec{x}\right)\right].
\label{data_processing_ENFS_fluttuazione_intensita}
\end{equation}
Because of the conservation of the total intensity during the
scattering process, 
by averaging $\left\{I\left(\vec{x}\right)\right\}$ over the whole plane,
we obtain $I_0$:
\begin{equation}
\left<\left\{I\left(\vec{x}\right)\right\}\right> = E_0^2 = I_0.
\end{equation}
We can now evaluate the heterodyne signal
$i\left(\vec{x}\right)$, subtracting the the stray light 
contribution:
\begin{equation}
i\left(\vec{x}\right) =
\frac{I\left(\vec{x}\right) - \left\{I\left(\vec{x}\right)\right\}}
{\left<\left\{I\left(\vec{x}\right)\right\}\right>}.
\label{eq_ENFS_valutazione_ottimale}
\end{equation}

\section{Correction for finite samples.}

The quantity $\left\{I\left(\vec{x}\right)\right\}$ should ideally be evaluated
by averaging infinite images. We obtain a good evaluation of it
by averaging
a great number $N$ of images $I_n\left(\vec{x}\right)$,
typically one hundred:
\begin{equation}
\bar{I}\left(\vec{x}\right) =
\frac{1}{N} \sum{I_n\left(\vec{x}\right)} \approx 
\left\{I\left(\vec{x}\right)\right\}
\end{equation}
From this evaluation, we obtain $i\left(\vec{x}\right)$:
\begin{equation}
i_n\left(\vec{x}\right) =
\frac{I_n\left(\vec{x}\right) - \bar{I}\left(\vec{x}\right)}
{\left<\bar{I}\right> }
\end{equation}

The average value $\bar{I}\left(\vec{x}\right)$,
evaluated over a given number of images,
is sistematically different from the true mean value, in
the direction that reduces the evaluation of the root mean
square displacement from the mean. This problem is analogous
to the one that leads to the so called Bessel correction for the
evaluation of the variance $\sigma$ of a stochastic variable, from the
knowledge of a finite number of stochastic values.

We evaluate the correlation function of 
$i_n\left(\vec{x}\right)$ for each $n$, then we average them,
thus obtaining $C_i\left(\Delta \vec{x}\right)$.
Now we want to evaluate $\left\{C_i\left(\Delta \vec{x}\right)\right\}$,
that is the mean value over infinite samples, in order to correct
systematic errors:
\begin{equation}
\left\{C_i\left(\Delta \vec{x}\right)\right\} =
\frac{1}{\left<\bar{I}\right>^2}
\left\{
\frac{1}{N}\sum_{n=0}^N{\left<
\left[I_n\left(\vec{x}\right) - \frac{1}{N}\sum_{m=0}^N
{I_m\left(\vec{x}\right)}\right]
\left[I_n\left(\vec{x}+\Delta \vec{x}\right) - \frac{1}{N}\sum_{m=0}^N
{I_m\left(\vec{x}+\Delta \vec{x}\right)}\right]
\right>}
\right\}
\end{equation}
The symbol $\left< \cdot \right>$ means the average over $\vec{x}$.
We can write$\bar{I}\left(\vec{x}\right) + \delta 
I\left(\vec{x}\right)$ instead of $I_n\left(\vec{x}\right)$:
\begin{equation}
\left\{C_i\left(\Delta \vec{x}\right)\right\} =
\frac{1}{\left<\bar{I}\right>^2}
\left\{
\frac{1}{N}\sum_{n=0}^N{\left<
\left[\delta I_n\left(\vec{x}\right) - \frac{1}{N}\sum_{m=0}^N
{\delta I_m\left(\vec{x}\right)}\right]
\left[\delta I_n\left(\vec{x}+\Delta \vec{x}\right) - \frac{1}{N}\sum_{m=0}^N
{\delta I_m\left(\vec{x}+\Delta \vec{x}\right)}\right]
\right>}
\right\}
\end{equation}
Evaluating the products:
\begin{eqnarray}
\left\{C_i\left(\Delta \vec{x}\right)\right\}& = &
\frac{1}{\left<\bar{I}\right>^2}
\frac{1}{N}
\sum_{n=0}^N
{\left\{ \left<
\delta I_n\left(\vec{x}\right)
\delta I_n\left(\vec{x}+\Delta \vec{x}\right)
\right>\right\} }
-\\
&&\frac{1}{\left<\bar{I}\right>^2}
\frac{1}{N^2}
\sum_{n,m=0}^N
{\left\{ \left<
\delta I_n\left(\vec{x}\right)
\delta I_m\left(\vec{x}+\Delta \vec{x}\right)
\right>\right\} }
-
\frac{1}{\left<\bar{I}\right>^2}
\frac{1}{N^2}
\sum_{n,m=0}^N
{\left\{ \left<
\delta I_m\left(\vec{x}\right)
\delta I_n\left(\vec{x}+\Delta \vec{x}\right)
\right>\right\} }
+\\
&&\frac{1}{\left<\bar{I}\right>^2}
\frac{1}{N^3}
\sum_{n,m,l=0}^N
{\left\{ \left<
\delta I_m\left(\vec{x}\right)
\delta I_l\left(\vec{x}+\Delta \vec{x}\right)
\right>\right\} }
\end{eqnarray} 
Since $\bar{\delta I}=0$:
\begin{equation}
\left\{C_i\left(\Delta \vec{x}\right)\right\} = 
\frac{N-1}{N}
\frac{1}{\left<\bar{I}\right>^2}
\left\{ \left<
\delta I\left(\vec{x}\right)
\delta I\left(\vec{x}+\Delta \vec{x}\right)
\right>\right\}
\end{equation} 
Now we can use Eq. (\ref{data_processing_ENFS_fluttuazione_intensita}):
\begin{equation}
\left\{C_i\left(\Delta \vec{x}\right)\right\} = 
\frac{N-1}{N}C_E\left(\Delta \vec{x}\right)
\end{equation}

The correlation function evaluated on $N$ samples is proportional 
to the correlation function evaluated for $N \to \infty$. The
poportionality constant is the same of the well known Bessel
correction.

\section{ENFS data processing.}
\label{sezione_ENFS_processing_correlazione}

\begin{figure}
\begin{center}
\includegraphics[scale=0.4]{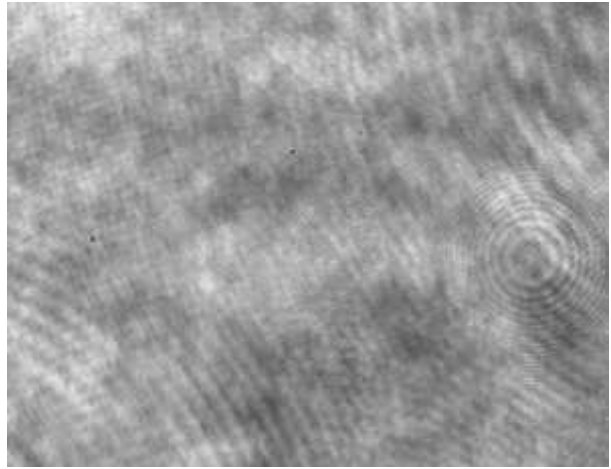}
\end{center}
\caption[ENFS background image.]{Background image.}
\label{ENFS_data_processing_fondo}
\end{figure}

Once the experimental apparatus has been built, as described in
Chapter \ref{capitolo_sistema_sperimentale}, in the absence of the
sample, the CCD should be illuminated in a quite uniform way. As a
matter of fact, the illumination is never completely uniform,
primarily because of the interference of the main beam with stray
light. A typical image is shown in
Fig. \ref{ENFS_data_processing_fondo}. We can easily see some sets of
concentric circles, each due to reflections inside a lens, along with
speckle patterns properly due to stray light.

\begin{figure}
\begin{center}
\includegraphics[scale=0.4]{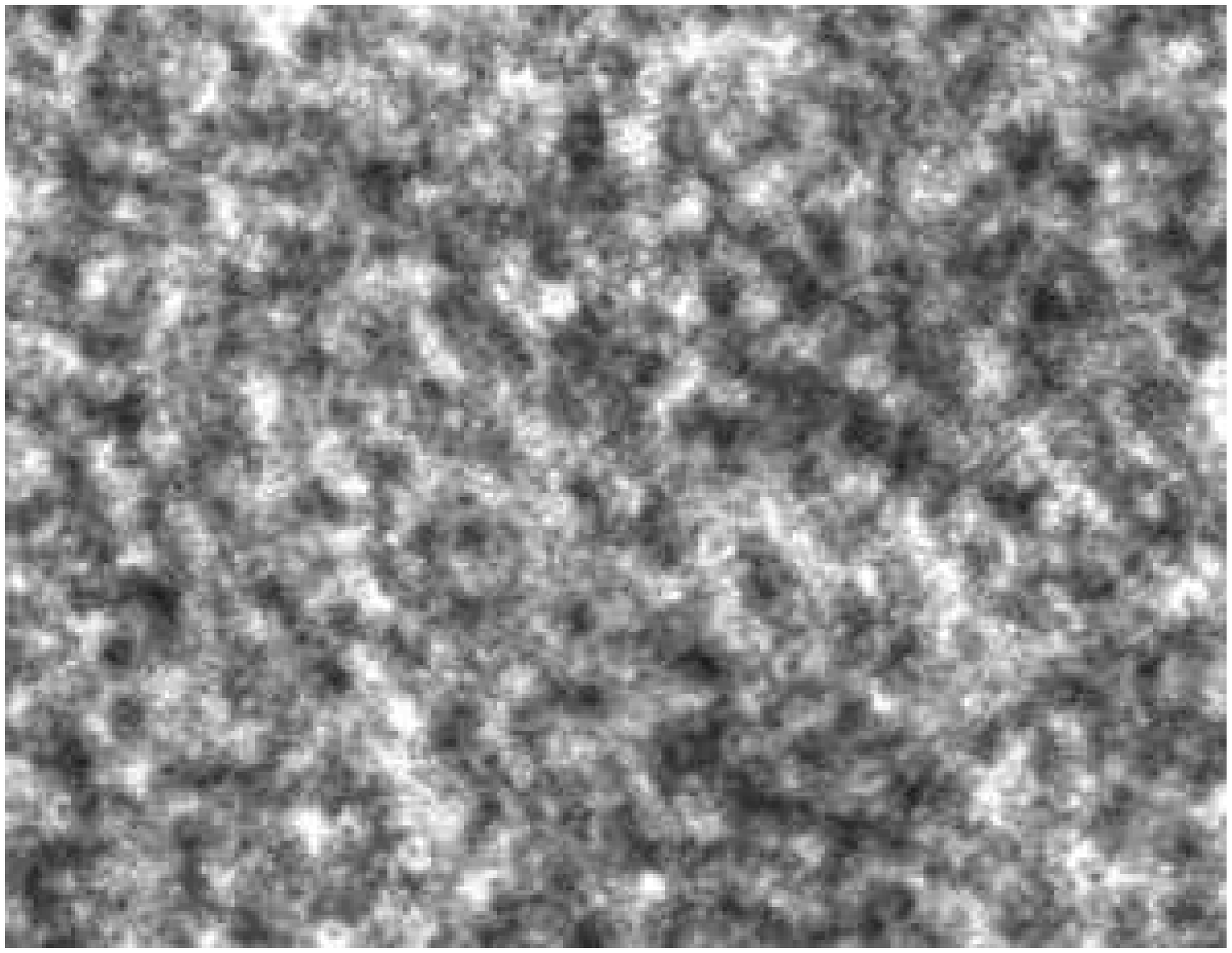}
\end{center}
\caption[ENFS $10.0\mathrm{\mu m}$ image.]{ENFS image of a $10.0\mathrm{\mu m}$
colloid.}
\label{ENFS_data_processing_imm_10um}
\end{figure}

\begin{figure}
\begin{center}
\includegraphics[scale=0.4]{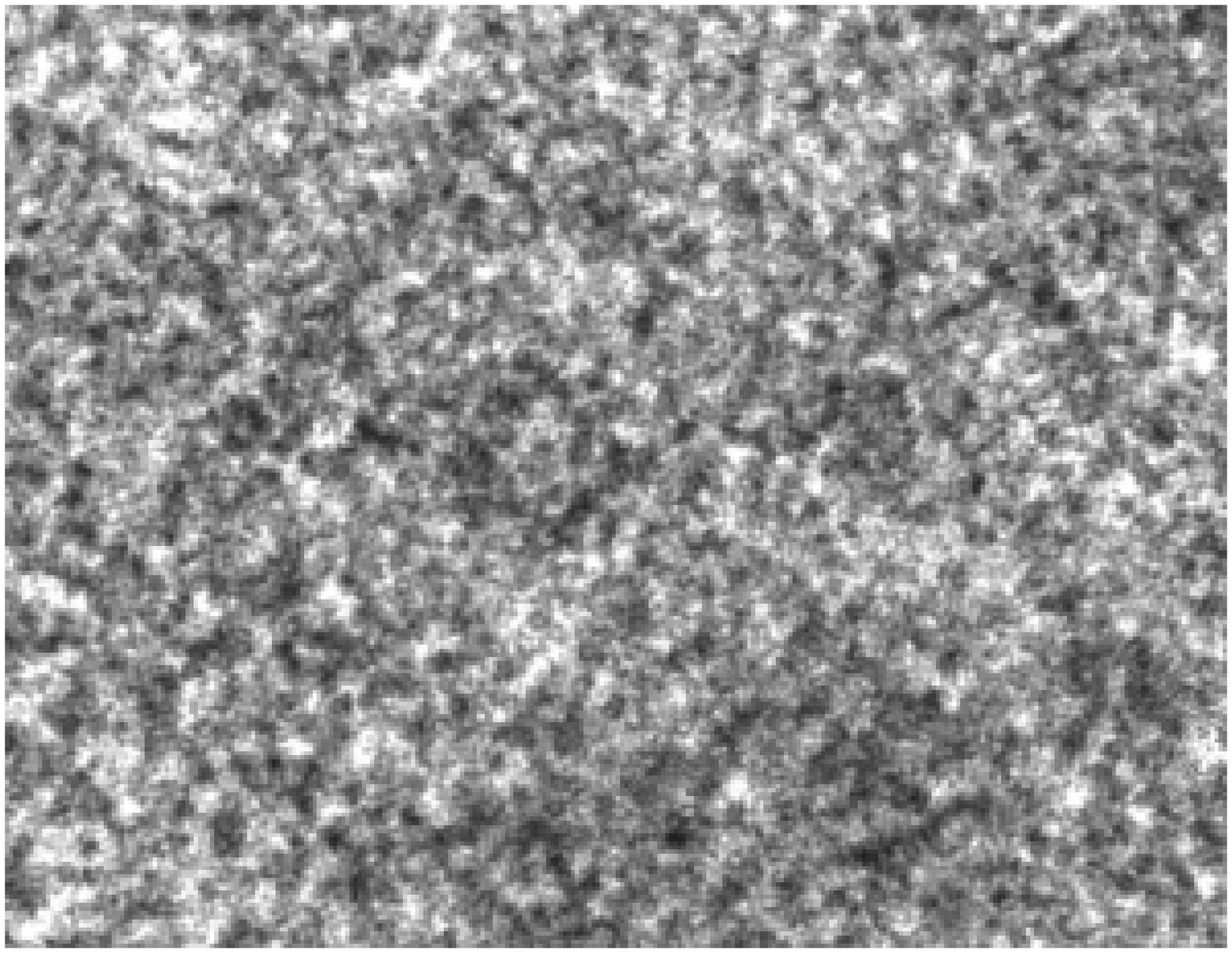}
\end{center}
\caption[ENFS $5.2\mathrm{\mu m}$ image.]{ENFS image of a $5.2\mathrm{\mu m}$
colloid.}
\label{ENFS_data_processing_imm_5um}
\end{figure}

When the the sample is placed in the right position, we acquire about
one hundred images for each measurement. The electronic shutter of 
the CCD and its interlacement time must be so short that no evident
evolution of the system happens during the exposure: for the samples
we studied, an interlacement delay of
$1/25\mathrm{s}$ is sufficient. Moreover, different images must be completely
uncorrelated. For a $10\mathrm{\mu m}$ colloid, images must be grabbed at
intervals longer that one minute, if only brownian movements are the
source of decorrelation, while for the non equilibrium fluctuations we
studied the images can be taken at intervals of $1\mathrm{s}$. In
figure \ref{ENFS_data_processing_imm_10um} and
\ref{ENFS_data_processing_imm_5um} we show two typical ENFS images,
generated by the interference of the main beam with the light
scattered by colloids of $5.2\mathrm{\mu m}$ and $10.0\mathrm{\mu m}$.
The images show a
mean intensity, modulated by the interference with the speckle
pattern. We can notice the different typical size of the
speckles. The set of concentric circles can be seen yet: the stray
light will be removed with the following step.

\begin{figure}
\begin{center}
\includegraphics[scale=0.4]{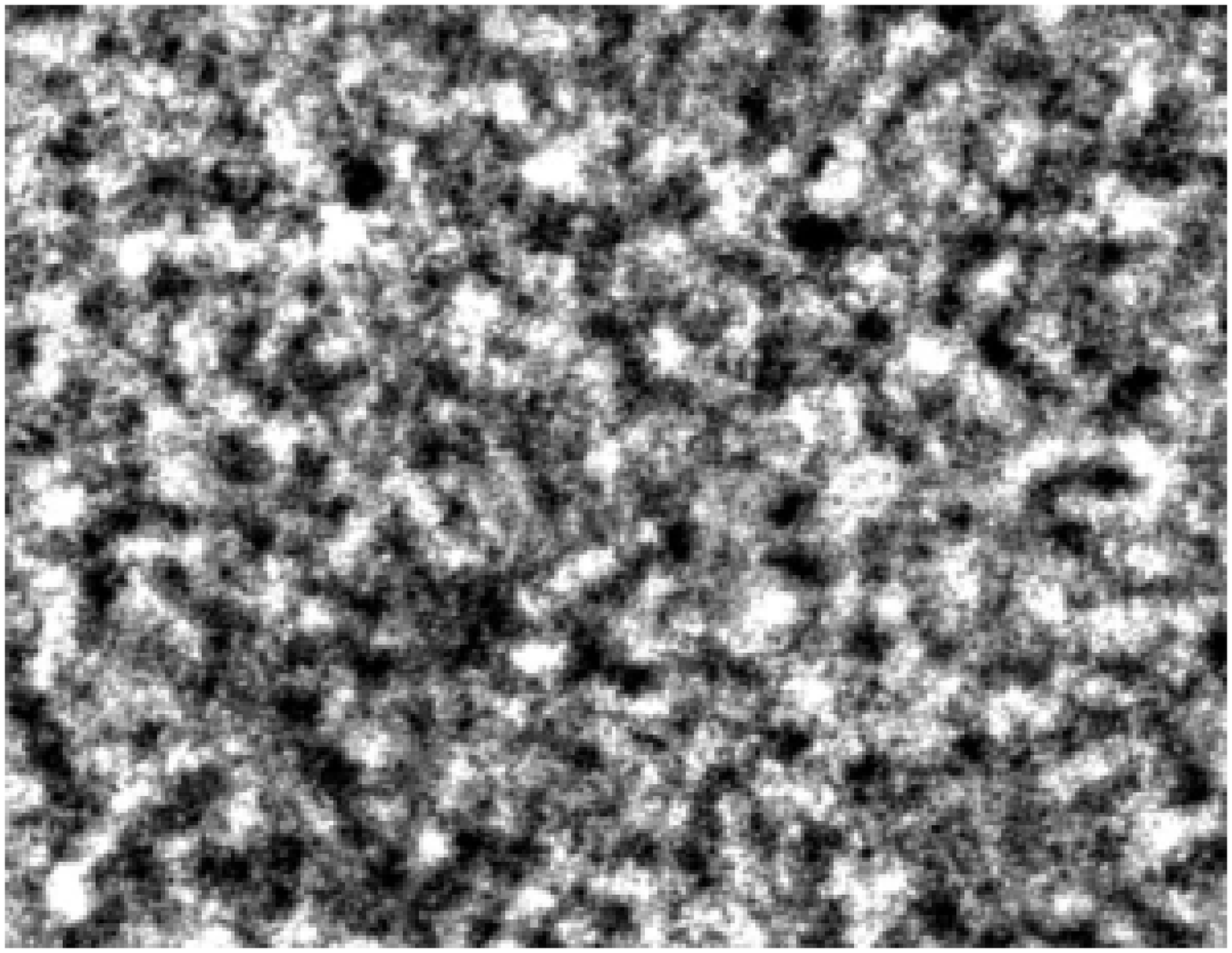}
\end{center}
\caption[ENFS $10.0\mathrm{\mu m}$ signal.]{ENFS signal of a $10.0\mathrm{\mu m}$
colloid.}
\label{ENFS_data_processing_segnale_10um}
\end{figure}

\begin{figure}
\begin{center}
\includegraphics[scale=0.4]{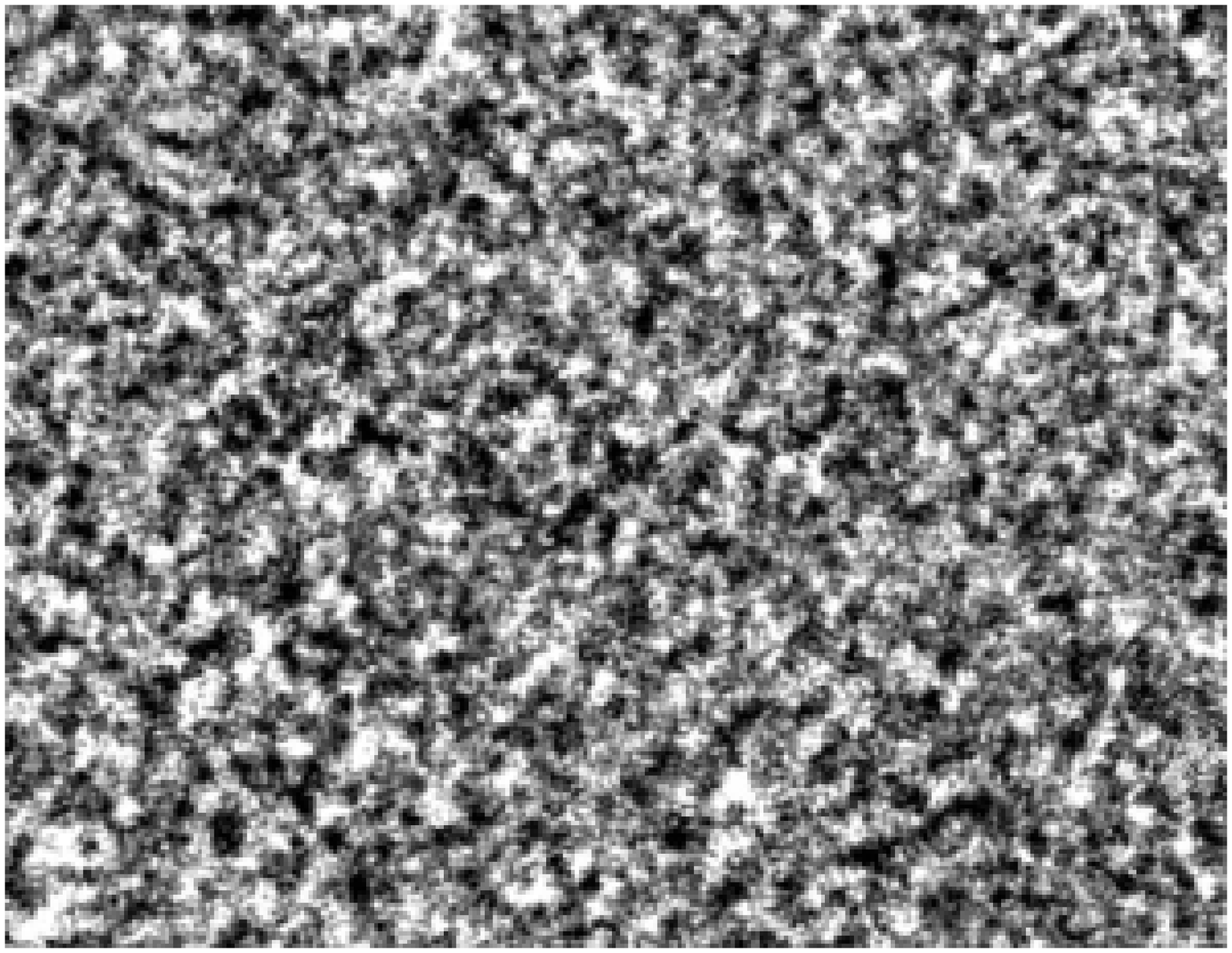}
\end{center}
\caption[ENFS $5.2\mathrm{\mu m}$ signal.]{ENFS signal of a $5.2\mathrm{\mu m}$
colloid.}
\label{ENFS_data_processing_segnale_5um}
\end{figure}

Once the images $I_n\left(\vec{x}\right)$ have been acquired, they are
averaged, in order to evaluate $\bar{I}\left(\vec{x}\right)$ and
$\left<\bar{I}\right>$. By using
Eq. (\ref{eq_ENFS_valutazione_ottimale}), we evaluate
$i\left(\vec{x}\right)$, the heterodyne signal. Figures
\ref{ENFS_data_processing_segnale_10um} and
\ref{ENFS_data_processing_segnale_5um} show the heterodyne signal:
since $i\left(\vec{x}\right)$ is negative, for some points, a constant
intensity has been added. The images thus simply represent the ENFS
images, cleaned from stray light and optical imprefections.

\begin{figure}
\begin{center}
\includegraphics{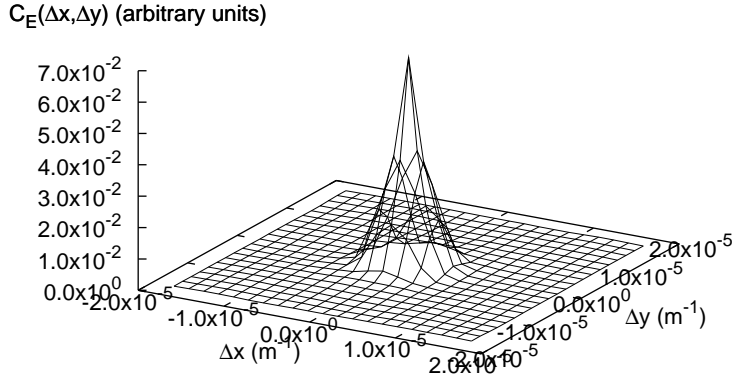}
\end{center}
\caption{ENFS measurement of the field correlation function, for a
$5.2\mathrm{\mu m}$ colloid.}
\label{ENFS_processing_imm_corr_5um}
\end{figure}

\begin{figure}
\begin{center}
\includegraphics{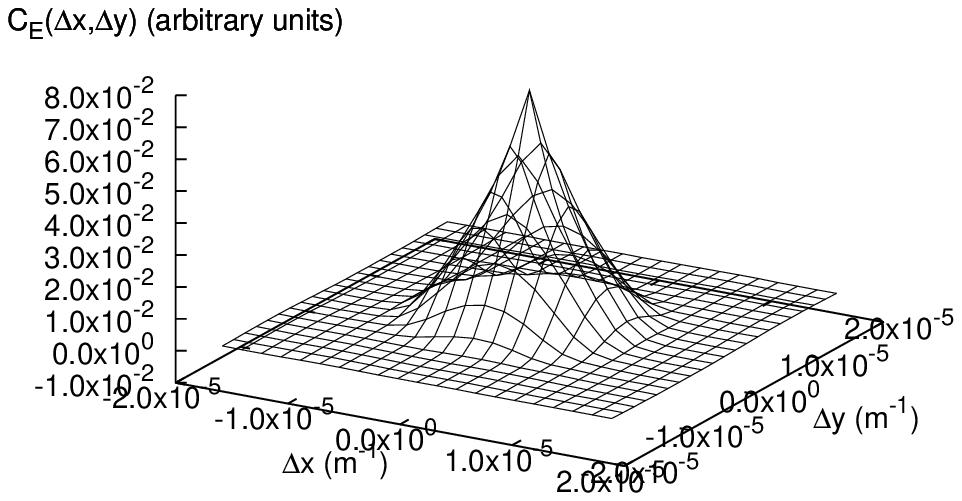}
\end{center}
\caption{ENFS measurement of the field correlation function, for a
$10.0\mathrm{\mu m}$ colloid.}
\label{ENFS_processing_imm_corr_10um}
\end{figure}

The heterodyne signal of each image is then elaborated in order to
obtain its power spectrum. Simple Fourier transforming of the signal
would be uncorrect, due to border effects. First of all, we evaluate
the correlation function. This operation is quite fast, since we can
use a Fast Fourier Transform (FFT) algorithm. An FFT algorithm allows
to evaluate the Fourier tranform of an $M\times N$ matrix, with a
number of arithmetic operations proportional to
$MN\log\left(MN\right)$. By using Perceval relation, we can obtain the
correlation function by doing an FFT, evaluating the square modulus,
and doing an Inverse FFT (IFFT). This only requires a number of
operation of the order of $MN\log\left(MN\right)$. By scanning every
value of $\Delta x$, and averaging over every $N\times M$ pixels, the
number of operations would be of the order of
$\left(MN\right)^2$. Using FFT, well known tricks can be used, in
order to correct the boundary effects \cite{num_recipes}. Figure
\ref{ENFS_processing_imm_corr_5um} and
\ref{ENFS_processing_imm_corr_10um} show the correlation functions thus
evaluated.

\begin{figure}
\begin{center}
\includegraphics{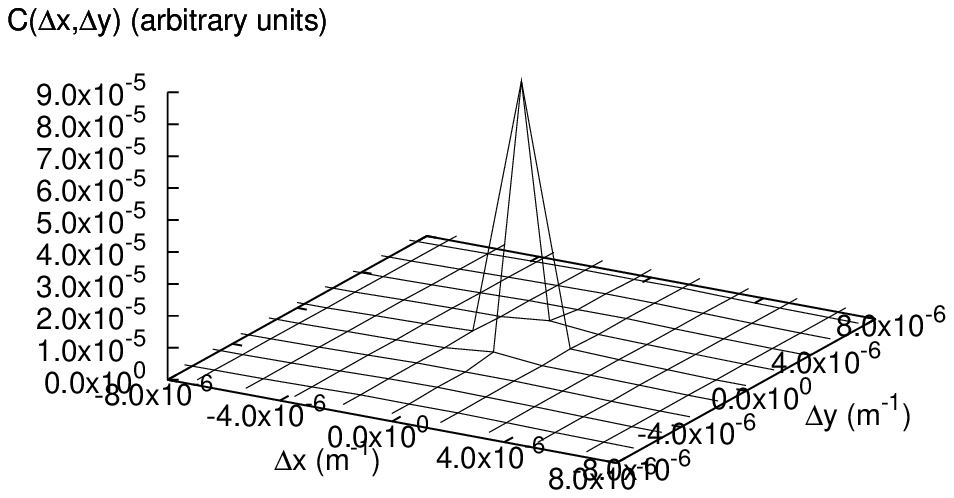}
\end{center}
\caption{Correlation function of the shot and read noise.}
\label{ENFS_processing_imm_corr_fondo}
\end{figure}

The correlation function evaluated following the above described algorithm
suffers from
shot and read noise, that is, for the noise due to the CCD light
measurement and acquisition systems. Since such a noise is not
correlated to the speckle field due to scattered light, the noise
correlation function sums to the speckle correlation function. In
order to evaluate the noise correlation function, we acquire a set of
about one hundred images, before putting the sample in the
system. Then, we apply the above described algorithm to the images, and
obtain the correlation function of the noise signal. Figure
\ref{ENFS_processing_imm_corr_fondo} shows the correlation function of
the noise signal. We can notice a marked peak in 0, quite narrow,
representing the correlation inside a row, and a correlation between
lines spaced by two pixels, due to interlacing.
The correlation function of the noise signal is then subtracted by the
overall correlation function.

\begin{figure}
\begin{center}
\includegraphics{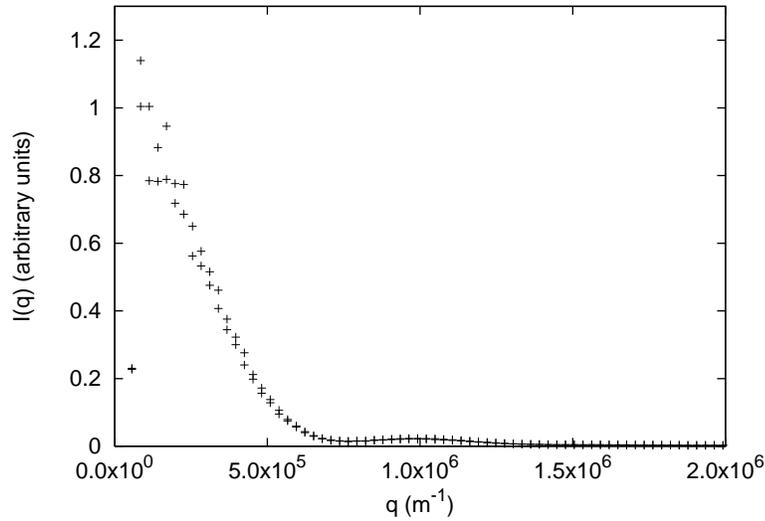}
\end{center}
\caption{ENFS measurement of the scattered intensity of a $10.0\mathrm{\mu m}$ colloid.}
\label{ENFS_data_processing_spettro_10um}
\end{figure}

\begin{figure}
\begin{center}
\includegraphics{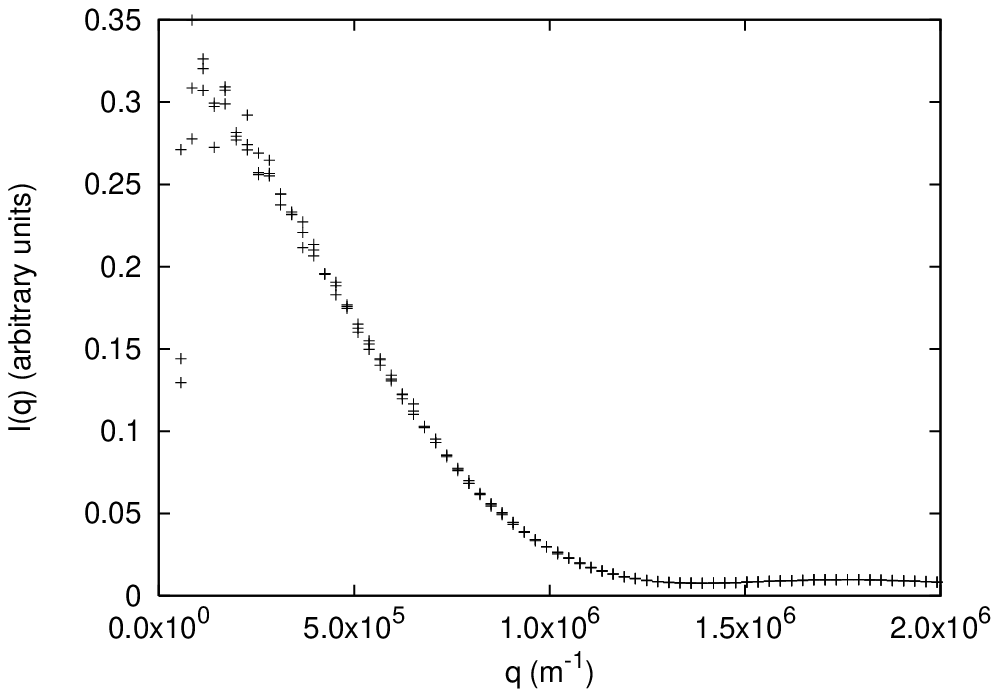}
\end{center}
\caption{ENFS measurement of the scattered intensity of a $5.2\mathrm{\mu m}$ colloid.}
\label{ENFS_data_processing_spettro_5um}
\end{figure}

Once the correlation function has been evaluated, through an FFT we
obtain the field power spectrum $S_E\left(q\right)$. Since our samples
are isotropic, we make an angular average of the power spectra, and
represent our data as a function of the modulus $q$ of $\vec{q}$. The
scattered intensity $I\left(q\right)$ is then obtained by using
Eq. (\ref{teoria_relazione_IQ_Ez_di_q}), that is, simply relating each
value of the power spectra, with wavelength $q$ to a value of
$I\left(Q\right)$, where the relation $Q\left(q\right)$ is given by
Eq. (\ref{teoria_trasferito_vettore_onda}). In
Fig. \ref{ENFS_data_processing_spettro_10um} and
\ref{ENFS_data_processing_spettro_5um} we show the measured
$I\left(q\right)$.

%
\chapter{Performances of ONFS and ENFS on two colloidal samples.}
\label{capitolo_confronto_ONFS_ENFS}

In this chapter we describe ONFS and ENFS measurements we made on two
colloids and discuss the results. The optical setup has already been
described in Sections \ref{sect_optical_setup_ENFS_colloidi}.

\section{The samples.}

The samples we used are monodisperse colloids made of polystyrene
spheres suspended in
water. In order to avoid sedimentation, we used a mixture of water and
heavy water with a weigh fraction of about 0.5. The diameters of the
two colloids we used are $5.2\mathrm{\mu m} \pm 0.5\mathrm{\mu m}$
and $10.0\mathrm{\mu m} \pm 0.3\mathrm{\mu m}$, whose polydispersity
is negligible. The diameters are quite large, since NFS gives
advantages with respect to classical LS for small wavevectors.

The colloids we measured are held in a cell with plane parallel
windows. The diameter is about $4\mathrm{cm}$,
since the sample and the beam intensity must be uniform on a
length $D$, where $D$ is given by Eq. 
(\ref{teoria_eq_condizione_diametro}). The thickness is
about $2\mathrm{mm}$. We selected the tickness and the particle
concentration in order to have a suitable attenuation of the
main beam, about 1\%. For ONFS measurements, the thickness of the cell
and the volumetric particle density are enough to fulfill
Eq. (\ref{teoria_eq_condizione_gaussianita}).

The liquid is held between the two windows by an O-ring; the parallelism
between the windows is not critical, nor the optical quality of them.
Since the measured scattered light comes from different regions of the sample,
we must provide that it is homogeneous. This implies that the thickness
must be uniform, but an optical quality allignment is far beyond
what is needed.

\section{Measurements.}

\begin{figure}
\begin{center}
\includegraphics{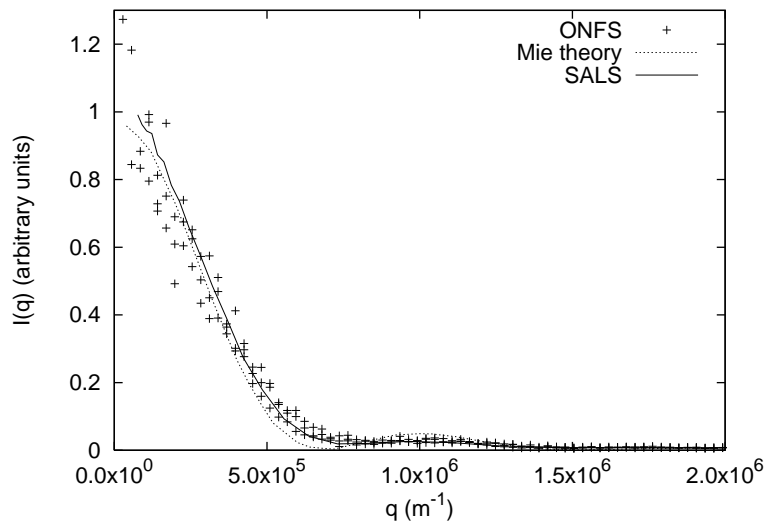}
\end{center}
\caption[ONFS measurement of the $10.0\mathrm{\mu m}$ colloid.]{ONFS measurement
of the $10.0\mathrm{\mu m}$ colloid. A SALS measurement and a theoretical
evaluation based on Mie theory are shown.}
\label{results_ONFS_10um}
\end{figure}

\begin{figure}
\begin{center}
\includegraphics{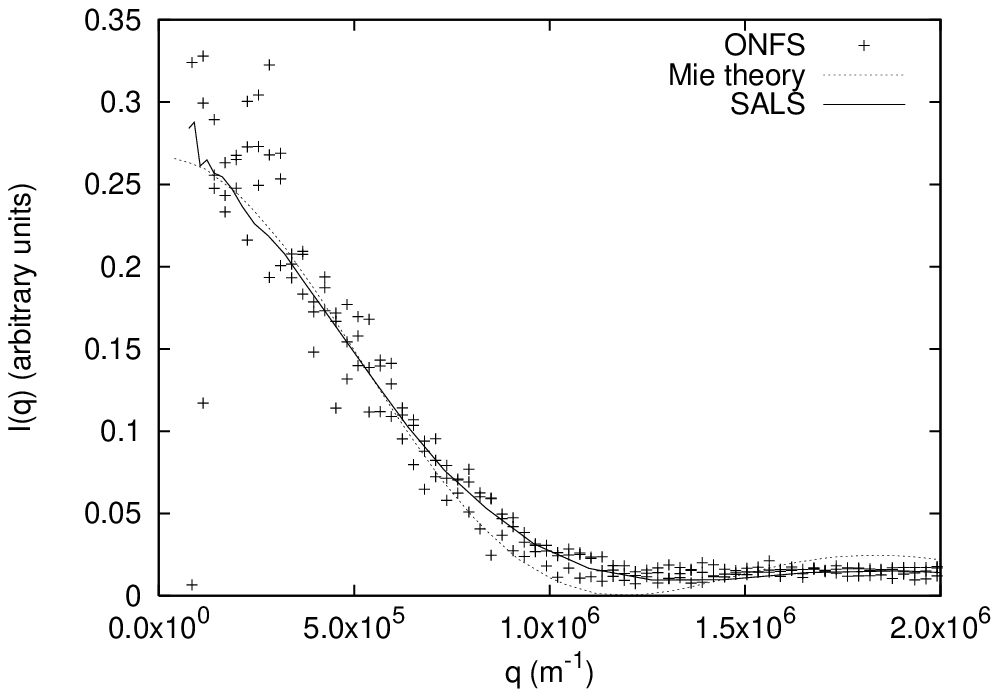}
\end{center}
\caption[ONFS measurement of the $5.2\mathrm{\mu m}$ colloid.]{ONFS measurement
of the $5.2\mathrm{\mu m}$ colloid. A SALS measurement and a theoretical
evaluation based on Mie theory are shown.}
\label{results_ONFS_5um}
\end{figure}

\begin{figure}
\begin{center}
\includegraphics{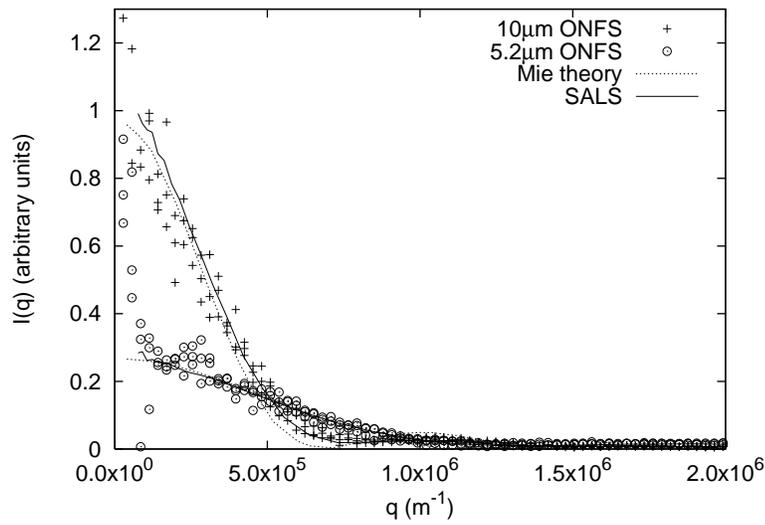}
\end{center}
\caption[ONFS measurements of the two colloids.]{ONFS measurements
of the two colloids. SALS measurements and theoretical
evaluations based on Mie theory are shown.}
\label{results_ONFS_5_10um}
\end{figure}

In Fig. \ref{results_ONFS_10um}, \ref{results_ONFS_5um} and 
\ref{results_ONFS_5_10um} are shown the results of ONFS
measurement. The same samples have been analyzed also by a SALS instrument;
the results are shown in the graphs.

\begin{figure}
\begin{center}
\includegraphics{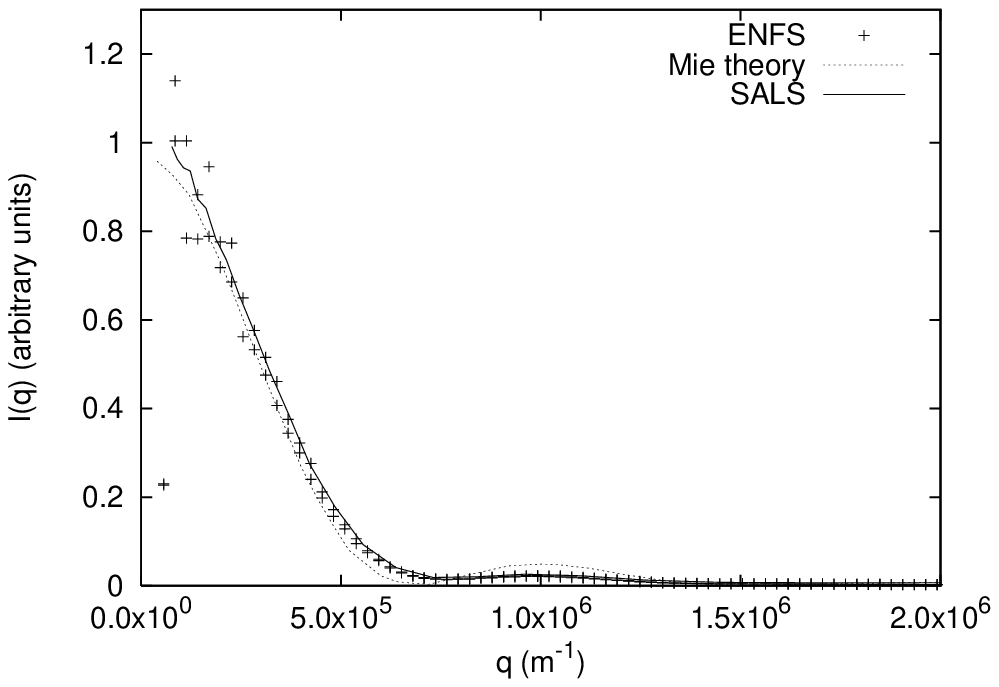}
\end{center}
\caption[ENFS measurement of the $10.0\mathrm{\mu m}$ colloid.]{ENFS measurement
of the $10.0\mathrm{\mu m}$ colloid. A SALS measurement and a theoretical
evaluation based on Mie theory are shown.}
\label{results_ENFS_10um}
\end{figure}

\begin{figure}
\begin{center}
\includegraphics{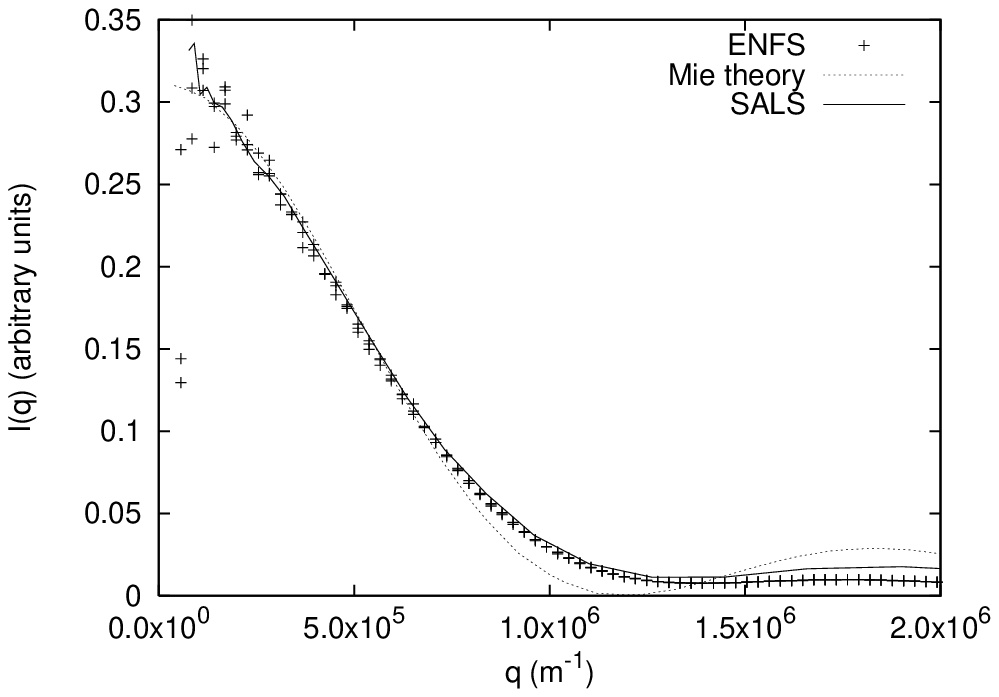}
\end{center}
\caption[ENFS measurement of the $5.2\mathrm{\mu m}$ colloid.]{ENFS measurement
of the $5.2\mathrm{\mu m}$ colloid. A SALS measurement and a theoretical
evaluation based on Mie theory are shown.}
\label{results_ENFS_5um}
\end{figure}

\begin{figure}
\begin{center}
\includegraphics{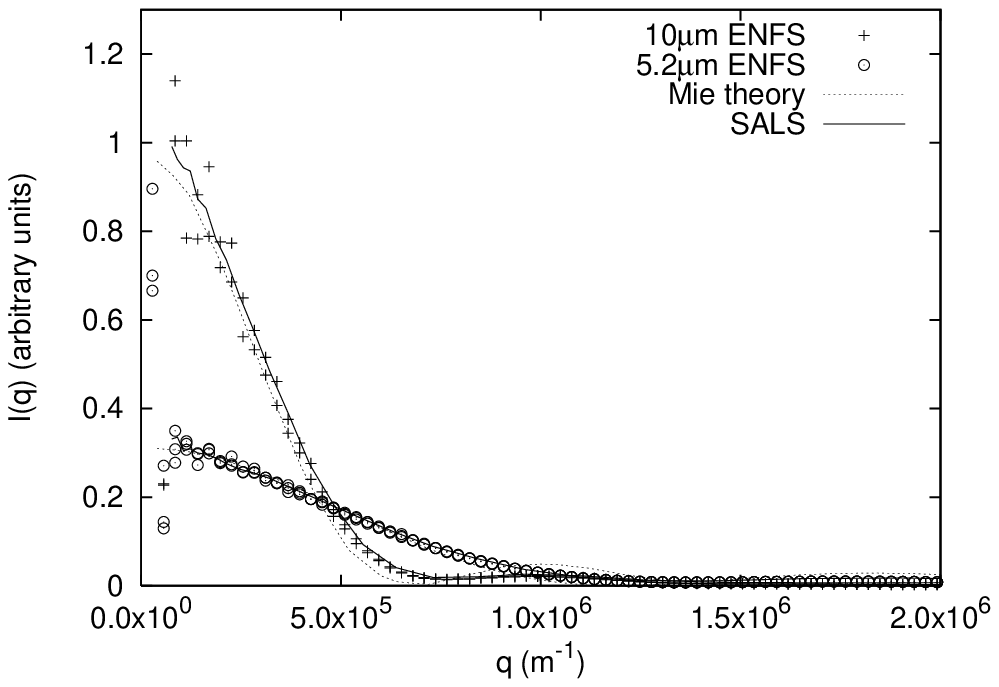}
\end{center}
\caption[ENFS measurements of the two colloids.]{ENFS measurements
of the two colloids. SALS measurements and theoretical
evaluations based on Mie theory are shown.}
\label{results_ENFS_5_10um}
\end{figure}

In Fig. \ref{results_ENFS_10um}, \ref{results_ENFS_5um} and 
\ref{results_ENFS_5_10um} are shown the results of ENFS
measurement. The same samples have been analyzed also by a SALS instrument;
the result is shown in the graphs.

Both ONFS and ENFS measurements agree with the SALS measurement,
but ONFS gives less accuracy.

\section{What is the main source of error?}

The ONFS data processing is based on Siegert relation; the correlation
function is evaluated by Eq. (\ref{stray_light_sottrazione_sl_bessel}).
First, we average the correlation functions of each image. The error in the
evaluation of the intensity correlation decreases as the square root of the
number of the samples, and this dependence is intrinsic in the stochastical
nature of the technique. Then, we evaluate the field correlation function,
by extracting the square root of the difference
between the mean intensity correlation and the square mean intensity.
Thus we obtain a quantity which converges to the field correlation function
as the fourth root of the number of samples.

On the contrary, ENFS gives directly the correlation function without any
extraction of square root. We can notice that the plateau of Fig. 
\ref{ENFS_processing_imm_corr_5um} and
\ref{ENFS_processing_imm_corr_10um} are more plain than in Fig. 
\ref{results_colloid_2d_c_5um} and \ref{results_colloid_2d_c_10um}.
The noise on the plateau is then spread across all the power spectrum
when the Fourier transform is performed.

The situation is worst than in dynamic light scattering. The reason is that
we are working in two dimensions: in order that the power spectrum
can be evaluated, the correlation function must approach 0 faster than
$1/r^3$, while in one dimension, it must be faster than $1/r^2$. We
can perform an angular mean, but we gain only a term $r^{-1/2}$.

\begin{figure}
\begin{center}
\includegraphics{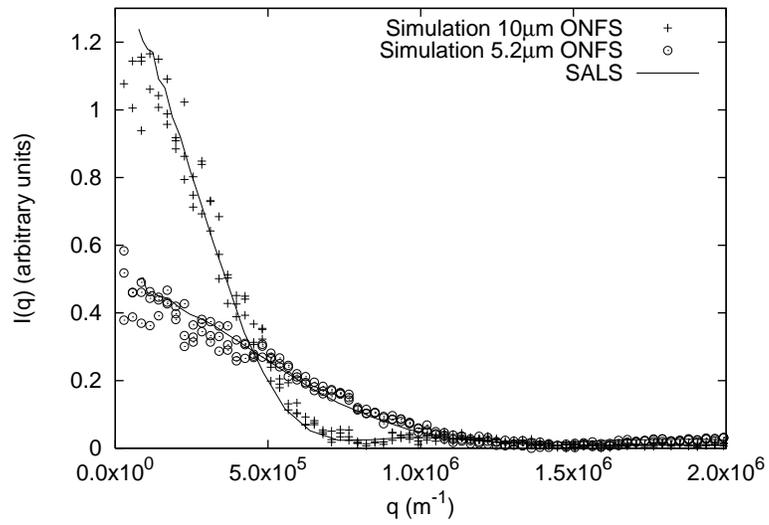}
\end{center}
\caption[ONFS simulations of the two colloids.]{Simulations of ONFS
measurements of the two colloids. SALS measurements are also shown.}
\label{simulazioni_ONFS_5_10um}
\end{figure}

\begin{figure}
\begin{center}
\includegraphics{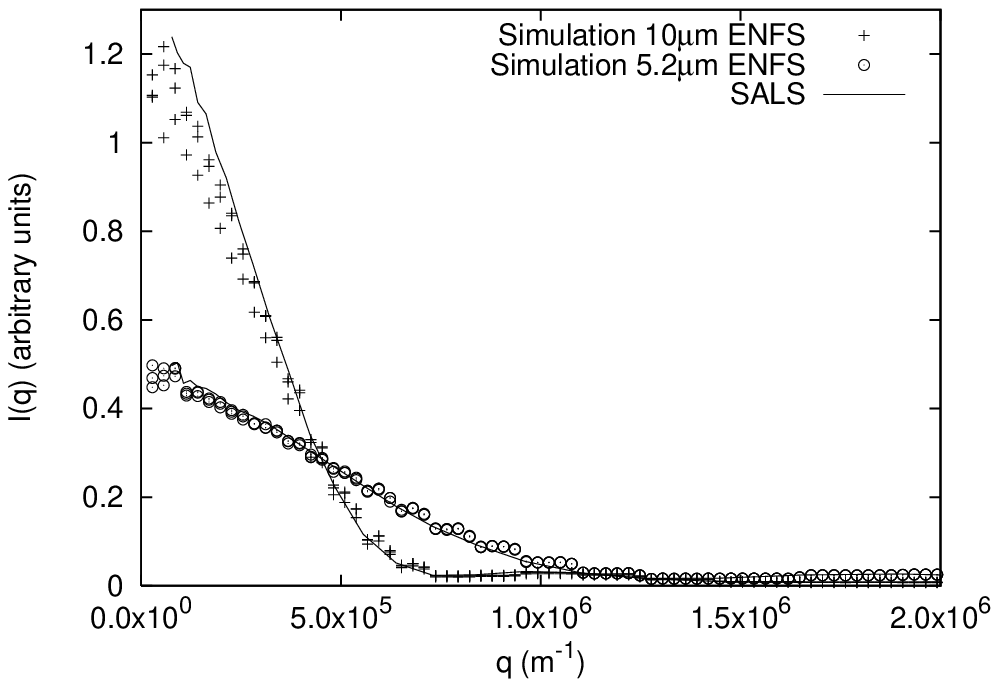}
\end{center}
\caption[ENFS simulations of the two colloids.]{Simulations of ENFS
measurements of the two colloids. SALS measurements are also shown.}
\label{simulazioni_ENFS_5_10um}
\end{figure}

These considerations should explain why ONFS data are much less accurate
than ENFS and SALS ones: the problem is the slow statistical convergence
as the number of samples increases. In order to test this explanation, we
performed some numerical simulations. We used the SALS data to simulate
a power spectrum. We created one hundred random fields, 
with gaussian probability and the given power spectrum.
We obtained the homodyne and heterodyne signals, thus creating images
similar to the ones we acquired during the experiments. Last, we
processed the data with the above described algorithms. 
The simulations are only affected
by the statistics: they are virtually free from any experimental error.
In Fig. \ref{simulazioni_ENFS_5_10um} and Fig. \ref{simulazioni_ONFS_5_10um}
we present, respectively, the results for the ONFS and
ENFS simulations. They look quite similar to the corresponding
experimental measurements of Fig. \ref{results_ENFS_5_10um} and
\ref{results_ONFS_5_10um}. This confirms that the main source
of error, for our ONFS measurements, is the poor statistical quality of
the samples. Since the quality increases as the fourth power of the number
of the samples, we cannot make ONFS measurements better than ENFS ones,
unless we process at least one million images: this is, for the moment,
impossible.

%
\chapter {Particle sizing with ENFS.}
\label{capitolo_particle_sizing_ENFS}

One of the most important applications of Light Scattering technique,
from the industrial point of view, is particle sizing. Industrial
particle sizers generally include some sensors, which measure the scattered
intensity $I\left(q\right)$, both at small and high angles. Generally,
a mechanical system makes the powder or the colloid flow in a cell, so
that a good statistical sample can be obtained.
An algorithm, based on Mie theory, tries to find the distribution of particle
diameters, which best fits the measured scattered intensity.

\begin{figure}
\begin{center}
\includegraphics{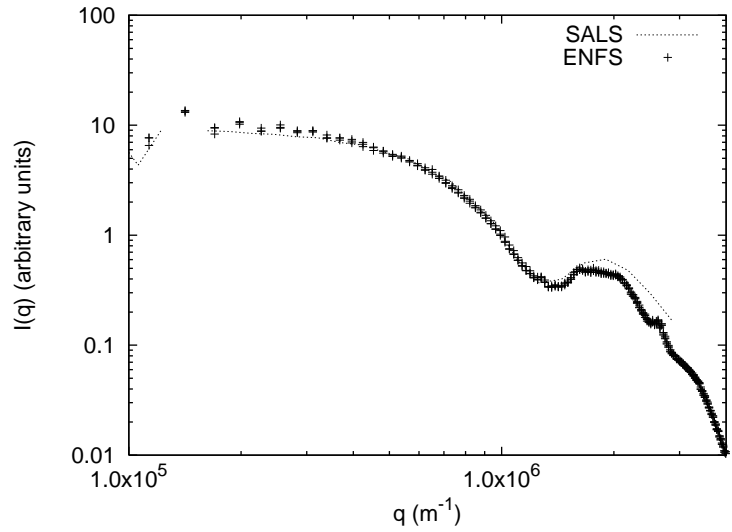}
\end{center}
\caption[Particle sizing: sample A, $5.2\mathrm{\mu m}$ colloid]
{Scattered light
intensity measurement of a $5.2\mathrm{\mu m}$ colloid (sample A).}
\label{particle_sizing_ENFS_fig_1x5um-0x10um}
\end{figure}

\begin{figure}
\begin{center}
\includegraphics{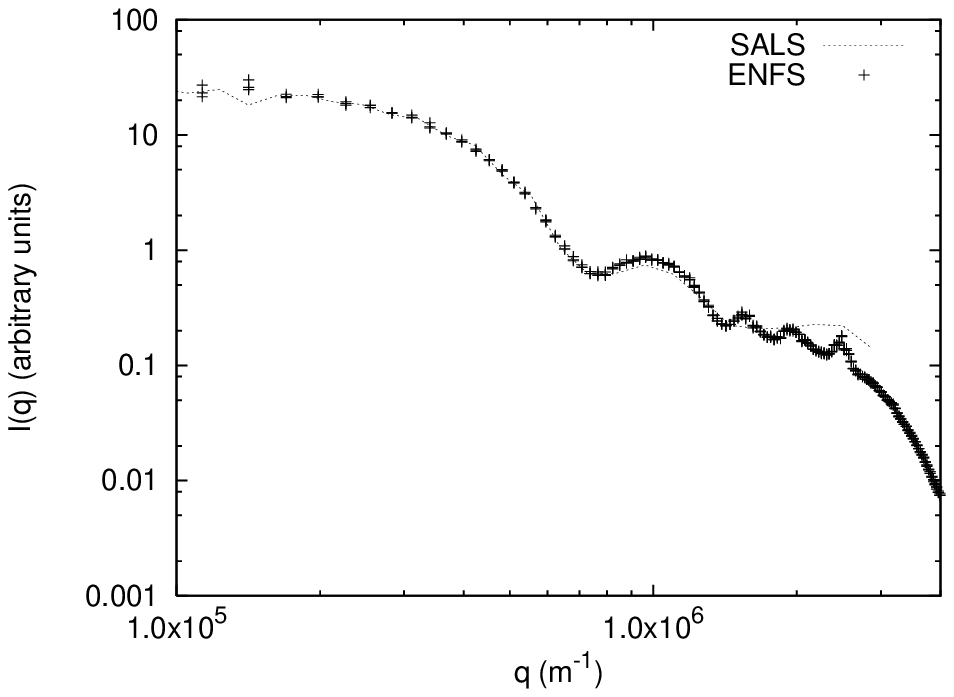}
\end{center}
\caption[Particle sizing: sample B, $10.0\mathrm{\mu m}$ colloid]
{Scattered
light intensity measurement of a $10.0\mathrm{\mu m}$ colloid (sample B).}
\label{particle_sizing_ENFS_fig_0x5um-1x10um}
\end{figure}

\begin{figure}
\begin{center}
\includegraphics{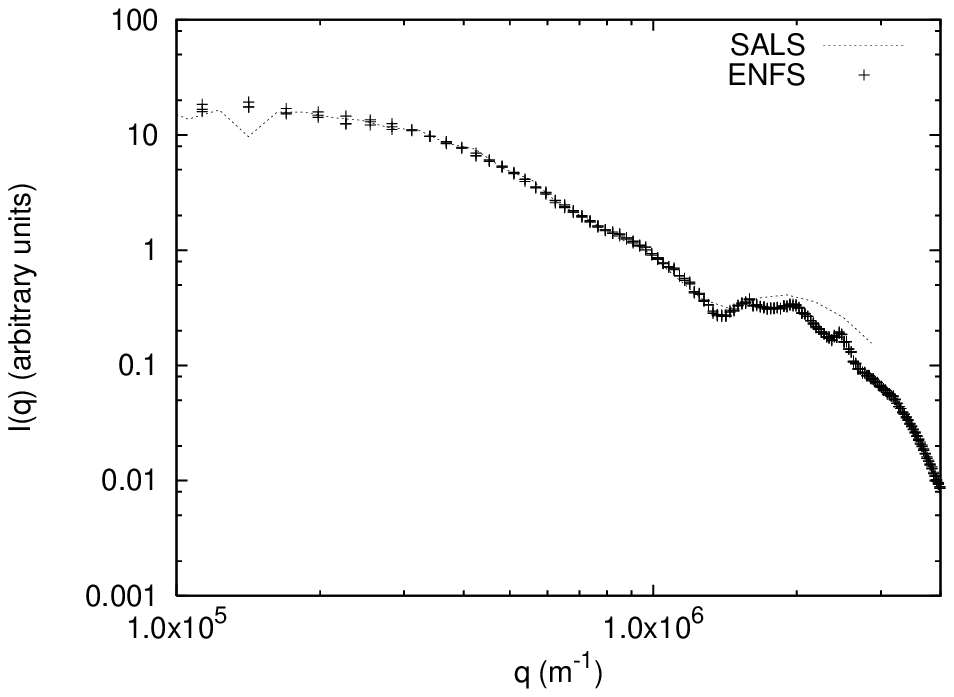}
\end{center}
\caption[Particle sizing: solution 1/2 A + 1/2 B] {Scattered light intensity
measurement of a mixture of the two samples. Volume fractions: 1/2 A,
1/2 B}
\label{particle_sizing_ENFS_fig_1x5um-1x10um}
\end{figure}

\begin{figure}
\begin{center}
\includegraphics{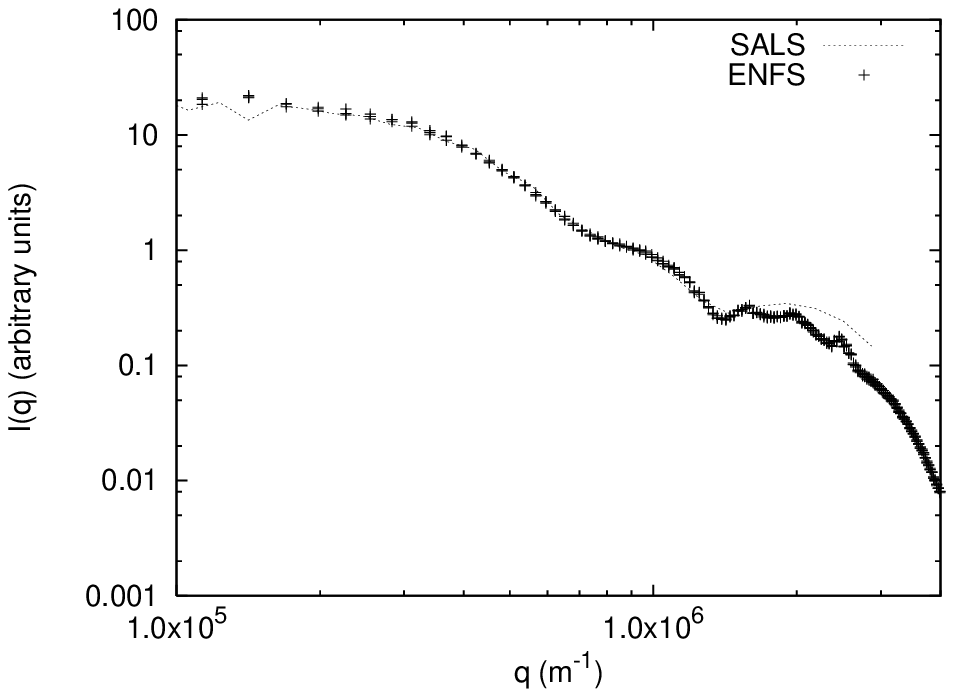}
\end{center}
\caption[Particle sizing: solution 1/3 A + 2/3 B] {Scattered light intensity
measurement of a mixture of the two samples. Volume fractions: 1/3 A,
2/3 B}
\label{particle_sizing_ENFS_fig_1x5um-2x10um}
\end{figure}

\begin{figure}
\begin{center}
\includegraphics{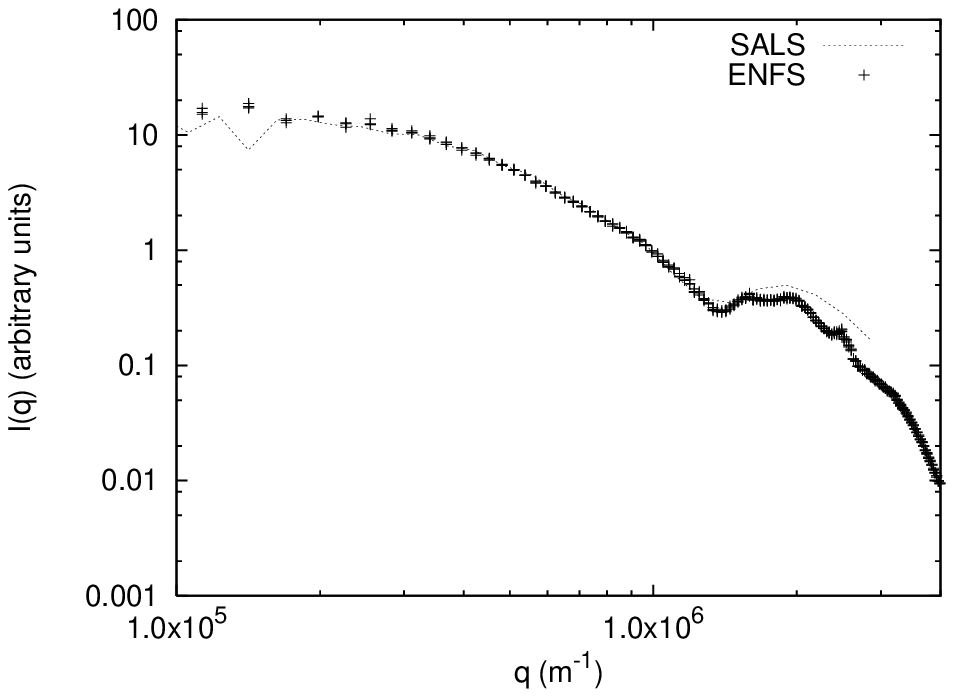}
\end{center}
\caption[Particle sizing: solution 2/3 A + 1/3 B] {Scattered light intensity
measurement of a mixture of the two samples. Volume fractions: 2/3 A,
1/3 B}
\label{particle_sizing_ENFS_fig_2x5um-1x10um}
\end{figure}

In order to asses the reliability of ENFS applied to particle sizing,
we analyzed some mixtures of two colloids. We prepared two colloidal
solution of polystyrene spheres. In order that the density
of the solvent matches the density of the colloid, we used a solution of equal
volumes of water and heavy water: the colloid was quite stable, and did
not sediment evidently over some hours. The diameters of the
two colloids are $5.2\mathrm{\mu m} \pm 0.5\mathrm{\mu m}$
and $10.0\mathrm{\mu m} \pm 0.3\mathrm{\mu m}$
(samples A and B). The refraction index of the polystyrene is $1.59$,
while the solvent has the refraction index of water, $1.33$. Then, we
prepared three mixtures of them, respectively with volume fractions of
1:1, 1:2, 2:1 of samples A and B. The scattered intensity was measured
both with ENFS and a state-of-the-art SALS instrument. The data are
presented in Figs. 
\ref{particle_sizing_ENFS_fig_1x5um-0x10um},
\ref{particle_sizing_ENFS_fig_0x5um-1x10um},
\ref{particle_sizing_ENFS_fig_1x5um-1x10um},
\ref{particle_sizing_ENFS_fig_1x5um-2x10um},
\ref{particle_sizing_ENFS_fig_2x5um-1x10um}

We define $\alpha$ and $\beta$ the volume fractions of samples A and B
in each mixture; the scattered intensity of the mixture with a given
$\alpha$ and $\beta$ is $I_{\alpha,\beta}\left(q\right)$.
The scattered intensities $I_{\alpha,\beta}\left(q\right)$,
obtained for the three mixtures, are compared with
the scattered intensities $I_A\left(q\right)$ and $I_B\left(q\right)$ of the 
two samples A and B. We evaluate the values of $\alpha'$ and $\beta'$
for which $I_{\alpha,\beta}\left(q\right) \approx 
\alpha' I_A\left(q\right)+ \beta' I_B\left(q\right)$, by looking for the minima
of the mean square deviation:
\begin{equation}
\left\{
\begin{array}{l}
\alpha'=\frac
{
\sum_q{I_{\alpha,\beta}\left(q\right)I_A\left(q\right) }
\sum_q{I_B^2\left(q\right)}
-
\sum_q{I_{\alpha,\beta}\left(q\right)I_B\left(q\right) }
\sum_q{I_A\left(q\right)I_B\left(q\right) }
}
{
\sum_q{I_A^2\left(q\right)}
\sum_q{I_B^2\left(q\right)}
-
\left[\sum_q{I_A\left(q\right)I_B\left(q\right) }\right]^2
}
\\
\beta'=\frac
{
\sum_q{I_{\alpha,\beta}\left(q\right)I_B\left(q\right) }
\sum_q{I_A^2\left(q\right)}
-
\sum_q{I_{\alpha,\beta}\left(q\right)I_A\left(q\right) }
\sum_q{I_A\left(q\right)I_B\left(q\right) }
}
{
\sum_q{I_A^2\left(q\right)}
\sum_q{I_B^2\left(q\right)}
-
\left[\sum_q{I_A\left(q\right)I_B\left(q\right) }\right]^2
}
\end{array}
\right.
\end{equation}
The values of $\alpha'$ and $\beta'$ are the measured colloid concentrations,
and must be compared with $\alpha$ and $\beta$. 
Table \ref{part_sizing_tab_frazioni_misurate} shows the measured values,
$\alpha'$ and $\beta'$, compared with the actual ones, $\alpha$ and $\beta$.
\begin{table}
\begin{center}
\begin{tabular}{|l|l|}
\hline
Measured&Actual\\
$\alpha'$,$\beta'$&$\alpha$,$\beta$\\
\hline
$0.52 A, 0.46 B$&$1/2 A, 1/2 B$\\
$0.69 A, 0.32 B$&$2/3 A, 1/3 B$\\
$0.31 A, 0.68 B$&$1/3 A, 2/3 B$\\
\hline
\end{tabular}
\end{center}
\caption[Measured ratio of sample A and B in the mixtures]
{Measured and actual values of volume concentrations of colloid A and B
in the three mixtures.}
\label{part_sizing_tab_frazioni_misurate}
\end{table}
The agreement is
quite good: this shows that ENFS is suited for particle sizing.

\begin{figure}
\begin{center}
\includegraphics{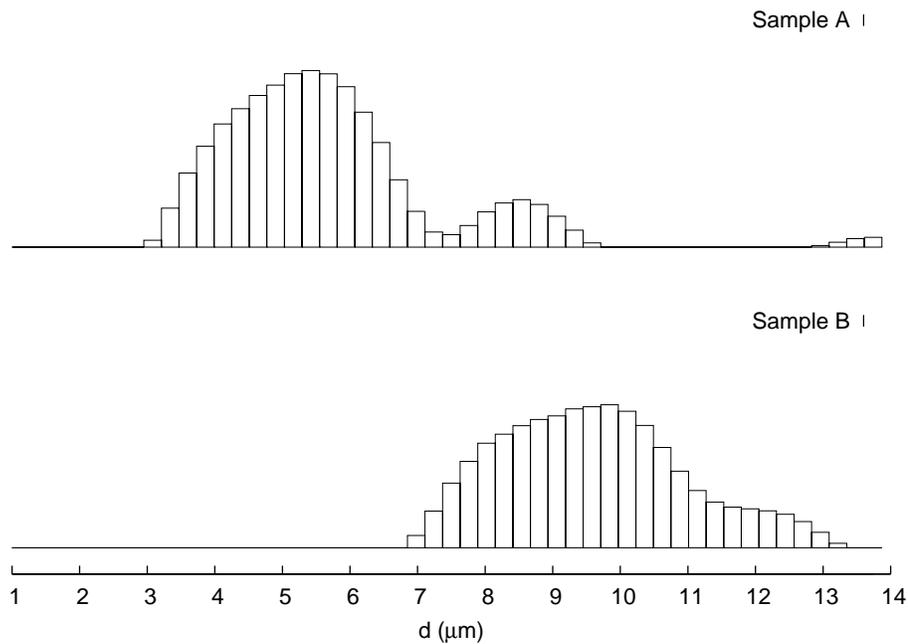}
\end{center}
\caption[Diameter distribution of two monodispese colloids measured by ENFS]
{Diameter distribution of the two colloidal samples measured by ENFS, obtained
by an inversion algorithm based on Mie theory. The height of the bars
is proportional to the intensity of light scattered by the particles
in the range covered by the horizontal extension of the bar.
Sample A is a $5.2\mathrm{\mu m}$ colloid, and sample B is a
$10\mathrm{\mu m}$ colloid. The two peaks are evident. Sample A shows
a small peak centered around $8\mathrm{\mu m}$: it corresponds to the
scattering of the dymers.
}
\label{part_sizing_istogrammi_inversione_campioni}
\end{figure}

\begin{figure}
\begin{center}
\includegraphics{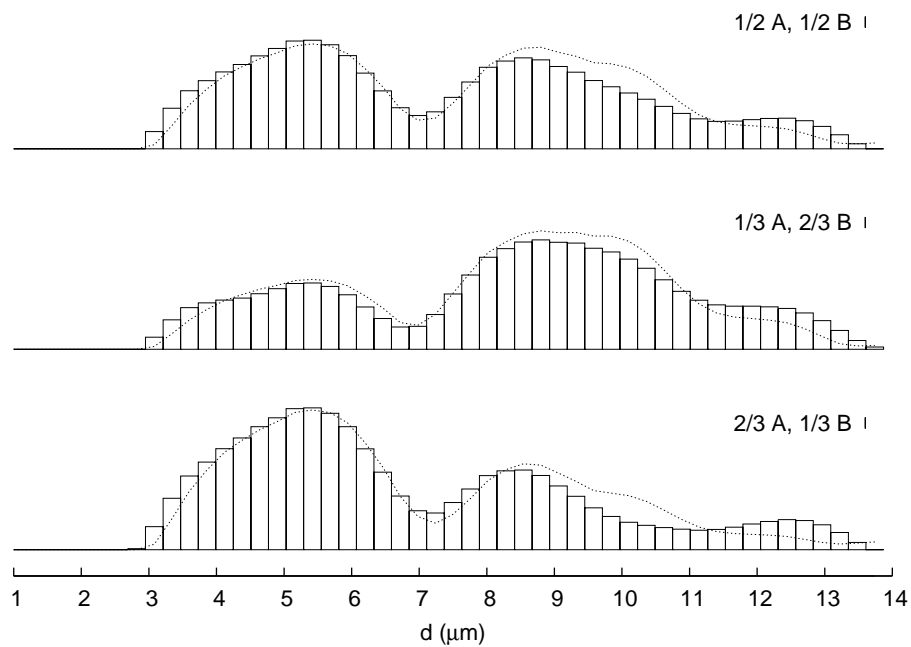}
\end{center}
\caption[Diameter distribution of the mixtures measured by ENFS]
{Diameter distribution of the mixtures of colloidal samples, measured
by ENFS, obtained 
by an inversion algorithm based on Mie theory. The height of the bars
is proportional to the intensity of light scattered by the particles
in the range covered by the horizontal extension of the bar. The
dotted curves are obtained by combining the values measured for samples
A and B, shown in Fig. \ref{part_sizing_istogrammi_inversione_campioni}
with coefficients given by the volume fractions of the two samples.}
\label{part_sizing_istogrammi_inversione_miscele}
\end{figure}

The scattering data has been analyzed by an inversion algorithm based
on Mie theory. Mie theory allows to evaluate the scattered intensity
$I\left(q\right)$ generated by
a given diameter distribution $\rho\left(d\right)$ 
of dielectric spheres; the inversion
algorithm looks for the distribution $\rho\left(d\right)$ which 
gives the best approximation to the measured $I\left(q\right)$.
The results are shown in
Figs. \ref{part_sizing_istogrammi_inversione_campioni} and
\ref{part_sizing_istogrammi_inversione_miscele}.
Two peaks are quite
evident: they are centered on the diameters of $5\mu m$ and $10\mu
m$. The small peak centered around $8\mathrm{\mu m}$ in the histogram
of Sample A corresponds to the scattering of the dymers: the colloid
is partially aggregated. The height of the peaks in
Fig. \ref{part_sizing_istogrammi_inversione_miscele} change
accordingly to the fraction of the samples A and B in the mixture.

It should be noticed that ENFS measures the intensity of the scattered
beams with reference to the main beam. This allows to evaluate the
particle concentration, and not only the relative concentration of
different particles. This is accomplished by using a single sensor; on
the contrary, with SALS, the transmitted and the scattered beams must
be measured by independent sensors,
because the intensities are generally extremely different. This
difference comes from the fact that SALS sensors measure the intensity
of scattered beams, while ENFS measures the interference of them. For
example, consider a sample that generates a single scattered beam,
whose intensity is $1/10000$ than the transmitted one. For SALS, we
need two sensors, one for measuring the scattered beam and one for the
transmitted beam, and they require an accurate calibration. A single
sensor could be used without calibration, but its dynamic range should
cover 4 decades, and in this range it should be quite linear. For
ENFS, the interference of the two beams generates a modulation of
about $1/100$. A single CCD array can easily measure such a
modulation.

%
\chapter {Non-equilibrium fluctuations in a free diffusion experiment 
studied with SNFS}
\label{capitolo_dinamico_SNFS}

In this chapter, we describe a measurement of the power spectrum of the
nonequilibrium fluctuations that arise during a free diffusion
experiment. In Sect. \ref{misura_snfs_sezione_teoria} we discuss the
origin of such fluctuations; in Sect. \ref{misura_snfs_sezione_sistema} we
describe the physical system we studied;
the results are shown in 
Sect. \ref{misura_snfs_sezione_risultati}. The optical setup has been
already described in Sect. \ref{misura_snfs_sezione_opt_setup}.

\section{Nonequilibrium fluctuations in free diffusion processes.}
\label{misura_snfs_sezione_teoria}

Diffusion is the fundamental mass transfer mechanism in many natural
and technological processes. The diffusive transport can be
interpreted by the simple molecular random walk model. A more refined
description requires the understanding of direct interaction between
the diffusing particles and possibly hydrodynamic interactions. Both
types of interactions may produce appreciable changes in the magnitude
of the effective diffusion coefficient $D$ but, at any extent,
diffusion is believed to give rise to an intimate and homogeneous
remixing on matter. The general belief is that while the process
occurs over quite microscopic distances, nothing peculiar should occur
at any other lengthscale, except the molecular one where the random
molecular diffusion takes place.

It has been recently shown that, quite unexpectedly, gigant
fluctuations are present during the diffusive remixing of two miscible
phases of a binary mixture not too far from its critical point 
\cite{vailati1997}. A fluctuating hydrodynamic description has been
developed \cite{vailati1998}, which indicates that gigant
nonequilibrium fluctuations should be present during the diffusive
remixing of fluids in general; moreover, it has been shown that the
fluctuations can be considered the origin of the whole Fick flow
\cite{brogioli2001}.

The presence of the fluctuations
has been detected experimentally during the free diffusive remixing
occurring in ordinary liquid mixtures and in macromolecular solutions
\cite{brogioli2000,brogioli2000_2}. The measurements concerned an
ordinary, low molecular weight liquid mixture, an aqueous solution of
a low molecular weight solid, a polymer
solution and a protein solution, thus giving evidence that these
anomalous fluctuations are a universal feature associated with
spontaneous diffusion across a macroscopic gradient. 

A free diffusion experiment begins filling a cell with the two
liquids, with the denser solution in the lower part to avoid
convective instability. The two horizontal layers are initially
separated by a fairly sharp meniscus. As soon as the two liquids came into
contact, the diffusive remixing begins, and the meniscus repidly becomes
smeared. The concentration profile inside the sample, initially a step
function as a function of the height $z$, gradually evolves into an
\emph{s}-shaped function \cite{cussler}, until eventually, after a few
days, the concentration becomes uniform throughout the sample.

During the free diffusion process described above, intense
fluctuations arise. Their power spectrum $S\left(\vec{q}\right)$,
with $\vec{q}$ in the horizontal direction, is given by:
\begin{equation}
S\left(q\right)=
S_0\frac{1}{1+\left[\frac{q}{q_{RO}}\right]^4}.
\label{SNFS_misure_eq_spettro_flutt}
\end{equation}
The roll-off wave vector $q_{RO}$ is given by:
\begin{equation}
q_{RO}=\sqrt[4]{\frac{\beta g \nabla c}{\nu D}}
\label{snfs_eq_roll_off}
\end{equation}
where $g$ is the gravity acceleration, $\nu$ is the kinematic viscosity,
$D$ is the diffusion coefficient, and $\beta=\frac{1}{\rho}{
\frac{\partial \rho}{\partial c}}_p,T$ is the solutal expansion
coefficient, a quantity that increases as increases the mismatching of
the two liquids. The gradient $\nabla c$ can be assumed roughly
connstant in the region between the fluids, where diffusion takes
place, and vanishes outside. The sample-dependent prefactor in
Eq. (\ref{SNFS_misure_eq_spettro_flutt}) is given by:
\begin{equation}
S_0=K_B T\left(\frac{\partial n}{\partial c}\right)^2
\frac{\Delta c}{\rho \beta g}
\label{snfs_eq_altezza_plateau}
\end{equation}
where $\Delta c$ is the total concentration difference across the
sample.

The power spectrum $S\left(q\right)$ dislpays a $q^{-4}$ power low
divergence at large wave vectors, $q\gg q_{RO}$,
and a saturation at a constant value at small wave vectors,
$q\ll q_{RO}$. The $q^{-4}$ power low is interpreted as the result of
of the coupling of velocity fluctuations with concentration
fluctuations, while the saturation is due to a stabilizing effect of
gravity on long wavelength fluctuations.

Moreover, the roll-off wave
vector where the transition between the two regimes occurs gets
smaller as $\beta g \nabla c$, and the the low wave vector value of
the power spectrum $S_0$ is roughly constant as free diffusion takes
place, since the concentration near the upper and lower windows of the
cell are initially constant.

The nonequilibrium concentration fluctuations are originated from the
coupling of velocity fluctuations with concentration fluctuations, due
to the presence of a macroscopic concentration gradient. This can be
understood by simple naive arguments, discussed in detail in
\cite{vailati1998} and \cite{weitz1997}. Suppose that a small parcel
of fluid of linear size $a$ undergoes a velocity fluctuation. This
fluctuation will displace the parcel until the viscous drag will stop
it in a time given approximately by $\tau_{visc}=a^2/\nu$, $\nu$ being
the kinematic viscosity. If the displacement of the parcel occurs in a
direction parallel to the macroscopic concentration gradient, the
parcel will be surrounded by fluid with different concentration. The
life time of this concentration fluctuation is $\tau_{diff}=a^2/D$,
and is much larger than the viscous time $\tau_{visc}$, as
$D\ll\nu$. Thus, in the presence of a macroscopic gradient, the effect
of a short living velocity fluctuation is to induce a long lasting
concentration fluctuation. Once a concentration fluctuation has been
created, two mechanisms may contribute to its relaxation: diffusion
and buoyancy. If the spatial extent of the fluctuation is small, then
the fluctuation will soon disappear due to diffusion. This mechanism
gives rise to the $q^{-4}$ divergence of the static power spectrum at
high wavevectors. As the wavevector increases, the velocity
fluctuation lives for a shorter time, and can displace the parcel of a
smaller amount, and this gives a factor $q^{-2}$; moreover, the
displaced parcel will be dissipated as $q^{-2}$.
However, if the fluctuation is large enough, the buoyancy force acting
on it will be able to restore the fluctuation in the layer of fluid
having the same density in a time shorter than the diffusive one. This
gives rise to the frustration of the $q^{-4}$ divergence at smaller
wavevectors.

\section{The cell.}
\label{misura_snfs_sezione_sistema}

We have investigated the free diffusion process that takes place when
two miscible fluids are brought in contact, the mixing between
adjoining regions being kept as little as possible before a
measurement sequence. The liquid sample we used was an aqueous
solution of glycerol, with a weight fraction of 0.3. It was diffused
into pure water. The cell was filled with the two liquids, with the
denser solution in the lower part to avoid convective instability. The
two horizontal layers are initially separated by a fairly sharp
meniscus. As SNFS is an image forming technique, at least for big
objects, we were able to
thoroughly check the sample for spurious disturbances at the interface
before starting collecting data. As soon as the two liquids came into
contact, diffusion takes place, and the nonequilibrium fluctuations
arise.

\begin{figure}
\includegraphics[scale=0.7]{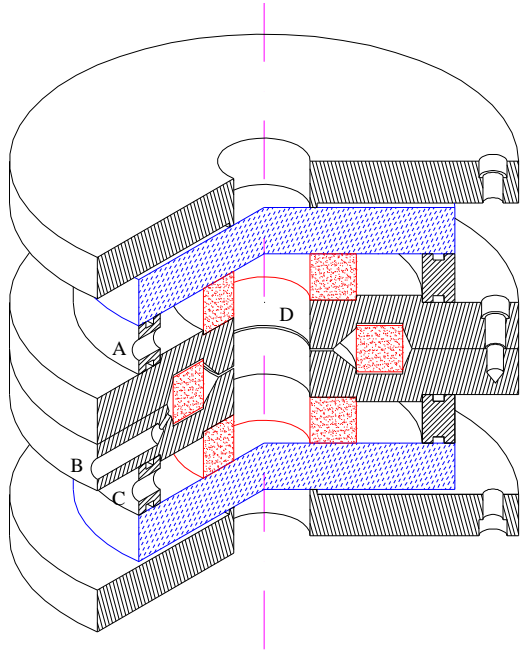}
\caption[Flowing Junction Cilindrical Cell]
{The Flowing Junction Cylindrical Cell, developed for measurements
of non equilibrium fluctuations in free diffusion experiments in
microgravity. The gray parts are made of perspex. The glass windows
are blue in the drawing. The two liquids are injected through the holes:
water in hole A and a solution of water and glycerol in hole C.
They fill, respectively, the upper and the lower part of the cell.
The two liquids come into contact in the middle of the cell, and the
solution they form flows through the slit D and is extracted from the
hole C. The porous rings, red in the drawing, make the flow more regular.}
\label{snfs_imm_cella_fjcc}
\end{figure}

\begin{figure}
\includegraphics[scale=1.0]{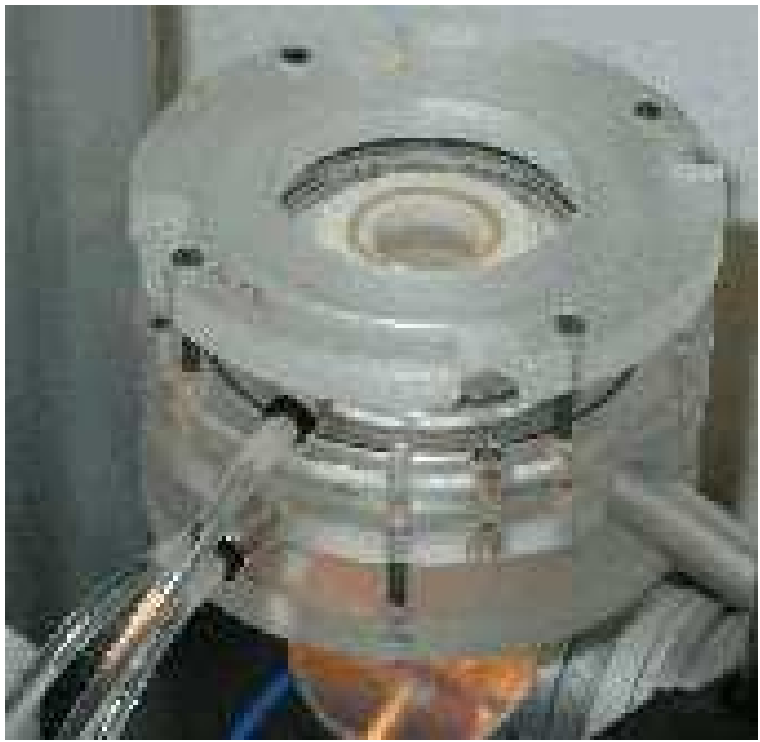}
\caption[Flowing Junction Cylindrical Cell.]
{A picture of the Flowing Junction Cylindrical Cell.}
\label{snfs_imm_cella_foto}
\end{figure}

The main difficulty is to fill the cell, keeping the interface between
the two liquids as regular as possible. We used a Flowing Junction
Cylindrical Cell (FJCC), a prototype developed for the study of
nonequilibrium fluctuations in microgravity \cite{croccolo_mth}.
From Eq. (\ref{snfs_eq_roll_off}) we see that the roll off depends on
the intensity of $g$, the gravitational acceleration. As $g$
decreases, gravity acts at increasingly shorter wavevectors, and the
divergence of fluctuations at small $q$ becomes more evident.
From Eq. (\ref{snfs_eq_altezza_plateau}) we see that the
intensity of the power spectrum for small wave vectors increases
linearily in $1/g$. This divergence of the intensity of fluctuations
on $g$ will be studied in an experiment performed on the Intarnational
Space Station. A drawing of the prototype cell is shown in
Fig. \ref{snfs_imm_cella_fjcc}; a picture can be seen in
Fig. \ref{snfs_imm_cella_foto}.
Two pipes feed the cell with the two liquids, in the present study
water and a solution of water and glycerol, with a small pressure. The
liquids enter in two ring-shaped chambers, from which they flow in
the cylindrical cell passing throug porous elements. The flow is quite
symmetric, due to the presence of the porous rings. The two liquids
fill the cell, water on the top and glycerol on the bottom; they come
into contact in the middle of the cell, and are pushed out the cell
through a circular slit. The outgoing liquid is collected in a third
chamber, passes through another porous ring, and is collected by a pipe.

The FJCC can be filled also in microgravity, since it is based on the
flow of liquids. However, gravity greatly simplifies this task: since
the denser fluid is in the lower part of the cell, 
big fluctuations, created by macroscopic motions, relax due to
buoyancy, while small fluctuations disappear quickly due to
diffusion.

\section{Results.}
\label{misura_snfs_sezione_risultati}

Although low angle light scattering techniques are very suitable to
study long range correlated fluids, their sensitivity is hampered by
the divergence of stry light at small wave vectors. The data presented
in \cite{brogioli2000,brogioli2000_2} were collected by using the
shadowgraph projection technique. Shadowgraphy has traditionally been
used to obtain a qualitative mapping of inhomogeneities in the index
of refraction. However, very recently the technique has been
reintroduced as a powerful quantitative tool to assess the features of
long wavelength fluctuations in fluids \cite{cannell1995,cannell1996}.

The main problem of shadowgraph is the oscillatory behaviour of its 
transfer function: see
Eq. (\ref{teoria_eq_transfer_function_shadowgraph}). The scattering
intensities around the zeroes of the transfer function cannot be
measured: for example, in Fig. 3 of \cite{brogioli2000}, the values of
$S\left(q\right)$ are missing for $q\approx 2\cdot 10^4\mathrm{m}^{-1}$.
Moreove, the region in which the zeroes are too frequent cannot be considered
in the data analysis. The overall wavevector range covered about one
decade. 

\begin{figure}
\begin{center}
\includegraphics{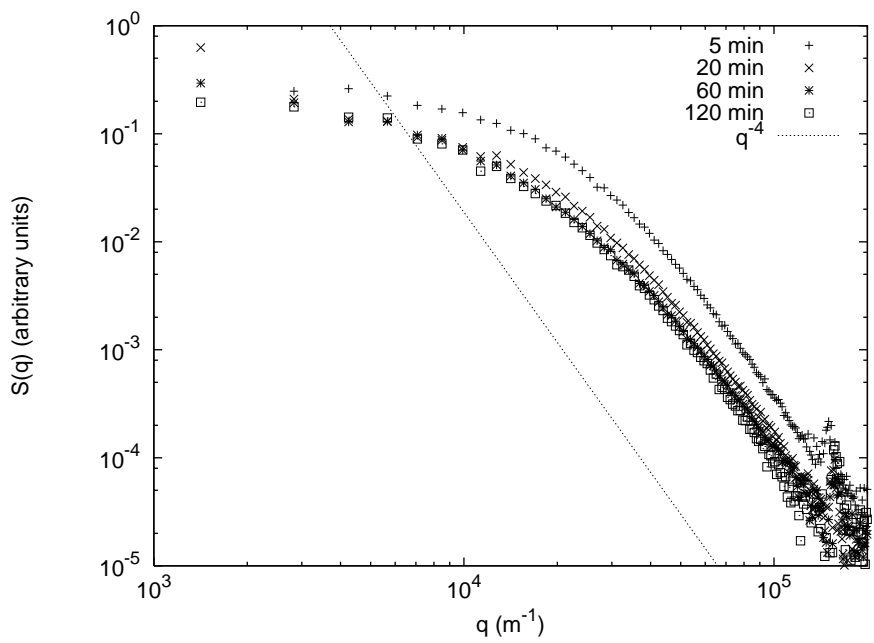}
\end{center}
\caption[Measured power spectrum of non equilibrium fluctuations.]
{Measured power spectrum of non equilibrium fluctuations in the
free diffusion process of glycerol in water.}
\label{snfs_imm_grafico_spettro}
\end{figure}

We measured the scatterig intensities at different times after the
beginning of the diffusion process. 
The power spectra measured with SNFS are shown in
Fig. \ref{snfs_imm_grafico_spettro}. They show the $q^{-4}$ divergence
and the saturation for small wavevectors. The roll off wavevector is
about $10^{4}\mathrm{m}^{-1}$, and is compatible with the value given
by Eq. (\ref{snfs_eq_roll_off})

The data cover about two decades in wavevectors: about ten times the
range covered with shadowgraph. The quality must be
compared with data shown in Fig. 2 of \cite{vailati1997}, obtained
with SALS on a similar, but quite peculiar system: the wavevector
range covered by SNFS is slightly more wide. Moreover, it should be
noticed that SALS gives no reliable results for the present system,
that is for the nonequilibrium fluctuations in the free diffusion of
simple fluids, due to stray light, since the scattering is weak and the
wavelength associated to the process is quite long. The results we
present are the best obtained for such a system up to now.

\appendix

%
\chapter{Three dimensional intensity correlation function.}
\label{capitolo_correlazione_tridimensionale}

The speckle field generated by a stocastic sample is formed by
speckles extending in both the orthogonal and parallel direction
with respect to the direction of propagation of the wave. The intensity
measured in a plane perpendicular to the direction of propagation 
varies as the plane is moved; small movements
of the plane will give small variations in the intensities. As a matter
of facts, the speckles appear and disappear as the plane is moved. This
allows us to speack of the three dimensional appearence of the speckles.
We will show that the speckles are elongated in the direction
of the propagation of the light. If the diamer is $\alpha$ times a wave
length $\lambda$, their length is $\alpha^2$ times $\lambda$.

In the following sections, we will show that the three dimensional
correlation function of the intensity of the scattered light gives 
more informations than the two dimensional one; in some cases it
is possible to determine the sign of the field correlation function,
thus determining it completely. Moreover, in analogy to the quadratic
relation between the diameter and length of a speckle, the longitudinal
frequencies should be related to the square root of the frequencies of
the sample: measuring the longitudinal correlations should double 
the dynamic of the system.

\section{Evolution equation of the field correlation.}

For $q\ll k$, Eq. (\ref{teoria_evoluzione_z_campo}) can be
approximated by:
%
%
\begin{equation}
E_z\left(\vec{q}\right) = E_0\left(\vec{q}\right)
e^{\displaystyle i k z}
e^{\displaystyle - i \frac{q^2}{2k} z} 
\end{equation}
In this approximation, neglecting the phase term $\exp\left(i k z\right)$,
the field follows a Schr\"oedinger equation:
%
%
\begin{equation}
\label{tree-d_corr_schroedinger_E}
i\frac{\partial}{\partial z} E\left(\vec{x},z\right) =
-\frac{1}{2k}\nabla^2 E\left(\vec{x},z\right)
\end{equation}
The three dimensional field correlation is defined as follows:
%
%
\begin{equation}
C_E\left(\Delta \vec{x},\Delta z\right) =
\frac{1}{S}\int_S{E\left(\vec{x},z\right) 
E\left(\vec{x}+\Delta \vec{x},z+\Delta z\right) \mathrm{d}\vec{x}\mathrm{d}z}
\end{equation}
In order to obtain an evolution equation for 
$C\left(\Delta \vec{x},\Delta z\right)$, as $\Delta z$ increases, we evaluate
the first derivative of the correlation function:
%
%
\begin{equation}
\frac{\partial}{\partial \Delta z} C_E\left(\Delta \vec{x},\Delta z\right) = 
\frac{1}{S}\int_S{E\left(\vec{x},z\right) \frac{\partial}{\partial \Delta z}
E\left(\vec{x}+\Delta\vec{x},z+\Delta z\right)\mathrm{d}\vec{x}\mathrm{d}z}
\end{equation}
Using eq. (\ref{tree-d_corr_schroedinger_E}):
%
%
\begin{equation}
\frac{\partial}{\partial \Delta z} C_E\left(\Delta \vec{x},\Delta z\right) = 
\frac{1}{S}\int_S{E\left(\vec{x},z\right) \frac{i}{2k}\nabla^2
E\left(\vec{x}+\Delta\vec{x},z+\Delta z\right)\mathrm{d}\vec{x}\mathrm{d}z}
\end{equation}
The operator $\nabla$ acts on the first argument of $E\left(\vec{x},z\right)$,
thus it can be considered as acting on $\Delta \vec{x}$:
%
%
\begin{equation}
\frac{\partial}{\partial \Delta z} C_E\left(\Delta \vec{x},\Delta z\right) =
\frac{i}{2k} \nabla^2_{\Delta \vec{x}}
\frac{1}{S}\int_S{E\left(\vec{x},z\right)
E\left(\vec{x}+\Delta\vec{x},z+\Delta z\right)\mathrm{d}\vec{x}\mathrm{d}z}
\end{equation}
This proves that the evolution equation for 
$\left(\Delta \vec{x},\Delta z\right)$, as $\Delta z$ increases,
is a Schr\"oedinger equation:
%
%
\begin{equation}
i\frac{\partial}{\partial\Delta z}C_E\left(\Delta\vec{x},\Delta z\right)=
-\frac{1}{2k} \nabla^2
C_E\left(\Delta \vec{x},\Delta z\right)
\end{equation}
This equation can easily be solved in Fourier space:
%
%
\begin{equation}
\label{tree-d_corr_soluzione_evoluzione_CE}
C_E\left(\vec{q},z\right)=C_E\left(\vec{q},z=0\right)
e^{\displaystyle -i\frac{q^2 z}{2k}}
\end{equation}

We can now extend eq. (\ref{teoria_eq_siegert_2d})
to the three dimensional case:
%
%
\begin{equation}
\label{tree-d_corr_siegert}
C_{I}\left(\Delta \vec{x},\Delta z\right) = \left<
I\left(\vec{x},z\right) I\left(\vec{x}+\Delta \vec{x},z+\Delta z\right)
\right> = \left< I \right>^2 + \left| 
C_E\left(\Delta \vec{x},\Delta z\right)
\right|^2
\end{equation}

\section{Gaussian speckles.}

In this section we consider gaussian speckles, and
we evaluate their three dimensional correlation function. 

Far field speckles are often generated by scattering a gaussian beam,
so that the far field speckles have a gaussian correlation function.
We consider gaussian speckles in near field, since the case is 
analitically solvable, and involves some calculations
used in quantum mechanics.

The field correlation function of the scattered light, in the plane
orthogonal to $z$, is gaussian:
%
%
\begin{equation}
C_E\left(\Delta \vec{x},\Delta z=0\right)= C e^{\displaystyle
-\frac{\Delta\vec{x}^2}{2\sigma^2}}.
\end{equation}
In the Fourier space:
%
%
\begin{equation}
C_E\left(\vec{q},\Delta z=0\right)= 2\pi\sigma^2C e^{\displaystyle
-\frac{1}{2}\sigma^2q^2}.
\end{equation}
Using eq. (\ref{tree-d_corr_soluzione_evoluzione_CE}):
%
%
\begin{equation}
C_E\left(\vec{q},z\right)= 2\pi\sigma^2C e^{\displaystyle
-\frac{1}{2}\sigma^2q^2 -i\frac{q^2z}{2k}}.
\end{equation}
Coming back to real space:
%
%
\begin{equation}
C_E\left(\vec{x},z\right)= C\frac{\sigma^2} {\sigma^2+iz/k}
e^{\displaystyle -\frac{x^2}{2\left(\sigma^2+iz/k\right)}}.
\end{equation}
Now we evaluate the modulus of the field correlation function,
the quantity needed in eq. (\ref{tree-d_corr_siegert}) to determine
the intensity correlation function:
%
%
\begin{equation}
\left|C_E\left(\vec{x},z\right)\right|^2= C^2
\frac{\sigma^4} {\sigma^4+z^2/k^2}
e^{\displaystyle -\frac{x^2\sigma^2}{\sigma^4+z^2/k^2}}.
\end{equation}

We can now evaluate the intensity correlation function for $\vec{x}=0$:
%
%
\begin{equation}
C_I\left(\vec{x}=0,z\right)= C^2 \left(1+
\frac{\sigma^4} {\sigma^4+z^2/k^2}\right),
\end{equation}
and for $z=0$:
%
%
\begin{equation}
C_I\left(\vec{x},z=0\right)= C^2 \left(1+
e^{\displaystyle \frac{x^2}{\sigma^2}}\right).
\end{equation}

While the transverse correlation function follows a gaussian
law, the longitudinal one is a Lorentzian, The diameter of the
speckles is about $\sigma$, while their length is $\sigma^2k$.

\section{Determination of the sign of the field correlation function.}

The power spectrum, that is $C_E\left(\vec{q},z=0\right)$, is real.
If the sample is isotropic, it is symetric with
respect to the origin, and then the correlation function 
$C_E\left(\vec{x},z=0\right)$ is real. The knowledge of the intensity
correlation function with $\Delta z=0$ gives the absolute value of the field
correlation function. The sign of the field correlation function does
not affect the intensity correlation function with $\Delta z=0$, but it
can affect its value for $\Delta z\ne0$. 

\begin{figure}
\includegraphics[angle=-90]{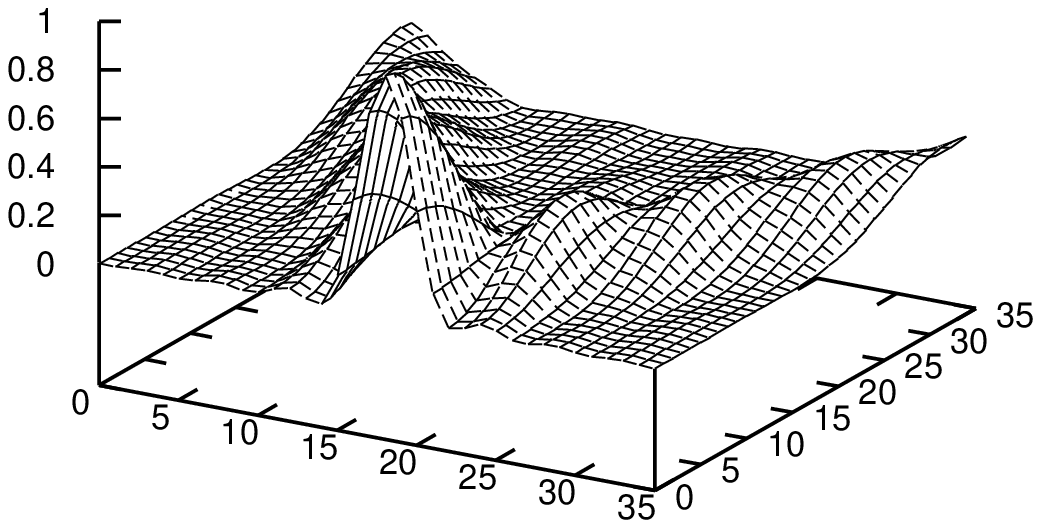}
\caption[Effect of the phase.]{Comparison between two square correlation functions. 
$C_E\left(x,\Delta z=0\right)=\sin\left(x\right)/x$}
\label{corr_3d_visib_sinc}
\end{figure}

\begin{figure}
\includegraphics[angle=-90]{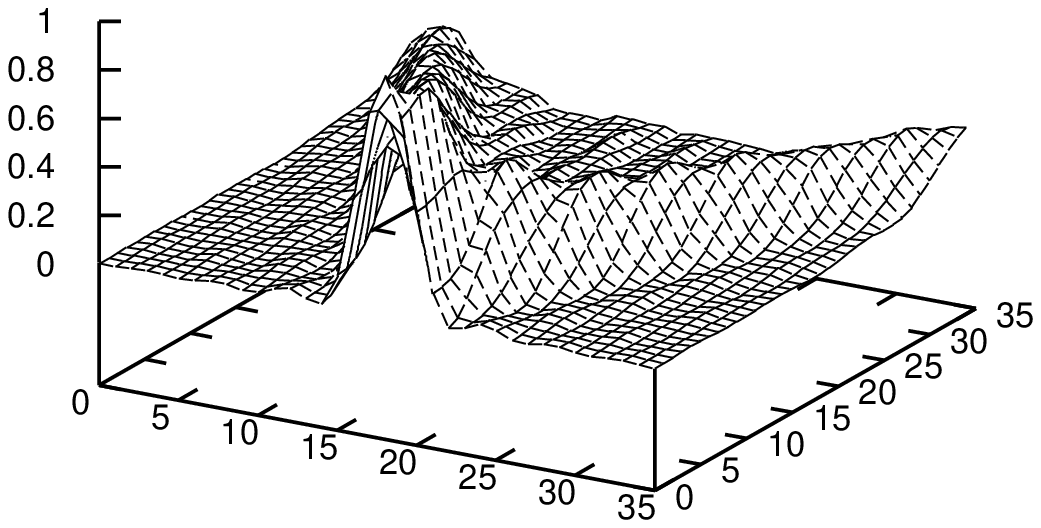}
\caption[Effect of the phase.]{Comparison between two square correlation functions. 
$C_E\left(x,\Delta z=0\right)=\left|\sin\left(x\right)/x\right|$}
\label{corr_3d_visib_mod_sinc}
\end{figure}

In figure \ref{corr_3d_visib_sinc} and \ref{corr_3d_visib_mod_sinc} we see
an example of this effect. The figures show the graphs of 
the square correlation functions. The first is such that 
$C_E\left(x,\Delta z=0\right)=\sin\left(x\right)/x$;
in the second, the correlation function has the
same absolute value, but alwais positive sign, for $\Delta z=0$.
For $\Delta z=0$ the square correlation functions are equal; their
evolution for other values of $\Delta z$ are different. We can explain
this fact considering the evolution of the positive and negative parts of the 
correlation function. The two parts evolve, and overlap, as $\Delta z$ 
increases. The interference of the two parts depends on the initial phase.
%
%
%
%

The sign of the correlation function is alwais possible, in principle. The presence
of errors could limit this possibility.

\section{Longitudinal correlation.}

We want to derive the field correlation along the $z$ axis.
We consider its Fourier transform:
%
%
\begin{equation}
C_E\left(\Delta \vec{x}=0,q_z\right)=
\frac{1}{\left(2 \pi\right)^2}\int{
C_E\left(\vec{q},z\right)e^{\displaystyle -iq_zz}
\mathrm{d}\vec{q}\mathrm{d}z}
\end{equation}
Using eq. (\ref{tree-d_corr_soluzione_evoluzione_CE}):
%
%
\begin{equation}
C_E\left(\Delta \vec{x}=0,q_z\right)=
\frac{1}{\left(2 \pi\right)^2}\int{
C_E\left(\vec{q},z=0\right)e^{\displaystyle 
-i\frac{q^2z}{2k} -iq_zz} \mathrm{d}\vec{q}\mathrm{d}z}
\end{equation}
The integration over $z$ gives a Dirac delta:
%
%
\begin{equation}
C_E\left(\Delta \vec{x}=0,q_z\right)=
\int{C_E\left(\vec{q},z=0\right)\delta\left(
\frac{q^2}{2k} + iq_z\right) \mathrm{d}\vec{q}}
\end{equation}
In radial coordinates:
%
%
\begin{equation}
C_E\left(\Delta \vec{x}=0,q_z\right)=
\int{C_E\left(q,\varphi,z=0\right)q\delta\left(
\frac{q^2}{2k} + iq_z\right) \mathrm{d}q\mathrm{d}\varphi}
\end{equation}
If the sample is isotropic:
%
%
\begin{equation}
C_E\left(\Delta \vec{x}=0,q_z\right)=2\pi
\int{C_E\left(q,z=0\right)q\delta\left(
\frac{q^2}{2k} + iq_z\right) \mathrm{d}q}
\end{equation}
The integral can be evaluated:
%
%
\begin{equation}
C_E\left(\Delta \vec{x}=0,q_z\right)=2\pi
C_E\left(\sqrt{2kq_z},z=0\right)\sqrt{2kq_z}
\end{equation}

The dynamic of an instrument measuring the longitudinal
correlation function is twice that obtained with
the transversal one. This facts closely mirrors
the quadratic relation between the diameter and the length of
the speckles.

%
\chapter{Definitions}

Fourier transform:
%
\begin{equation}
f\left(q\right) = \int
{f\left(x\right) e^{\displaystyle - i q x} \mathrm{d}x}
\end{equation}

Inverse Fourier transform:
%
\begin{equation}
f\left(x\right) = \frac{1}{\left(2\pi\right)^n} \int
{f\left(q\right) e^{\displaystyle i q x} \mathrm{d}q}
\end{equation}

Convolution:
%
\begin{equation}
\begin{array}{l}
g\left(x\right)=\left|f\left(x\right)\right|^2\\
g\left(q\right)=\frac{1}{\left(2\pi\right)^n}
\int
{f\left(q'\right)
f^*\left(q'-q\right)
\mathrm{d}q'}
\end{array}
\end{equation}

NFS Near Field Scattering

ONFS hOmodyne Near Field Scattering

ENFS hEterodyne Near Field Scattering

SNFS Schlieren-like Near Field Scattering

LS Light Scattering

SALS Small Angle Light Scattering

IFS Intensity Fluctuation Spectroscopy

CCD Charge Coupled Device


\bibliography{biblio}

\tableofcontents

\listoffigures

\listoftables

\end{document}